\documentclass[acmsmall,screen]{acmart}
\AtBeginDocument{%
  }

\setcopyright{acmlicensed}
\copyrightyear{2026}
\acmYear{2026}
\acmDOI{XXXXXXX.XXXXXXX}
\acmISBN{978-1-4503-XXXX-X/2026/06}

\usepackage[table]{xcolor}
\usepackage{xspace}
\usepackage{enumitem}
\usepackage{multirow}
\usepackage{multicol}
\usepackage{listings}
\usepackage{xcolor}
\usepackage{makecell}
\usepackage{siunitx}
\usepackage{subcaption}
\usepackage{colortbl}
\usepackage{hyperref}
\usepackage{amsmath}
\usepackage{listings}
\usepackage{backnaur}
\usepackage{algorithm}
\usepackage{algpseudocode}

\definecolor{headercolor}{RGB}{245,245,245}
\newcommand{\tool}{ARC\xspace}

\newcommand{\STATE}{\State}
\newcommand{\REQUIRE}{\Require}
\newcommand{\ENSURE}{\Ensure}
\newcommand{\IF}[1]{\If{#1}}
\newcommand{\ENDIF}{\EndIf}

\newcommand{\WHILE}[1]{\While{#1}}
\newcommand{\ENDWHILE}{\EndWhile}
\newcommand{\FORALL}[1]{\ForAll{#1}}
\newcommand{\ENDFOR}{\EndFor}
\newcommand{\PROCEDURE}[2]{\Procedure{#1}{#2}}
\newcommand{\ENDPROCEDURE}{\EndProcedure}
\renewcommand{\algorithmiccomment}[1]{\hfill $\triangleright$ #1}
\newcommand{\COMMENT}[1]{\algorithmiccomment{#1}}
\newcommand{\RETURN}{\STATE \textbf{return}}

\begin{document}

%%
%% The "title" command has an optional parameter,
%% allowing the author to define a "short title" to be used in page headers.
\title{Compiling Large Multi-Modal Requirement Documents into Runnable Software Systems: From an Agentic Test-Driven Perspective}

%%
%% The "author" command and its associated commands are used to define
%% the authors and their affiliations.
%% Of note is the shared affiliation of the first two authors, and the
%% "authornote" and "authornotemark" commands
%% used to denote shared contribution to the research.
\author{Weiyu Kong}
\email{kwy160034@sjtu.edu.cn}
\affiliation{%
  \institution{Shanghai Jiao Tong University}
  \city{Shanghai}
  \country{China}
}

\author{Yun Lin}
\authornote{Corresponding author.}
\email{lin_yun@sjtu.edu.cn}
\affiliation{%
  \institution{Shanghai Jiao Tong University}
  \city{Shanghai}
  \country{China}
}

\author{Xiwen Teoh}
\email{xiwen.teoh@u.nus.edu}
\affiliation{%
  \institution{National University of Singapore}
  \city{Singapore}
  \country{Singapore}
}

\author{Duc-Minh Nguyen}
\email{minh.nguyen@sjtu.edu.cn}
\affiliation{%
  \institution{Shanghai Jiao Tong University}
  \city{Shanghai}
  \country{China}
}

\author{Ruofei Ren}
\email{renruofei0120@sjtu.edu.cn}
\affiliation{%
  \institution{Shanghai Jiao Tong University}
  \city{Shanghai}
  \country{China}
}

\author{Jiaxin Chang}
\email{cjx001234@sjtu.edu.cn}
\affiliation{%
  \institution{Shanghai Jiao Tong University}
  \city{Shanghai}
  \country{China}
}

\author{Haoxu Hu}
\email{andyHu42@sjtu.edu.cn}
\affiliation{%
  \institution{Shanghai Jiao Tong University}
  \city{Shanghai}
  \country{China}
}

\author{Haoyu Chen}
\email{chy-sunday@sjtu.edu.cn}
\affiliation{%
  \institution{Shanghai Jiao Tong University}
  \city{Shanghai}
  \country{China}
}

%%
%% By default, the full list of authors will be used in the page
%% headers. Often, this list is too long, and will overlap
%% other information printed in the page headers. This command allows
%% the author to define a more concise list
%% of authors' names for this purpose.
\renewcommand{\shortauthors}{Kong et al.}

%%
%% The abstract is a short summary of the work to be presented in the
%% article.
\begin{abstract}
    Large Language Models (LLMs) have significantly improved programming efficiency by parsing natural language into code snippets.
    However, their performance degrades significantly as requirements scale; when faced with multi-modal documents containing hundreds of scenarios, LLMs often produce incorrect implementations or omit crucial constraints.
    Observing LLMs' ever-evolving capability and their persistent stochastic hallucination, 
    we raise a question, i.e., whether it is possible to make LLM-based agentic programming go beyond \textit{code generation} to \textit{requirement compilation}, i.e., whether programmers can produce a runnable system by only accomplishing (non-trivial) requirement documents?

    In this work, we take a first step by proposing the \tool (Agentic Requirement Compilation) technique to parse a multi-modal requirement document, describing hundreds of scenarios in a DSL format, into a runnable web system.
    In addition to the source code, \tool also generates software engineering artifacts including
    (1) a modular design that spans the user interface, API interface, and database,
    (2) enriched test cases for each interface (including unit tests, modular tests, and integration tests), and
    (3) detailed traceability across all artifacts for software maintenance. 
    Our approach employs a bidirectional test-driven agentic loop: (1) a \textbf{top-down architecture phase} that decomposes requirements into UI, API, and database interfaces, each of which is equipped with verifiable test suites, and (2) a \textbf{bottom-up implementation phase} where agents generate code that must satisfy the generated tests. 
    Throughout this process, ARC maintains strict traceability across requirements, design, and code to facilitate intelligent asset reuse and follow-up maintenance.
    We evaluate \tool on two complementary benchmarks, i.e., a depth-oriented benchmark of 6 runnable web systems spanning 50--200 requirement scenarios, and the breadth-oriented AppForge benchmark comprising 101 Android app generation tasks.
    Across 3 independent trials, \tool outperforms all state-of-the-art LLM-based baselines, with the generated web systems passing on average 50.6\% more GUI tests, and achieving 100\% compile success and 68.3\% test case pass rate on AppForge.
    In addition, a user study with 21 participants shows that participants with limited programming experience successfully write DSL-based documents consisting of 50 to 174 scenarios, within 5.6 hours on average,
    to generate a runnable system such as a real-world ticket-booking system of around 10K lines of code with maintainable architecture.
\end{abstract}

%%
%% The code below is generated by the tool at http://dl.acm.org/ccs.cfm.
%% Please copy and paste the code instead of the example below.
%%


%%
%% Keywords. The author(s) should pick words that accurately describe
%% the work being presented. Separate the keywords with commas.
%% A "teaser" image appears between the author and affiliation
%% information and the body of the document, and typically spans the
%% page.

\received{29 Jan 2026}
\received[revised]{22 May 2026}
\received[accepted]{}

%%
%% This command processes the author and affiliation and title
%% information and builds the first part of the formatted document.

% ================
% revision-config.tex
% Master toggle: set showrevision to true/false
% true  → color-coded diff enabled, revision index shown
% false → all coloring disabled, revision index hidden
\newtoggle{showrevision}
\togglefalse{showrevision}
% ================

% Add a new toggle for the in-text highlights
\newtoggle{showhighlights}
\togglefalse{showhighlights}

% Toggle color-coded diff
\newtoggle{color_metareview}
\newtoggle{color_review}
\iftoggle{showhighlights}{   % ← was showrevision
  \toggletrue{color_metareview}
  \toggletrue{color_review}
}{
  \togglefalse{color_metareview}
  \togglefalse{color_review}
}

\newtoggle{color_done}
\togglefalse{color_done}

\newcommand{\metareview}[1]{%
  \iftoggle{color_metareview}{\textcolor{blue}{#1}}{#1}%
}

\newcommand{\review}[1]{%
  \iftoggle{color_review}{\textcolor{purple}{#1}}{#1}%
}

% Revision reference
\newcommand{\metareviewref}[3][undone]{%
  \iftoggle{color_metareview}{%
    \hyperref[revision:#2]{\textcolor{blue}{\textbf{M#2:}}}
    \ifthenelse{\equal{#1}{done}}{%
      \iftoggle{color_done}{\textcolor{green!60!black}{#3}}{#3}%
    }{#3}%
  }{}%
}

% Revision label
\newcommand{\metareviewlabel}[1]{%
  \phantomsection\label{revision:#1}%
  \iftoggle{color_metareview}{%
    \textcolor{blue}{\textbf{(M#1)}~}%
  }{}%
}

% Others reference
\newcommand{\reviewref}[3][undone]{%
  \iftoggle{color_review}{%
    \hyperref[others:#2]{\textcolor{purple}{\textbf{R#2:}}}
    \ifthenelse{\equal{#1}{done}}{%
      \iftoggle{color_done}{\textcolor{green!60!black}{#3}}{#3}%
    }{#3}%
  }{}%
}

% Others label
\newcommand{\reviewlabel}[1]{%
  \phantomsection\label{others:#1}%
  \iftoggle{color_review}{%
    \textcolor{purple}{\textbf{(R#1)}~}%
  }{}%
}

\robustify{\metareviewlabel}
\robustify{\reviewlabel}
\iftoggle{showrevision}{%
    \pagestyle{empty}

{\setlength{\parskip}{0.4em}

\section*{Major Revision Change Index}

\subsection*{Meta Review}

\begin{itemize}[leftmargin=*,itemsep=1em, topsep=0.2em]
    \item \metareviewref{1}{Since this is an approach-oriented paper, there must be some argument (some experiments) that demonstrate generalization across diverse, existing benchmarks is essential. Hence we require that the authors conduct more comprehensive evaluations on well-established datasets and benchmarks using widely accepted metrics.}
    \begin{itemize}[leftmargin=*,itemsep=1em, topsep=1em]
        \item \metareviewref[done]{1.1}{Inclusion of AppForge~\cite{appforge} as an additional benchmark, with justification of its complementary evaluation dimensions relative to our primary benchmark.}
        \item \metareviewref[done]{1.2}{Description of how \tool and all baselines were adapted to 
        AppForge for a fair and controlled comparison across methods.}
        \item \metareviewref[done]{1.3}{Description of what implementation stack was used on AppForge.}
        \item \metareviewref[done]{1.4}{Adoption of widely accepted evaluation metrics from
        AppForge, specifically Compile Succ., Test Suite Pass,
        Test Case Pass, and Final Pass, alongside Test Pass Rate for comparison.}
        \item \metareviewref[done]{1.5}{Extend RQ1 (Effectiveness) to include comparison against baselines on AppForge.}
        \item \metareviewref[done]{1.6}{Extend RQ2 (Overhead Analysis) to include comparison against baselines on AppForge.}
        \item \metareviewref[done]{1.7}{Extend RQ3 (Ablation Study) to include performance of \tool with and without enhanced requirements on AppForge.}
    \end{itemize}
    \item \metareviewref[done]{2}{Also, since you are evaluating new benchmarks, more care should be taken in the paper to argue for the reasonableness of those benchmarks.}
    \item \metareviewref[done]{3}{The revision must clearly describe the process followed to create the implementations using the baseline approaches.}
    \item \metareviewref[done]{4}{The revision must discuss the threat to validity of relying on students to create artifacts, and how this threat was mitigated.} 
\end{itemize}

\subsection*{Reviewers' Comments}

\subsubsection*{Reviewer A}
\begin{itemize}[leftmargin=*,itemsep=1em, topsep=0.2em]
    \item \reviewref[done]{A-1}{How easy would it be for an existing requirements document to be transformed into the DSL format?}
    \item \reviewref[done]{A-2}{What language is the system generated in? What format (i.e., testing frameworks) are
the generated tests in?}
    \item \reviewref[done]{A-3}{How were the requirement documents created? Was there validation by non-students?}
    \item \reviewref[done]{A-4}{What is the risk of using graduate students to create the requirement documents? How
much professional experience did they have?}
    \item \reviewref[done]{A-5}{Add more information about the process for creating the GUI test suites. Who created
them? How were they created?}
    \item \reviewref[done]{A-6}{What LLM was used in ARC? What settings were used for the LLM?}
    \item \reviewref[done]{A-7}{How were the baselines used to create implementations? It is not clear what input was
provided or how they were employed.}
    \item \reviewref[done]{A-8}{Was only a single trial performed? Discussion of non-determinism is needed.}
    \begin{itemize}[leftmargin=*,itemsep=1em, topsep=1em]
        \item \reviewref[done]{A-8.1}{Revise experiments to have multiple independent trials.}
        \item \reviewref[done]{A-8.2}{Discussion of outcome of independent trials.}
        \item \reviewref[done]{A-8.3}{Discussion of non-determinism in threats to validity.}
    \end{itemize}
    \item \reviewref[done]{A-9}{The user study only had undergraduate students. It would have been nice to also have
practitioners.}
    \item \reviewref[done]{A-10}{Reliance on commercial LLMs is a threat to validity, as models can be replaced without
disclosure.}
\end{itemize}

\subsubsection*{Reviewer B}
\begin{itemize}[leftmargin=*,itemsep=1em, topsep=0.2em]
    \item \reviewref[done]{B-1}{The evaluation is conducted on only six tasks. How do the authors ensure the results
are statistically significant and representative of broader scenarios?}
    \item \reviewref[done]{B-2}{How exactly are the evaluation test cases generated? Are they produced by ARC itself,
or independently constructed? If the former, how to avoid evaluation bias?}
    \item \reviewref[done]{B-3}{Why were no established benchmarks used for evaluation? Would the approach generalize to other benchmarks such as AppForge?}
    \item \reviewref[done]{B-4}{What are the exact LLM backends and decoding settings used for ARC and each baseline?}
    \item \reviewref[done]{B-5}{Were all methods given comparable computational budgets? If not, how might additional
resources affect baseline performance?}
    \item \reviewref[done]{B-6}{Why was Test Pass Rate chosen instead of a more conventional correctness metric such
as problem-level pass rate?}
    \item \reviewref[done]{B-7}{The replication package could be better documented. The dataset could not be down-
loaded.}
\end{itemize}

\subsubsection*{Reviewer C}
\begin{itemize}[leftmargin=*,itemsep=1em, topsep=0.2em]
    \item \reviewref[done]{C-1}{Can you clarify the resources required for each part of ARC?}
    \item \reviewref[done]{C-2}{What is the ground truth of those tests? Are those tests “right”?}
\end{itemize}
}

% end this page, start fresh
\clearpage
\pagestyle{standardpagestyle}
\setcounter{page}{1}%
}{}%

\maketitle

\section{Introduction}
\label{sec:intro}
% \highlight{TODO: Thorough language polishing pass -- (1) reduce overly strong adjectives (rigorous, reliable, novel, etc.); (2) change method section from present tense to past tense; (3) fix all grammatical issues listed by Reviewer A}{(C15)}

The software engineering landscape has entered a transformative era where Large Language Models (LLMs) have evolved from simple autocompletion assistants into proactive coding partners capable of planning and executing complex development workflows \cite{xi2025rise, shethiya2024engineering}. 
The integration of these models into the software development life cycle has broadened access to code creation, shifting the developer's role from a low-level code producer to a high-level curator and architect \cite{nguyen2022empirical, terragni2024futuresoftwareengineeringaidriven}. 
However, while LLMs excel at generating localized code snippets, their performance faces a ``complexity wall'' when tasked with large-scale projects. 
As requirement documents transition from simple text to multi-modal specifications spanning hundreds of scenarios, 
traditional agentic generative approaches frequently suffer from requirement drift and omission and persistent hallucinations \cite{liu2024lost, hou2024large, jiang2024survey}.
This widening gap between intent and implementation raises a fundamental question in the age of agentic programming, i.e., 
\textit{how far are we from shifting the paradigm from mere code generation to a more rigorous process of requirement compilation?}

The research community has primarily explored the following paradigms to address large-scale code generation from textual descriptions \cite{jiang2024survey}:
\begin{itemize}[leftmargin=*]
    \item \textbf{Role-Based Orchestration:} These frameworks \cite{hong2023metagpt, qian2024chatdev} such as MetaGPT simulate a traditional software company by assigning specialized roles (e.g., Product Manager, Architect, Programmer) to different agents. By following Standard Operating Procedures (SOPs), they decompose high-level requirements into intermediate artifacts like PRDs and system designs before implementation. While their waterfall approaches \cite{hong2023metagpt} can output a complete repository structure, they remain procedural rather than verifiable, i.e., a misinterpretation in the initial PRD generates a structurally sound but functionally broken repository.
    \item \textbf{Reactive Action-Loops (OpenHands \cite{wang2024openhands} / SWE-agent \cite{yang2024swe}):} These frameworks focus on tight environment interaction. OpenHands, in particular, utilizes an event-driven architecture where a centralized ``EventStream'' coordinates interactions between the user, the agent, and a secure docker-based runtime. By this means, these tools enable the agents to autonomously search complex directory structures, execute shell commands, and iteratively refine code based on real-time feedback from compilers or test execution results.
\end{itemize}

Despite these sophisticated structures, generating a runnable system from a multi-modal document remains difficult for the following reasons.
First, most existing tools treat requirements as natural language prompts. However, natural language is inherently ambiguous and lacks the explicit hierarchy needed for repository-scale dependency tracking. MetaGPT’s waterfall approach \cite{hong2023metagpt} is procedural but less trustworthy in practice. For example, if a "Product Manager" agent misinterprets a requirement node, that error propagates into the architecture and code without an early-stage gate to stop the cascade.
Second, reactive agents (e.g., OpenHands \cite{wang2024openhands}) are excellent at ``fixing'' but poor at ``compiling''. They often struggle with a blank-slate repository because they cannot proactively build a global data contract (e.g., among UI, API, and database) across hundreds of scenarios. This results in fragmented implementations where individual files are created but fail to integrate into a cohesive, runnable system.

To bridge this gap, we propose \tool (Agentic Requirement Compilation), a framework that shifts the paradigm from stochastic code generation to a rigorous process of requirement compilation. Unlike existing ``role-play'' or ``reactive'' systems, \tool treats multi-modal requirement documents as a high-level source language, structured via a light-weight, graph-based Domain Specific Language (DSL). By transforming dense specifications into a formal graph of requirement scenarios, we provide a deterministic blueprint which reduces requirement drift and implementation omission typical of natural language prompts. This structured foundation enables a novel bidirectional, test-driven agentic loop, 
i.e., 
\begin{itemize}[leftmargin=*]
    \item \textbf{Testable Architecture Construction:} A top-down phase that first decomposes the graph into verifiable UI, API, and database interfaces, accompanied by reliable tests for each interface.
    As a result, the architecture is modularized as a set of interactable interfaces, each of which is equipped with a set of tests in the form of unit tests, component tests, and integration tests.
    \item \textbf{Constrained Code Generation:} A bottom-up implementation phase where agents generate code that must satisfy these predefined test suites as constraints.
    Specifically, \tool implements the architecture by satisfying low-level unit tests first, then moves to satisfy component tests and integration tests in the higher level.
\end{itemize}

Generally, this dual-phase approach simulates the workflow of the classical V-model \cite{vmodel} in a test-driven manner,
which solves the ``complexity wall'' by enforcing a continuous feedback loop between design and code. By prioritizing the generation of interface tests before application logic, \tool creates a grounding mechanism that ``gates'' agent creative output, ensuring every generated module is strictly verified against its functional requirement. Furthermore, \tool maintains an explicit \textit{traceability record} across the requirements, designs, and tests, allowing the system to intelligently reuse existing assets and maintain global consistency even as project scale increases. This systematic alignment ensures that the final output is not just a collection of code snippets, but a cohesive web system that is consistent with the original multi-modal document.
Finally, the maintained traceability record can also facilitate software maintainability. 

We evaluated \tool by compiling 6 complex web systems from multi-modal requirement documents containing 50 to 200 distinct scenarios. \metareviewlabel{1}\metareview{To further assess generalizability beyond web systems, we additionally evaluated \tool on the AppForge benchmark \cite{appforge}, which comprises 101 Android app generation tasks.} We discuss the selection rationale in Section~\ref{sec:benchmark}. Our experimental results demonstrate that by replacing the stochastic one-shot generation of vanilla LLMs and the rigid waterfall roles of frameworks like MetaGPT with our bidirectional compilation loop, we enhance system reliability. Specifically, \tool outperforms state-of-the-art baselines by generating repositories that pass on average 50.6\% in GUI tests and successfully eliminating the fragmented implementation errors common in existing agentic workflows.
Furthermore, to assess the accessibility of our lightweight DSL, we conducted a user study involving 21 participants, including novice users with limited programming experience.  The study demonstrates that the structured, graph-based DSL serves as an effective medium for human-AI collaboration. Participants successfully specify an average of 107 scenarios per project within 5.6 hours, and existing requirement documents could be transformed into the DSL format with LLM assistance in under 10 minutes. %(from 50 to 174 scenarios). 
Finally, our contributions are as follows:
\begin{itemize}[leftmargin=*]
    \item \textbf{A Test-Driven Requirement Compilation Paradigm: } We introduce a novel agentic software engineering paradigm that shifts from stochastic code generation to rigorous requirement compilation. This approach leverages a lightweight, graph-based DSL to provide formal grounding and a bidirectional, test-driven loop that ensures high reliability. By maintaining an explicit traceability record across requirements, design, and implementation, our paradigm significantly enhances the long-term maintainability of AI-generated systems.
    \item \textbf{The \tool Framework:} We design and implement \tool, an agentic framework capable of transforming multi-modal documents into fully runnable, repository-scale web systems. The framework features a unique top-down architecture construction phase and a bottom-up implementation generation phase, creating a reliable feedback loop that eliminates requirement drift and omission.
    \item \textbf{Extensive Evaluation: } We conduct a comprehensive evaluation of \tool by generating 6 real-world web systems and additionally assessing its generalizability on the AppForge benchmark with 101 Android app generation tasks. Our results demonstrate significant improvements in runnability and GUI test pass rates (up to 108.1\%) compared to state-of-the-art agentic baselines. Furthermore, a user study with 21 participants confirms that the framework is accessible to novice users, enabling them to successfully compile production-grade repositories from scratch.
\end{itemize}

Given the space limit,  all the video, tool, and experiment details are available at \cite{arc}.

\section{Preliminaries}

\begin{figure}[htbp]
    \centering
    \includegraphics[width=\linewidth, trim=0 0 0 0, clip]{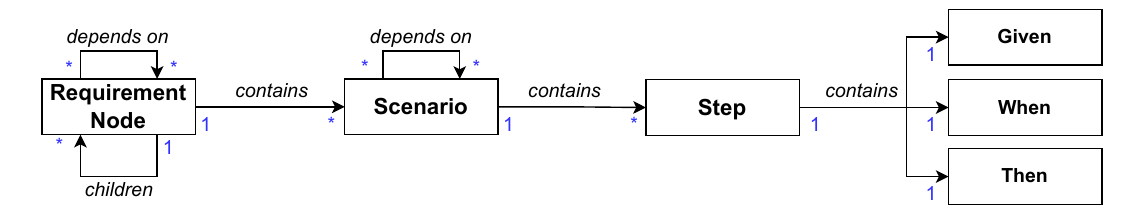}
    \caption{The meta-model (or schema) of the multi-modal requirement of \tool. Each requirement node is equipped with a multi-modal description (via text and picture). Each requirement node and scenario is assigned with an identifier for reference. Each step is described with action and expectation via \textit{Given}, \textit{When}, and \textit{Then} keywords.}
    \label{fig:requirement-metamodel}
\end{figure}

% To shift from stochastic code generation to a rigorous compilation process, we first formalize the source of truth: the multi-modal requirement document. This section defines our graph-based Domain Specific Language (DSL) and the underlying meta-model that enables deterministic system construction.

\subsection{Formal Definition and Meta-Model of Requirement}
We define a requirement document as a directed acyclic graph (DAG) of Requirement Nodes. As illustrated in our meta-model (see \autoref{fig:requirement-metamodel}), each node acts as a container for full-stack specifications, spanning from high-level context at the root to specific UI, API, and DB constraints at the leaves.
The core structure of our DSL is built upon three primary entities: 

\begin{itemize}[leftmargin=*]
    \item \textbf{Requirement Node}: Each node represents a unique feature. It explicitly defines multi-modal description, \emph{Dependencies} (prerequisite requirement) and \emph{Children} (hierarchical sub-features), ensuring the framework can track repository-scale impacts.
    \item \textbf{Scenario:} Each requirement node contains one or more workflows that represent end-to-end user journeys.
    One scenario can specify a set of prerequisite scenarios, indicating its dependence.
    \item \textbf{Step}: A scenario is decomposed into an ordered sequence of atomic steps expressed in a Gherkin-like manner. Each step captures the test context (\textit{Given}), the action under test (\textit{When}), and the expected, verifiable outcome (\textit{Then}), which are used to generate test cases.
\end{itemize}

The full formal grammar of our DSL is specified as follows:
\vspace{-5pt}
\[
\small
\boxed{
\begin{aligned}
\text{RequirementDoc} & ::= \text{Node} \\
\text{Node}           & ::= \mathbf{id:\ } \text{ID} \\
                      & \quad \mathbf{name:\ } \text{String} \\
                      & \quad \mathbf{description:\ } \text{MultiModalText} \\
                      & \quad \mathbf{dependencies:\ [} [\text{ID} \{ , \text{ID} \}] \mathbf{]} \\
                      & \quad \mathbf{scenarios:\ [} [\text{Scenario} \{ , \text{Scenario} \}] \mathbf{]} \\
                      & \quad \mathbf{children:\ [} [\text{Node} \{ , \text{Node} \}] \mathbf{]} \\
\\
\text{Scenario}       & ::= \mathbf{id:\ } \text{ID} \\
                      & \quad \mathbf{name:\ } \text{String} \\
                      & \quad \mathbf{prerequisites:\ [} [\text{ID} \{ , \text{ID} \}] \mathbf{]} \\
                      & \quad \mathbf{steps:\ [} [\text{Step} \{ , \text{Step} \}] \mathbf{]} \\
\\
\text{Step}           & ::= \mathbf{given:\ } \text{String} \quad \mathbf{when:\ } \text{String} \quad \mathbf{then:\ } \text{String} \\
\\
\text{MultiModalText} & ::= \{ \text{Text} \mid \text{ImageTag} \} \\
\text{ImageTag}       & ::= \mathbf{![image](} \text{Path} \mathbf{)}
\end{aligned}
}
\]
The design of this DSL is not merely for organization. 
It is the fundamental grounding mechanism that allows \tool to overcome the complexity wall of large-scale systems.
\begin{itemize}[leftmargin=*]
    \item \textbf{Deterministic Blueprinting}: By converting ambiguous natural language into a dependency-aware graph, we provide the agentic loop with a deterministic path for top-down Architecture Construction (see Section~\ref{sec:top-down}).
    \item \textbf{Test-Driven Grounding}: Every step requires a paired action and expectation; the DSL serves as a direct source for generating GUI and API test suites. These expectations act as the ``gates'' that prevent the requirement drift and hallucinations common in existing models.
    \item \textbf{Traceability}: The unique identifiers (IDs) within the DSL allow \tool to maintain a Traceability Record. This ensures that every line of generated code in the Bottom-Up phase (see Section~\ref{sec:bottom-up}) is provably linked back to a specific requirement expectation.
\end{itemize}

While formal, our DSL is designed to be lightweight and accessible to both human architects and AI agents.
In addition, we design a visualized tool to enable users to design the requirement graph in a more intuitive way.
Interested readers can check our anonymous website \cite{arc} for how the tool is used to draft requirement documents.

\subsection{An Example}

\definecolor{darkyellow}{RGB}{204,153,0}

\begin{figure}[htbp]
    \centering
    \includegraphics[width=\linewidth, trim=0 10 0 10, clip]{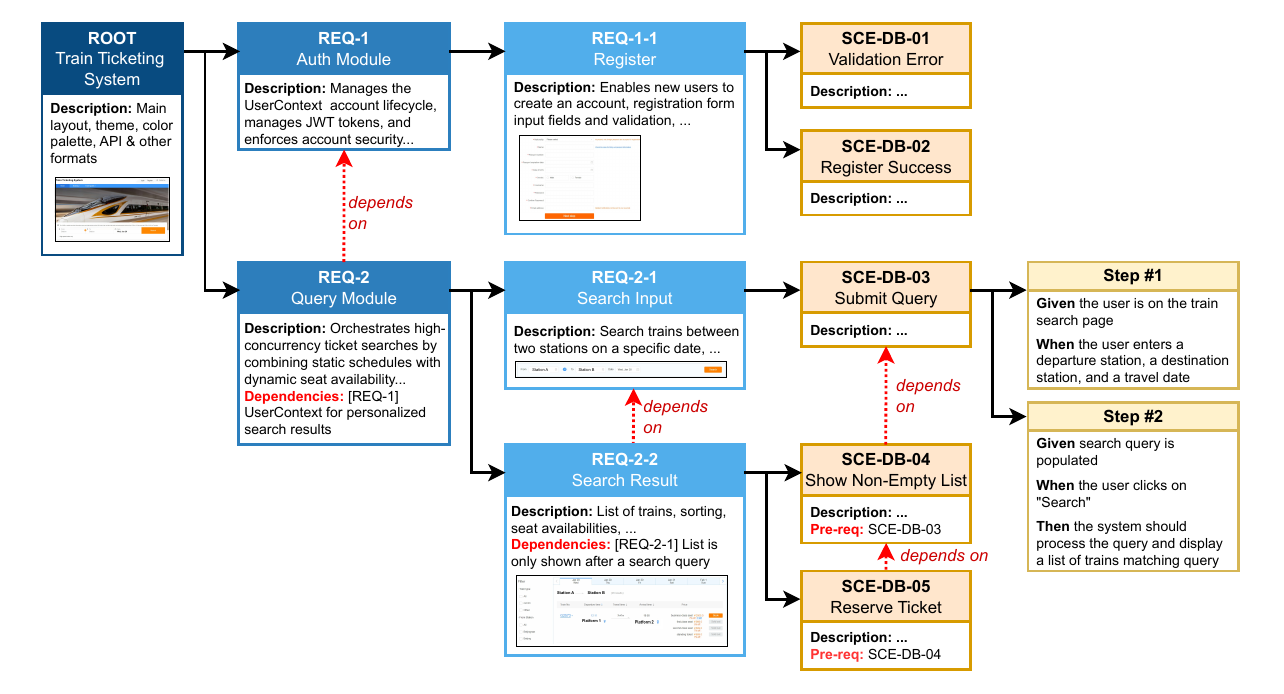}
    \caption{An instance of multi-modal requirement document conforming to our DSL. \textcolor{blue}{\textbf{Blue boxes}} (ROOT, REQ-) represent requirement nodes, \textcolor{orange}{\textbf{orange boxes}} (SCE-) represent scenarios, and \textcolor{darkyellow}{\textbf{yellow boxes}} represent steps.}
    \label{fig:req-example}
\end{figure}

\autoref{fig:req-example} shows an instance of a requirement document that is in accordance with the meta-model in \autoref{fig:requirement-metamodel}. It describes a train ticketing system with two modules: authentication (login \& registration) and query (ticket search). 
Each node includes multi-modal descriptions of appearance and behavior to support unit-level testing. At the \emph{system level}, 
the requirement specifies inter-feature dependencies. For example, accessing search results depends on submitting a query (\texttt{REQ-2-1}~$\to$~\texttt{REQ-2-2}), and performing a search requires a user account. These constraints can be translated into integration tests. 
At the \emph{use-case level}, the scenarios capture the required interaction order for achieving specific functionalities. For instance, a user must log in before searching for tickets, and must search before reserving a ticket (\texttt{SCE-DB-03}~$\to$~\texttt{SCE-DB-04}~$\to$~\texttt{SCE-DB-05}). Detailed operations are recorded as step sequences, which can further serve as system-level and acceptance-level test cases (e.g., end-to-end GUI tests).

\section{Problem Statement}
\label{sec:problem-statement}
Then, we define our problem statement as follows.
Let $\mathcal{G}_r = (V_r, E_r)$ be a multi-modal requirement graph, where $V_r$ represents the set of  nodes of requirements/scenarios and $E_r \subseteq V_r \times V_r$ represents the relations between them.
We define the Requirement Compilation Problem as synthesizing a realized software system $\mathcal{S}$ and a traceability mapping $\mathcal{M}$, defined as follows:

\noindent\textbf{The Realized System ($\mathcal{S}$).}
The system is a tuple $\mathcal{S} = (V_r, \mathcal{I}, \mathcal{T}, \mathcal{C}, \mathcal{E}_{call}, \mathcal{E}_{imp}, \mathcal{E}_{ver})$, where:
\begin{itemize}[leftmargin=*]
    \item $\mathcal{I}$ is the set of generated Interfaces (API signatures, UI components, DB schemas).
    \item $\mathcal{T}$ is the set of Test Suites, strictly derived from the expectations specified in $V_r$.
    \item $\mathcal{C}$ is the set of executable Code Implementations.
    \item $\mathcal{E}_{call} \subseteq \mathcal{I} \times \mathcal{I}$ represents the Call Graph, denoting dependency relations between interfaces.
    \item $\mathcal{E}_{imp} \subseteq \mathcal{I} \times \mathcal{C}$ represents the Implementation Relation, mapping an abstract interface to its concrete logic.
    \item $\mathcal{E}_{ver} \subseteq \mathcal{C} \times \mathcal{T}$ represents the Verification Relation, denoting that implementation $c$ is validated by test $t$.
\end{itemize}

\noindent\textbf{The Traceability Mapping ($\mathcal{M}$).}
We define $\mathcal{M}$ as a provenance relation
\[
\mathcal{M} \subseteq V_r \times \mathcal{I} \times \mathcal{T} \times \mathcal{C},
\] 
where each traceability tuple $(r, \mathcal{I}_r, \mathcal{T}_r, c_r) \in \mathcal{M}$ signifies that:
\begin{itemize}[leftmargin=*]
    \item $\mathcal{I}_r$ is the set of interfaces projected from a requirement node or scenario $r$.
    \item $\mathcal{T}_r$ is the set of test cases derived from $r$.
    \item $c_r \in \mathcal{C}$ is an implementation that conforms to an interface $i\in\mathcal{I}_r$ and satisfies the tests in $\mathcal{T}_r$.
\end{itemize}

\noindent\textbf{Objective.} The goal of \tool is to synthesize $\mathcal{S}$ and $\mathcal{M}$ to optimize two competing objectives:
\begin{itemize}[leftmargin=*]
    \item Maximize Alignment: Maximize the structural consistency between the requirement space $V_r$ and the test space $\mathcal{T}$ via the interface layer $\mathcal{I}$. 
    Intuitively, all the requirements are translated into tests runnable upon the designed interfaces.
    %Formally, this ensures that the induced dependency graph of the interfaces is homomorphic to the requirement graph (i.e., if $(r_1, r_2) \in E_r \implies (\text{map}(r_1), \text{map}(r_2)) \in \mathcal{E}_{call}$).
    \item Minimize Implementation Errors (Execution Phase): Minimize the cardinality of test failures in the generated code.$$\min \sum_{(c, t) \in \mathcal{E}_{ver}} \mathbb{1}(\text{ExecuteTest}(c, t) \neq \text{PASS})$$
\end{itemize}

\section{Approach}
\label{sec:top-down}

\begin{figure}
    \centering
    \includegraphics[scale=0.38]{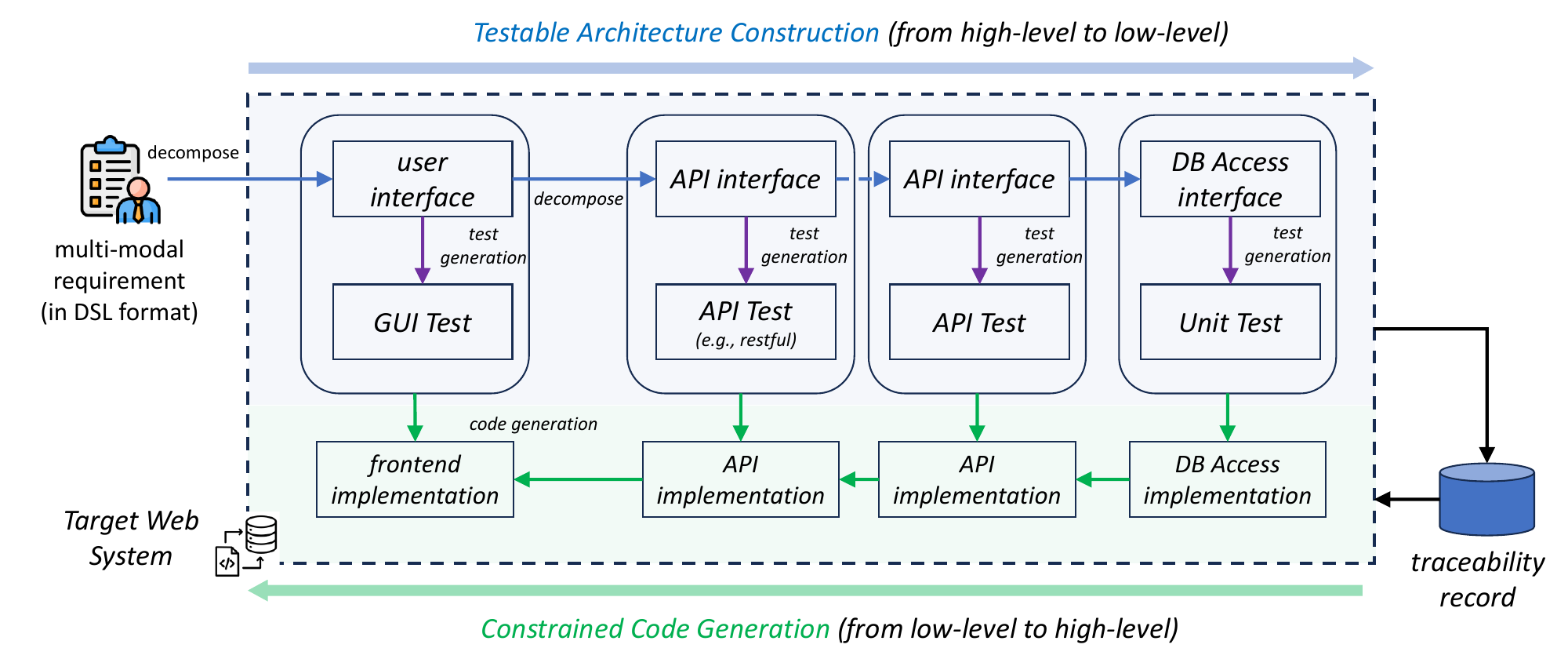}
    \caption{An overview of \tool to parse a multi-modal requirement into a runnable web system. In addition to the web system, the resultant software artifacts include 
    (1) a software architecture consisting of a set of interfaces (UI, API, and database),
    (2) a test suite for each generated interface, and
    (3) a traceability record to capture all the provenance from the requirement to the design, test, and implementation.
    }
    \label{fig:overview}
\end{figure}

\autoref{fig:overview} shows the overview of the \tool framework,
which is designed as a rigorous ``requirement compiler'' that transforms high-level, multi-modal specifications into validated software systems.
The framework takes a multi-modal requirement document, which has been structured into a lightweight, graph-based Domain Specific Language (DSL),
serving as the ``source code'' of the intent, capturing functional scenarios and visual constraints.
The final output is a fully runnable web system. Along with the system, \tool produces three critical artifacts: a software architecture consisting of UI, API, and DB interfaces, a comprehensive test suite for each interface, and a traceability record that maps the provenance from requirement to code.
\tool adopts a dual-phase, i.e., Testable Architecture Construction phase (in blue in \autoref{fig:overview}) and Constrained Code Generation phase (in green in \autoref{fig:overview}).

\begin{itemize}[leftmargin=*]
    \item The first phase follows a \textbf{top-down decomposition strategy} to translate abstract requirements into a blueprint for validation.
    Starting from the DSL-based requirements, \tool systematically decomposes the project into hierarchical layers: the User Interface (UI), the API interfaces, and the Database (DB) Access interfaces.
    For every interface defined, \tool immediately generates a corresponding test suite, including GUI Tests, RESTful API Tests, Integration Tests, and Unit Tests, before any application logic is written.
    \item The second phase operates in a \textbf{bottom-up implementation} direction, where agentic creative output is strictly ``gated'' by the previously defined architecture.
    Agents begin by implementing the lowest level, the DB Access layer, and move upward through the API implementation to the frontend implementation.
    Crucially, each generated code module is immediately executed against its predefined test suite. If a module fails its test, the agent enters a reactive loop to fix the code until it satisfies the interface contract.
\end{itemize}

Throughout both phases, \tool maintains a traceability record. This captures the link between a specific scenario in the original document and its corresponding test and code implementation. This record allows \tool to decide whether to reuse, modify, or create a new interface. 
In addition, such a record serves as a blueprint for the generated system, allowing developers to understand exactly why a piece of code exists and how it relates to the original multi-modal intent.

\subsection{Overall Algorithm}
To be more specific, we introduce the detailed algorithm to parse the DSL-based requirement document.
Algorithm \ref{alg:driver} serves as the central engine of the \tool framework, utilizing a recursive Depth-First Search (DFS) strategy to traverse the requirement graph $\mathcal{G}_r$\footnote{An illustrative animation of our algorithm is available at \cite{arc}.}. This traversal orchestrates the construction of the target software system $\mathcal{S} = \langle \mathcal{I}, \mathcal{T}, \mathcal{C} \rangle$ by managing the lifecycle of each requirement node $r$ through three distinct states:
\begin{itemize}[leftmargin=*]
    \item \textsc{Unprocessed}: The initial state for all nodes, indicating that neither the design nor the implementation for this requirement has been attempted.
    \item \textsc{Working} (Top-Down Phase): The node is currently being visited. In this state, the framework prioritizes Architecture Design, synthesizing the Interface ($\mathcal{I}_r$) and Test Suite ($\mathcal{T}_r$) before addressing dependencies. This prevents architectural drift by establishing a contract first.
    \item \textsc{Done} (Bottom-Up Phase): The node and all its dependencies have been successfully processed. The generated implementation ($c_r \in \mathcal{C}$) has been verified against $\mathcal{T}_r$, and integrated into $\mathcal{S}$.
\end{itemize}

\begin{algorithm}[t]
\caption{Main Compilation Driver}
\label{alg:driver}
\small
\begin{algorithmic}[1]
\REQUIRE Requirement Graph $\mathcal{G}_r = (V_r, E_r)$
\ENSURE System $\mathcal{S}$, Traceability $\mathcal{M}$

\STATE \textbf{Global} $\mathcal{S} \leftarrow (\emptyset, \dots, \emptyset)$, $\mathcal{M} \leftarrow \emptyset$
\STATE \textbf{Global} $State \leftarrow \{r: \text{\textsc{Unprocessed}} \mid \forall r \in V_r\}$

\PROCEDURE{CompileNode}{$r$}
    \STATE $State[r] \leftarrow \text{\textsc{Working}}$
    
    \STATE \COMMENT{\textbf{Phase 1: Top-Down Interface Design}}
    \STATE $\langle \mathcal{I}_r, \mathcal{T}_r \rangle \leftarrow \text{SynthesizeInterface}(r)$ \COMMENT{See Alg. \ref{alg:design}}
    \STATE $\mathcal{S}.\mathcal{I} \leftarrow \mathcal{S}.\mathcal{I} \cup \mathcal{I}_r; \quad \mathcal{S}.\mathcal{T} \leftarrow \mathcal{S}.\mathcal{T} \cup \mathcal{T}_r$
    
    \STATE \COMMENT{Recursive Step \& Call Graph Construction}
    \FORALL{$v \in \text{Children}(r)$}
        \IF{$State[v] = \text{\textsc{Unprocessed}}$}
            \STATE \textsc{CompileNode}($v$)
        \ENDIF
        \STATE $\mathcal{S}.\mathcal{E}_{call} \leftarrow \mathcal{S}.\mathcal{E}_{call} \cup \{ (i_r, i_v) \mid i_r \in \mathcal{I}_r,~i_v \in v.\text{interfaces} \}$
    \ENDFOR

    \STATE \COMMENT{\textbf{Phase 2: Bottom-Up Implementation}}
    \STATE $c_r \leftarrow \text{GenerateImplementation}(\mathcal{I}_r,~\mathcal{T}_r,~\mathcal{S}.\mathcal{E}_{call},~\mathcal{M})$ \COMMENT{See Alg. \ref{alg:impl}}
    
    %\IF{$\text{passed}$}
    \STATE $\mathcal{S}.\mathcal{C} \leftarrow \mathcal{S}.\mathcal{C} \cup \{c_r\}$
    \STATE $\mathcal{M} \leftarrow \mathcal{M} \cup \{ (r, i, t, c_r) \mid i \in \mathcal{I}_r, t \in \mathcal{T}_r \}$
    %\ENDIF
    \STATE $State[r] \leftarrow \text{\textsc{Done}}$
\ENDPROCEDURE

\STATE \textbf{Start:}
\FORALL{$root \in \text{Roots}(\mathcal{G}_r)$}
    \STATE \textsc{CompileNode}($root$)
\ENDFOR
\RETURN $\langle \mathcal{S}, \mathcal{M} \rangle$
\end{algorithmic}
\end{algorithm}

The compilation process proceeds as follows.
Upon visiting an \textsc{Unprocessed} node $r$, the algorithm marks it as \textsc{Working}. It immediately invokes the SynthesizeInterface subroutine (Algorithm \ref{alg:design}) to generate the interfaces $\mathcal{I}_r$ and test suite $\mathcal{T}_r$. These artifacts are added to the global system sets $\mathcal{S}.\mathcal{I}$ and $\mathcal{S}.\mathcal{T}$, effectively ``locking in'' the design constraints before any code is written (Lines 6-8).
The algorithm then iterates through the children of $r$. For each child $v$, it recursively calls CompileNode. Crucially, as the recursion unwinds, we explicitly populate the call graph $\mathcal{S}.\mathcal{E}_{call}$ by linking the parent interface $i_r$ to the child interface $i_v$. This ensures structural alignment between the requirement graph dependencies and the software's internal call structure (Lines 10-15).
Once all children are processed (i.e., their interfaces are stable), the algorithm enters the implementation phase. It calls GenerateImplementation (Algorithm \ref{alg:impl}), passing the now-defined call graph $\mathcal{S}.\mathcal{E}_{call}$ to ensure the generated code $c_r$ correctly invokes its dependencies. If the code passes verification, it is added to $\mathcal{S}.\mathcal{C}$, and the provenance is recorded in $\mathcal{M}$. Finally, the node is marked as \textsc{Done}, signaling to its own parents that it is ready for integration (Lines 17-23).

\subsection{Top-down: Requirement \& Design Decomposition}

In this section, we describe the top-down phase in more detail (Algorithm \ref{alg:design}). We refer to a node's requirements as an aggregation of its multi-modal descriptions, scenarios, and step-level behaviors. Before generating new interfaces, \tool queries the node’s ancestors via GetAncestors (line 1 in Algorithm \ref{alg:design}) to check whether existing interfaces can be adapted to satisfy part or all of the node’s requirements using AdaptInterfaceSignature (line 5 in Algorithm \ref{alg:design}). If applicable, it will modify/reuse these interfaces and update their test cases, producing a set of adapted interfaces $\mathcal{I}_\text{adapted}$. Finally, GenInterfaceSignature (line 11 in Algorithm \ref{alg:design}) takes the node’s requirements and $\mathcal{I}_\text{adapted}$ to decide whether new interfaces are needed, generating $\mathcal{I}_r$ and calling GenTestSuite (line 12 in Algorithm \ref{alg:design}) to produce their corresponding test suite $\mathcal{T}_r$.

\subsubsection{Generating Interfaces} In GenInterfaceSignature (line 11 in Algorithm \ref{alg:design}),
\tool follows a formalized principle to determine what interfaces to add. 
\tool uses \textit{events} to track data flows between two interfaces that interact (e.g., via function calls).
%, i.e., to determine which interfaces communicate with each other. 
An \emph{event} is described as 
\(e = (\textit{name},~ \textit{payload})\), where \(\textit{name}\) identifies the event, and \(\textit{payload}\) contains structured data in JSON standard (e.g., \texttt{\{username: string, password: string}\}). 
We first formally define the behaviors of each type of interface:

\begin{itemize}[leftmargin=*]
    \item \textbf{UI:} Defined as both a producer and a consumer of events, 
    \(i_{\mathbf{UI}} = (E_{\text{out}}, E_{\text{in}})\), where 
    \(E_{\text{out}}\) is the set of events generated by user actions and 
    \(E_{\text{in}}\) is the set of events received from APIs.
    
    \item \textbf{API:} Defined as mapping events to operations and emits events, 
    \(i_{\mathbf{API}} := (e_{\text{in}}, \mathcal{O}, e_{\text{out}})\), where 
    \(e_{\text{in}}\) is the incoming event, \(\mathcal{O}\) is a data operation, and 
    \(e_{\text{out}}\) is the event emitted to the \textbf{UI} or other \textbf{APIs}.
    
    \item \textbf{DB:} Defined as a set of data entities, 
    \(i_{\mathbf{DB}} := \{D_i\}\), where each entity 
    \(D_i = (\text{name}, \{a_1, a_2, \dots, a_n\})\) consists of a name and a set of attributes.
\end{itemize}

To determine the interfaces required by a node $r$, \tool parses both textual descriptions and images (via an image captioning vision-language model). Based on the formal definitions of interface types, it generates a set of interfaces $i\in\mathcal{I}_r$ that collectively capture all requirements. Each interface signature specifies the type (\textbf{UI}, \textbf{API}, or \textbf{DB}), the name, and the expected location in the system $\mathcal{S}$. For \textbf{UI} interfaces, the signature includes layout information inferred from images, as well as the events produced and consumed. For \textbf{API} interfaces, it includes the incoming events, the data operations performed (expressed as function headers with parameters), and the outgoing events emitted to other interfaces. For \textbf{DB} interfaces, it includes the data entity name and its attributes. When two interfaces reference the same event (i.e., one produces it and another consumes it), a data flow is established between them, which serves as a reference for the bottom-up phase.

Using the login feature in \autoref{fig:req-example} as an example, the agent generates a \textbf{UI} interface that produces a \texttt{LoginClicked(username, password)} event and consumes a \texttt{LoginResponse(success, message)} event, with layout information inferred from the screenshot. It also generates an \textbf{API} interface that maps the incoming \texttt{LoginClicked} event to the \texttt{authService.login(username, password)} data operation (function header) and emits the corresponding \texttt{LoginResponse} event. Finally, the \textbf{DB} interface defines a \texttt{User} entity with attributes \texttt{username} and \texttt{passwordHash}.
The detailed agent prompt and more detailed animated demonstrations are available at \cite{arc}.

\subsubsection{Generating Test Suites} 

After generating interfaces, \tool calls GenTestSuite to produce test scripts for the node $r$. The inputs to this process include: (1) The scenarios in the node, $r.\text{scenarios}$, and their steps, $scenario.\text{steps}$. (2) Interface signatures $i \in \mathcal{I}_r$ to be tested. (3) Ancestor tests $\mathcal{T}_a'$, which can serve as test fixtures.

For generating the type of tests, the agent follows a formal principle:
\begin{enumerate}[leftmargin=*]
    \item \textbf{End-to-end tests} validate \textbf{UI}~$\leftrightarrow$~\textbf{API} interactions.
    \item \textbf{Integration tests} validate \textbf{API}~$\leftrightarrow$~\textbf{API} and \textbf{API}~$\leftrightarrow$~\textbf{DB} interactions.
    \item \textbf{Unit tests} validate the behavior of individual \textbf{UI}, \textbf{API}, \textbf{DB}, and the data operators of \textbf{API}.
\end{enumerate}

To generate the content of tests, the agent interprets each scenario step:
\begin{enumerate}[leftmargin=*]
    \item \textbf{Given} establishes test fixtures or prerequisites, potentially reusing ancestor tests $\mathcal{T}_a'$.
    \item \textbf{When} specifies the actions or events to perform in the test.
    \item \textbf{Then} specifies the expected outcomes, which the agent translates into concrete assertions or verification statements in the generated test scripts.
\end{enumerate}

Using the same example, \tool generates end-to-end tests that verify a \textbf{UI}-initiated interaction (i.e., through submitting a login UI form) corresponding to \texttt{LoginClicked(username, password)} results in a response consistent with \texttt{LoginResponse(success, message)}. Integration tests validate the \textbf{API}~$\leftrightarrow$~\textbf{DB} interaction by ensuring that the \textbf{API} correctly queries the \texttt{User} entity, verifies the stored \texttt{passwordHash}, and properly handles both successful and failed login scenarios. Unit tests focus on the \textbf{API} logic by exercising \texttt{authService.login(username, password)} under varied inputs, while \textbf{DB} unit tests validate the schema and constraints of the \texttt{User} entity.

\begin{algorithm}[t]
\caption{Phase 1: Synthesize Interface (Top-Down)}
\label{alg:design}
\small
\begin{algorithmic}[1]
\REQUIRE Node $r$
\ENSURE Interfaces $\mathcal{I}_r$, Test Suite $\mathcal{T}_r$

\STATE $Ancestors \leftarrow \text{GetAncestors}(r)$  

\STATE $AncestorInterfaces \leftarrow \bigcup_{a \in Ancestors} a.\text{interfaces}$  
\COMMENT{Collect interfaces synthesized in ancestors}

\STATE $\mathcal{I}_{\text{adapted}} \leftarrow \emptyset$  

\FORALL{$i_a \in AncestorInterfaces$}
    \STATE $\langle i_a',\mathcal{T}_a'\rangle \leftarrow \text{AdaptInterfaceSignature}(i_a, r)$  
    \COMMENT{Try to modify or reuse ancestor interface}
    \IF{$i_a' \neq \text{null}$} 
        \STATE $\mathcal{I}_{\text{adapted}} \leftarrow \mathcal{I}_{\text{adapted}} \cup \{i_a'\}$  
    \ENDIF
\ENDFOR

\STATE\COMMENT{\parbox[t]{0.4\linewidth}{Generate new interfaces for any parts of $r$ not covered by adapted interfaces}}
\STATE $\mathcal{I}_r \leftarrow \text{GenInterfaceSignature}(r,~\mathcal{I}_{\text{adapted}})$  

\STATE $\mathcal{T}_r \leftarrow \text{GenTestSuite}(r.\text{scenarios},~\mathcal{T}_a',~~\mathcal{I}_r)$  
\COMMENT{Generate tests for the new interfaces}

\RETURN~$\langle \mathcal{I}_r, \mathcal{T}_r \rangle$
\end{algorithmic}
\end{algorithm}

\subsection{Bottom-up: Constrained Code Generation}
\label{sec:bottom-up}

The second phase of the \tool framework operates in a bottom-up direction, transforming the testable architecture into a realized software system. As described in Algorithm~\ref{alg:impl}, this phase ensures that agentic creative output is strictly gated by the predefined interfaces and test suites, preventing architectural drift. The implementation process follows the recursive unwinding of the DFS traversal, distinguishing between leaf and non-leaf nodes.

\paragraph{Implementation of Leaf Nodes} 
For a leaf node $r \in V_r$ (where $\text{Children}(r) = \emptyset$), the implementation focus is localized. The input to the agent consists of the set of interfaces $I_r$ generated during the top-down phase. The objective is to synthesize a code implementation $c_r$ that fulfills the contract defined by $I_r$ and satisfies its local test suite $\mathcal{T}_r$. In this case, the dependency set is empty:
\begin{equation}
\mathcal{D}_r = \emptyset
\end{equation}
The agent is prompted to generate code $c_r$ such that $c_r \models \mathcal{T}_r$.

\paragraph{Implementation of Non-Leaf Nodes}
For a non-leaf node $r$, the implementation begins only after all its children $v \in \text{Children}(r)$ have been processed. At this stage, \tool must not only implement the interfaces $I_r$ directly associated with the current node but also correctly invoke the realized interfaces generated by its children. Let the set of dependency interfaces be defined as:
\begin{equation}
\mathcal{D}_r = \bigcup_{v \in \text{Children}(r)} I_v
\end{equation}
The implementation $c_r$ is thus conditioned on both the target interfaces $I_r$ and the implemented child interfaces in $\mathcal{D}_r$. This ensures structural alignment between the requirement hierarchy and the software's internal call structure.

\paragraph{The Reactive Verification Loop}
The implementation phase is essentially a test-driven development (TDD) loop that functions as a grounding gate for agentic output. As shown in Algorithm~\ref{alg:impl}, the generation process is constrained by the local test suite $\mathcal{T}_r$ and a maximum iteration budget $b$:
\begin{equation}
\text{s.t.} \quad 
\begin{cases} 
c_r \models \mathcal{T}_r \\ 
\text{Attempts} \le b 
\end{cases}
\end{equation}
In each iteration, \tool invokes \textit{GenCode} to produce a candidate implementation $c_r$ based on the target interfaces $I_r$, the dependency interfaces $\mathcal{D}_r$ from children, and the current traceability context $\mathcal{M}$. The candidate is immediately verified by \textit{ExecuteTest}. If the tests fail, the resulting \textit{Feedback} (comprising compiler error logs or test assertion failures) is fed back into the next generation cycle as a refinement prompt. To ensure compilation efficiency and prevent infinite optimization loops for highly complex scenarios, we limit the number of attempts to the budget $b$. If the budget is exhausted before passing all tests, the framework retains the last version of $c_r$ to maintain the workflow. This iterative mechanism ensures that the generated code is not only architecturally compliant but also functionally verified against the original requirement expectations.

More details about the agent's prompt are provided on our project website~\cite{arc}.

% \begin{algorithm}[t]
% \caption{Phase 2: Generate Implementation (Bottom-Up)}
% \label{alg:impl}
% \small
% \begin{algorithmic}[1]
% \REQUIRE Interfaces $\mathcal{I}_r$,~Test Suite $\mathcal{T}_r$,~Call Graph $\mathcal{E}_{call}$,~Traceability Record $\mathcal{M}$
% \ENSURE Code $c_r$, Boolean $\text{passed}$

% \STATE $b \leftarrow \text{MaxBudget}$
% \STATE $passed \leftarrow \text{False}$
% \STATE $Dependencies \leftarrow \bigcup_{i_r \in \mathcal{I}_r} \{ i_v \mid (i_r, i_v) \in \mathcal{E}_{call} \}$
% \STATE $Feedback \leftarrow\emptyset$

% \WHILE{$b > 0 \land \neg passed$}
%     \STATE\COMMENT{\parbox[t]{0.4\linewidth}{Generate candidate code conditioned on interfaces, dependencies, and feedback}}
%     \STATE $c_r \leftarrow \text{GenCode}(\mathcal{I}_r,~\mathcal{M}, ~Dependencies,~Feedback)$
    
%     \STATE \COMMENT{Verify against expectations in test suite}
%     \STATE $passed,~Feedback\leftarrow \text{ExecuteTest}(c_r, \mathcal{T}_r)$
    
%     \STATE $b \leftarrow b - 1$
% \ENDWHILE

% \RETURN $\langle c_r, passed \rangle$
% \end{algorithmic}
% \end{algorithm}

\begin{algorithm}[t]
\caption{Phase 2: Generate Implementation (Bottom-Up)}
\label{alg:impl}
\small
\begin{algorithmic}[1]
\REQUIRE Interfaces $\mathcal{I}_r$,~Test Suite $\mathcal{T}_r$,~Call Graph $\mathcal{E}_{call}$,~Traceability Record $\mathcal{M}$
\ENSURE Code $c_r$, Boolean $passed$

\STATE $b \leftarrow \text{MaxBudget}$
\STATE $passed \leftarrow \text{False}$
\STATE $\mathcal{D}_r \leftarrow \bigcup_{i_r \in \mathcal{I}_r} \{ i_v \mid (i_r, i_v) \in \mathcal{E}_{call} \}$ \COMMENT{Retrieve realized interfaces from children}
\STATE $Feedback \leftarrow \emptyset$

\WHILE{$b > 0 \land \neg passed$}
    \STATE\COMMENT{\parbox[t]{0.4\linewidth}{Generate candidate code conditioned on target interfaces, child dependencies, and feedback}}
    \STATE $c_r \leftarrow \text{GenCode}(\mathcal{I}_r,~\mathcal{D}_r,~\mathcal{M},~Feedback)$

    \STATE\COMMENT{\parbox[t]{0.4\linewidth}{Verify implementation against the node's test suite}}
    \STATE $passed, Feedback \leftarrow \text{ExecuteTest}(c_r, \mathcal{T}_r)$
    
    \STATE $b \leftarrow b - 1$
\ENDWHILE

\RETURN $\langle c_r, passed \rangle$
\end{algorithmic}
\end{algorithm}

\section{Evaluation}
\label{sec:evaluation}

We design our experiments to answer the following research questions:

\begin{itemize}[leftmargin=*]
    \item \textbf{RQ1: Effectiveness.} Can \tool correctly implement web systems of different scales from complex multi-modal requirements, in comparison to state-of-the-art approaches?
    \item \textbf{RQ2: Overhead Analysis.} What is the overhead introduced by \tool, measured in token consumption and development time?
    \item \textbf{RQ3: Ablation Study.} How much does each individual component (DFS search strategy, test suite $\mathcal{T}$, traceability record $\mathcal{M}$) contribute to the overall performance of \tool?
\end{itemize}

\subsection{Benchmark}
\label{sec:benchmark}

\reviewlabel{B-1}\metareviewlabel{2}\reviewlabel{B-3}\metareviewlabel{1.1}\metareview{
We evaluate \tool on two complementary benchmarks, i.e., 
(1) a depth-oriented benchmark consisting of six complex web systems, each specified by hundreds of requirement scenarios, and 
(2) a breadth-oriented benchmark, AppForge~\cite{appforge}, consisting of 101 Android apps, each involving fewer than 10 scenarios. 
The former stresses repository-scale synthesis under deeply hierarchical, interdependent requirements, whereas the latter emphasizes generalization across a large number of smaller and more diverse app-generation tasks.}

\subsubsection{Depth-oriented Web Systems}
We constructed a benchmark that consists of a set of systems $\mathcal{S} = \{\mathcal{S}_1, \ldots, \mathcal{S}_n\}$. For each system $\mathcal{S}_i$, 
we prepared (1) a multi-modal requirements document $\mathcal{G}_r^i$ that specifies the intended functionalities, and (2) a GUI test suite $\mathcal{T}^i$ used to validate the correctness of the implemented system $\mathcal{S}_i$. 

\noindent\textit{Selection of Target Systems.} 
In this study, we selected six benchmark web systems based on two criteria: (1) \emph{Maturity}, the system shall be in operation for at least 10 years; and 
(2) \emph{Popularity}, the system shall have either at least 100 million users or be adopted by companies among the top 100 by revenue. 
For each system, we evaluated a subset of its visible and functional features (see \autoref{table:benchmark} and our anonymous website \cite{arc} for details).

\begin{enumerate}[leftmargin=*]
    \item \textbf{BookStack.} \cite{bookstack} A wiki platform supporting hierarchical content organization and collaborative CRUD operations, representing content-heavy knowledge-management systems.
    \item \textbf{Keep.} \cite{keep} A note-taking system supporting multimodal inputs and synchronized user interactions, representing responsive user-centric applications.
    \item \textbf{Stack Overflow.} \cite{stackoverflow} A community Q\&A platform supporting search, voting, commenting, authentication, and reputation management, representing socially interactive multi-user systems.
    \item \textbf{PrestaShop.} \cite{prestashop} An e-commerce platform supporting shopping carts, product catalogs, payment processing, and order management, representing transaction-heavy applications.
    \item \textbf{12306.} \cite{12306} A train ticketing platform supporting route, schedule, and seat booking workflows, representing large-scale high-concurrency transactional systems.
    \item \textbf{Ctrip.} \cite{Ctrip} A travel booking platform integrating flights, hotels, and train services, representing heterogeneous multi-service applications.
\end{enumerate}

% More details about these systems are shown in \autoref{table:benchmark}. Overall, our selection is reasonably representative of standard enterprise and transactional web systems.

% \reviewlabel{B-1}\metareviewlabel{2}\metareview{We frame our evaluation as \emph{depth- and complexity-oriented} rather than breadth-oriented. Although our benchmark includes six systems, evaluation reliability in our setting is determined by \emph{within-system structural complexity} rather than the number of systems. Each system contains hundreds of interdependent requirements spanning authentication, search, and multi-step workflows, hierarchically structured into up to five levels of decomposition (e.g., a “booking” feature decomposed into selection, configuration, confirmation, and state updates) and coupled through long-range dependencies across modules (e.g., authentication or user state affecting ordering, payment, and notifications). This makes each system a high-dimensional evaluation environment with many coupled constraints for synthesis. Combined with the high construction cost (approximately 200 man-hours per system), our benchmark prioritizes realism and engineering complexity over statistical generalization across many homogeneous tasks, which aligns with the evaluation with repository-scale synthesis goals.}

\begin{table}[!t]
\caption{Overview of web systems in the benchmark.}
\label{table:benchmark}
\centering
\scriptsize
\renewcommand{\arraystretch}{1.2}
\setlength{\tabcolsep}{6pt}
\begin{tabular}{l l c c p{3cm} p{4cm}}
\toprule
\textbf{Scale} & \textbf{Web System} & \textbf{\#REQ} & \textbf{\#TEST} & \textbf{Link} & \textbf{Type} \\
\midrule
Small & BookStack & 32 & 30 & \href{https://demo.bookstackapp.com/}{demo.bookstackapp.com}   & Knowledge Base \\
& Keep & 35  & 45 & \href{https://keep.google.com/}{keep.google.com} & Interactive Productivity Tool \\
Medium & Stack Overflow & 66 & 60 & \href{https://stackoverflow.com/}{stackoverflow.com} & Community Q\&A Platform \\
& PrestaShop & 86 & 75 & \href{https://demo.prestashop.com/}{demo.prestashop.com} & E-commerce Platform \\
Large & 12306 & 142 & 80  & \href{https://www.12306.cn/}{12306.cn} & Public Ticketing System \\
& Ctrip & 152 & 80 & \href{https://ctrip.com/}{ctrip.com} & Service Aggregation Platform \\
\bottomrule
\end{tabular}
\end{table}

\noindent\textit{Construction of Requirement Documents.}
For each selected system $\mathcal{S}_i$, we manually constructed its corresponding requirement graph $\mathcal{G}_r^i$.
We recruited six annotators (graduate students with computer science backgrounds) to construct the DSL-based requirement documents.
Instead of manually writing the DSL format, we provided them with a development requirement modeling tool (available on our project website \cite{arc}) to visualize hierarchical requirement nodes and scenarios. \reviewlabel{A-3}\review{
We also provided detailed guidelines and examples for the quality of the documents (see \cite{arc} for details).
For each system, one annotator spent two weeks on exploring the system and writing the DSL-based requirement documents.
All documents were then reviewed in group meetings and further refined based on feedback from another two annotators. 
The full construction process took approximately from 1200 man-hours.}

\noindent\textit{Construction of Test Suites}
\label{sec:collection-of-test-suites}
\reviewlabel{A-5}\reviewlabel{B-2}\review{
We evaluate functionalities of generated systems by constructing end-to-end GUI tests for each web system.
%The GUI test suites were manually constructed by the same annotators responsible for the requirement graphs, but are independent of \tool and any LLM-based generation. 
The annotators were instructed to create one GUI test case for each leaf-node scenario in the requirement documents. 
\reviewlabel{C-2} Each test case was \textit{manually} constructed, independent of using \tool, according to the actual function of real-world web system.}

\subsubsection{Breadth-oriented App Systems}
\metareviewlabel{1.2}\metareview{
To complement our depth-oriented web benchmark, we additionally evaluate \tool on AppForge~\cite{appforge}, a breadth-oriented benchmark consisting of 101 Android app development tasks derived from real-world applications. In contrast to our primary benchmark, which emphasizes repository-scale synthesis under deeply hierarchical and interdependent requirements, AppForge emphasizes cross-app diversity: each app is relatively small, typically involving fewer than 10 features, while collectively covering a broad spectrum of app functionalities and implementation challenges.}

\metareview{
To adapt AppForge to the input format of \tool, we first convert each natural-language app specification into our DSL by mapping the described functionalities into requirement nodes and their relations. This conversion is structural rather than semantic, i.e., it preserves the original content while making the requirement organization explicit for \tool. However, we observe that some AppForge specifications do not fully enumerate all functions, interaction scenarios, or intermediate execution constraints needed for test-driven generation. 
We therefore further perform \textit{requirement enhancement} by supplementing missing intermediate scenarios and finer-grained execution constraints.
As a result, for each AppForge app, we prepare two versions of the requirement document, i.e., the \textit{originally described} version obtained by structural conversion only, and the \textit{fully described} version after requirement enhancement. The fully described documents are used in RQ1 to evaluate effectiveness under more complete requirements, while both versions are used in RQ3 to study how requirement completeness affects the performance of \tool.
}

% \noindent\textit{Additional Benchmarks.} \reviewlabel{B-3}\metareviewlabel{1.1}\metareview{To assess generalizability, we additionally evaluate \tool on AppForge~\cite{appforge},
% an established benchmark comprising 101 mobile app generation tasks. Compared to our benchmark,
% which prioritizes repository-scale complexity with multi-modal requirements,
% AppForge offers three complementary dimensions: (1) \textit{breadth}, covering diverse app types
% rather than a few complex systems; (2) \textit{platform diversity}, targeting mobile apps rather
% than web systems; and (3) \textit{independent requirements}, sourced externally rather than
% constructed by our team.} 
% \metareviewlabel{1.2}\metareview{We converted AppForge's natural language task descriptions into our DSL format using an LLM 
% (configured as in Section~\ref{sec:llm-configuration}), mapping features into requirement nodes (i.e., pure structural transformation that preserves the original 
% content). We additionally performed \textit{requirement enhancement}, supplementing more scenarios to support test-driven generation. The same enhanced documents are 
% provided to all methods. The effect of this enhancement is studied in RQ3 (Section~\ref{sec:rq3}).}

\subsection{Baselines}\label{sec:baseline} 
We selected four state-of-the-art baselines. \reviewlabel{A-7}\metareviewlabel{3}\metareview{The baselines are described below, along with details of how each was used to generate web systems in our benchmark. }

\begin{enumerate}[leftmargin=*]
    \item \metareview{\textbf{MetaGPT} \cite{hong2023metagpt}: a multi-agent repository-level CLI framework that generates code from high-level requirements via role-based agent coordination.}
    \item \metareview{\textbf{OpenHands} \cite{wang2024openhands}: a multi-agent repository-level IDE-based framework that performs code generation through iterative planning, execution, and feedback-driven refinement.}
    \item \metareview{\textbf{Cursor} \cite{cursor}: a single-agent repository-level IDE framework operated in agent mode, supporting interactive, multi-step code generation and editing with repository-wide context.}
    \item \metareview{\textbf{Copilot} \cite{copilot}: a single-agent repository-level IDE framework operated in agent mode, performing localized code generation and refinement at the line or function level.}
\end{enumerate}

\metareview{
For Cursor and Copilot, we initialized either a Web/Android project template and the requirement documents. 
Then we opened a new chat session within IDE to prompt them with the requirement documents. 
For OpenHands, we used the OpenHands GUI to upload the project template and requirement document in a new session, and prompted the system to generate the corresponding implementation. For MetaGPT, we provided the requirement document as input to its official Docker implementation, which then generated artifacts via its built-in four-agent configuration. 
%We configured all baselines to use our designated LLM provider to ensure consistent underlying LLM capabilities. 
Stepwise setup documentation is available on our project website \cite{arc}.}

\subsection{Evaluation Metrics}
\label{sec:metrics}

Following prior benchmark studies \cite{swebench,prdbench}, we evaluated all methods using 
\textbf{Test Pass Rate (\%)}. For each benchmark system, the provided test suite $\mathcal{T}^i$ 
served as the ground-truth execution oracle (see Section~\ref{sec:collection-of-test-suites}), 
where $\mathcal{T}$ denotes the set of all test cases and $\mathcal{T}_{pass} \subseteq \mathcal{T}$ 
the passed test cases.
\[
\text{Test Pass Rate (\%)} = \frac{|\mathcal{T}_{pass}|}{|\mathcal{T}|} \times 100
\]
A test was considered \emph{passed} if and only if all its test steps executed successfully 
on the system under test and all assertions were satisfied. 
\reviewlabel{B-6}\review{
Since each leaf node in the requirement tree was mapped to 
one test, 
every test corresponded to a concrete function or scenario
to be implemented.
We adopted this metric 
for repository-scale evaluation because it 
allows us to precisely evaluate how many required functions are successfully generated.}

\metareviewlabel{1.4}\metareview{
Unlike web
systems where the system is still runnable (i.e., testable) regardless of partial compilation errors in some modules.
The generated Android APKs need to be compilable before running any tests.
Therefore, we use the following metrics for the evaluation. 
Let $\mathcal{S}$ be the set of all Android apps, $\mathcal{S}_{comp} \subseteq \mathcal{S}$ be the successfully compiled apps,
and $\mathcal{S}_{suite} \subseteq \mathcal{S}$ be the apps whose entire test suite passes (i.e., all test cases pass).
\[
\text{Compile Succ. (\%)} = \frac{|\mathcal{S}_{comp}|}{|\mathcal{S}|} \times 100
\qquad
\text{Test Suite Pass (\%)} = \frac{|\mathcal{S}_{suite}|}{|\mathcal{S}|} \times 100
\]
In addition,
we further evaluate the per-app test case pass rate to capture partial functional
completion. Let $\mathcal{T}^s$ and $\mathcal{T}^s_{pass}$ denote the test cases and passed test cases
for app $s$, respectively.
\[
\text{Test Case Pass (\%)} = \frac{1}{|\mathcal{S}_{comp}|} \sum_{s \in \mathcal{S}_{comp}} \frac{|\mathcal{T}^s_{pass}|}{|\mathcal{T}^s|} \times 100
\qquad
\text{Final Pass (\%)} = \frac{1}{|\mathcal{S}|} \sum_{s \in \mathcal{S}} \frac{|\mathcal{T}^s_{pass}|}{|\mathcal{T}^s|} \times 100
\]
Note that Final Pass (\%) $=$ Compile Succ.(\%) $\times$ Test Case Pass.(\%)}

\subsection{Experimental Setup} 
\label{sec:evaluation_setup}
%\subsubsection{LLM Configuration}
\label{sec:llm-configuration}

\reviewlabel{A-6}\reviewlabel{B-4}\review{\tool and all baselines used Google Gemini 3 Pro \cite{gemini} with temperature 1.0, top-$p$ 0.95, top-$k$ 64, and \textit{thinking level} as high.}
\reviewlabel{A-2}\review{We pre-define the project template for Web application by React (frontend), Node.js (backend), SQLite3 (database), Vitest (unit and integration testing), and Playwright (end-to-end GUI testing).} \metareviewlabel{1.3}
\metareview{We pre-define the project template for Android application by Java, Gradle (build system), and Robolectric (unit, integration, and end-to-end testing), 
with JUnit as the test execution framework.}
\reviewlabel{A-8.1}\review{Considering the non-determinism of LLMs, each experiment was run for 3 trials, and we reported the mean and standard deviation of the test pass rates.}

\subsection{RQ1: Effectiveness}
\label{sec:rq1}

\begin{table}[!t]
\caption{GUI test case pass rates (\%) over 3 independent trials (mean $\pm$ std) indicating the functional correctness of implemented systems. (\textcolor{teal}{$\uparrow x$}) denotes absolute improvement over the best baseline.}
\label{table:rq1}
\centering
\scriptsize
\renewcommand{\arraystretch}{1.2}
\setlength{\tabcolsep}{12pt}
\begin{tabular}{l r@{\,$\pm$\,}l @{\;}l r@{\,$\pm$\,}l r@{\,$\pm$\,}l r@{\,$\pm$\,}l r@{\,$\pm$\,}l}
\toprule
\multirow{2}{*}{\textbf{Benchmark System}} & \multicolumn{11}{c}{\textbf{Test Pass Rate (\%)}} \\
\cmidrule(lr){2-12}
& \multicolumn{3}{c}{\textbf{Ours}}
& \multicolumn{2}{c}{\textbf{MetaGPT}}
& \multicolumn{2}{c}{\textbf{OpenHands}}
& \multicolumn{2}{c}{\textbf{Cursor}}
& \multicolumn{2}{c}{\textbf{Copilot}} \\
\midrule
BookStack
  & $\mathbf{91.1}$ & $\mathbf{5.1}$  & (\textcolor{teal}{$\uparrow$14.4})
  & $\underline{76.7}$ & $10.0$
  & $55.6$ & $24.6$
  & $54.4$ & $12.6$
  & $52.2$ & $15.0$ \\
Keep
  & $\mathbf{91.9}$ & $\mathbf{5.6}$  & (\textcolor{teal}{$\uparrow$22.2})
  & $55.6$ & $13.3$
  & $40.0$ & $21.2$
  & $40.7$ & $26.9$
  & $\underline{69.6}$ & $4.6$ \\
Stack Overflow
  & $\mathbf{86.7}$ & $\mathbf{6.0}$  & (\textcolor{teal}{$\uparrow$33.3})
  & $\underline{53.3}$ & $6.7$
  & $37.2$ & $8.6$
  & $35.0$ & $15.5$
  & $36.1$ & $6.7$ \\
PrestaShop
  & $\mathbf{85.8}$ & $\mathbf{1.5}$  & (\textcolor{teal}{$\uparrow$17.8})
  & $48.0$ & $6.7$
  & $20.4$ & $12.6$
  & $\underline{68.0}$ & $9.3$
  & $34.7$ & $6.7$ \\
12306
  & $\mathbf{78.8}$ & $\mathbf{5.7}$  & (\textcolor{teal}{$\uparrow$28.0})
  & $41.3$ & $7.5$
  & $31.7$ & $11.6$
  & $\underline{50.8}$ & $10.6$
  & $45.0$ & $3.3$ \\
Ctrip
  & $\mathbf{75.4}$ & $\mathbf{7.6}$  & (\textcolor{teal}{$\uparrow$39.2})
  & $18.8$ & $6.3$
  & $15.0$ & $5.7$
  & $\underline{36.3}$ & $13.0$
  & $23.8$ & $11.9$ \\
\bottomrule
\end{tabular}
\end{table}
\begin{table}[!t]
\centering
\scriptsize
\begin{minipage}{0.62\textwidth}
\centering
\caption{Compile success rate, test suite pass rate, test case pass rate, and final pass rate (\%) over 3 independent trials (mean $\pm$ std) on AppForge.}
\label{table:rq1-appforge}
\renewcommand{\arraystretch}{1.2}
\setlength{\tabcolsep}{4pt}
\begin{tabular}{l r@{\,$\pm$\,}l r@{\,$\pm$\,}l r@{\,$\pm$\,}l r@{\,$\pm$\,}l}
\toprule
\textbf{Methods}
& \multicolumn{2}{c}{\makecell{\textbf{Compile Succ.} \\ \scriptsize \textit{(\%)}}}
& \multicolumn{2}{c}{\makecell{\textbf{Test Suite Pass} \\ \scriptsize \textit{(\%)}}}
& \multicolumn{2}{c}{\makecell{\textbf{Test Case Pass} \\ \scriptsize \textit{(\%)}}}
& \multicolumn{2}{c}{\makecell{\textbf{Final Pass} \\ \scriptsize \textit{(\%)}}} \\
\midrule
\textbf{Ours}      & $\mathbf{100.0}$ & $\mathbf{0.0}$ & $\mathbf{31.2}$ & $\mathbf{0.5}$ & $\mathbf{68.3}$ & $\mathbf{1.4}$ & $\mathbf{68.3}$ & $\mathbf{1.4}$ \\
\textbf{MetaGPT}   & 23.3 & 5.8 & 1.6 & 0.7 & 13.9 & 2.8 & 3.2 & 1.0 \\
\textbf{OpenHands} & 32.7 & 9.3 & 3.6 & 1.2 & 23.3 & 5.4 & 8.0 & 3.6 \\
\textbf{Cursor}    & 59.9 & 2.2 & 10.3 & 0.4 & $\underline{41.9}$ & 1.0 & 25.1 & 0.9 \\
\textbf{Copilot}   & $\underline{85.1}$ & 1.6 & $\underline{12.3}$ & 2.1 & 36.2 & 1.0 & $\underline{30.8}$ & 2.1 \\
\bottomrule
\end{tabular}
\end{minipage}
\hfill
\begin{minipage}{0.35\textwidth}
\centering
\caption{Average execution time ("Time") and token consumption ("Cost") on AppForge.}
\label{table:rq2-appforge}
\renewcommand{\arraystretch}{1.2}
\setlength{\tabcolsep}{6pt}
\begin{tabular}{p{2cm} c c}
\toprule
\textbf{Methods}
& \makecell{\textbf{Time} \\ \scriptsize \textit{(min)}}
& \makecell{\textbf{Cost} \\ \scriptsize \textit{(M)}} \\
\midrule
\textbf{Ours}      & 83.18 & 3.07 \\
\textbf{MetaGPT}   & 23.50 & \textbf{0.30} \\
\textbf{OpenHands} & \textbf{7.21} & 1.01 \\
\textbf{Cursor}    & 17.13 & 1.14 \\
\textbf{Copilot}   & 18.12 & 1.40 \\
\bottomrule
\end{tabular}
\end{minipage}
\end{table}

\autoref{table:rq1} reports the results where \tool consistently outperforms all four baselines regarding all the measurement with the test-driven design. 
\tool outperforms the state-of-the-art baseline by 14.4–39.2\% regarding
the pass rates where
\tool improves the test passing rate by on average 54.9\%. 
In addition, 
we observe that the pass rates decrease as the system complexity increases. 
Nevertheless, \tool demonstrates stronger robustness comparing to the baselines. 
For example, the performance of OpenHands drops from 55.6\% (BookStack) to 15.0\% (Ctrip).
In contrast, \tool still keeps its performance as 75.4\% in generating the Ctrip project.
%These results suggest that: (1) without enforcing the strict interface design and implementation constraints of \tool, baseline methods often fail to implement even half of the required functionality, and (2) prior knowledge or familiarity with the system (by the underlying code models) is insufficient to ensure successful development. 
\reviewlabel{A-8.2}\review{Across all trials, \tool exhibits consistently low variance (std $\leq$ 7.6\%), while baselines show substantially higher variance (e.g., Cursor std 26.9\% on Keep, OpenHands 24.6\% on BookStack), confirming that the test-driven design improves both the effectiveness and its performance stability.} 
\metareviewlabel{1.5}\metareview{\autoref{table:rq1-appforge} reports the results on AppForge where \tool consistently outperforms all baselines, achieving 100\% compile success, 31.2\% test suite pass rate, and 68.3\% test case and final pass rates. This demonstrates that the \tool framework also exhibits strong generalizability over diverse Android apps.}

\paragraph{Why can \tool outperform the baselines?} 
% \highlight{TODO: After adding baseline execution details, ensure that explanations like ``MetaGPT generates PRDs and system designs from coarse, lossy overviews'' and ``existing frameworks implement features without isolation'' are meaningful given the clarified baseline process}{(C3,C11)} 
First, we observe that baselines often compress requirements into unstructured summaries, 
the resultant context is usually long and noisy, potentially leading to hallucination by missing or mistaking the crucial implementation.
In contrast, \tool interprets each feature (i.e., requirement node) into a set of verifiable tests,
providing a more reliable signal for LLM to produce code.
%For example, MetaGPT generates PRDs and system designs from coarse, lossy overviews rather than original specifications, leading to repositories that are structurally sound but functionally incomplete. 
Second, existing frameworks and AI IDEs (e.g., Cursor and Copilot) are likely to repetitively modify implemented code, 
even if their features are independent in the requirement documents.
In contrast, \tool avoids these issues by parsing the requirements with the designed DSL where agents can strictly process one feature after another, isolating their implementation with minimal interference. 
As a result, \tool enforces orthogonality and non-regression in the process of code generation. 
%Given a set of existing implementations and passing tests, adding new tests and implementations should never affect prior functionality. 
% Last, a common failure mode in baselines is fragmented implementation. They might rely on stubs or mock data and fail to capture full system integration. \tool prevents this with test suites. Its tests verify not only the correctness of individual interfaces (\textbf{API}, \textbf{DB}, \textbf{UI}) but also their interactions (\textbf{API}~$\leftrightarrow$~\textbf{API},~\textbf{API}~$\leftrightarrow$~\textbf{DB},~\textbf{API}~$\leftrightarrow$~\textbf{UI}).

\paragraph{When can \tool fail to pass the GUI tests?} By investigating the GUI tests that \tool failed to pass, we identify two main causes: 
\begin{itemize}[leftmargin=*]
    \item \textbf{Implicit requirement scenarios are missing in documents.} \tool treats the requirements graph as the source of truth, 
    but it may not infer the implicit requirements well.
    For example, 
    our documents do not explicitly specify that the user ID cannot be duplicated when registering a new user account in the system of 12306.
    While it is common sense for human developers, we observe \tool can miss such code implementation, 
    leading to failing relevant test cases.
    It indicates that the requirement document processed by agents could be more detailed and explicit than the traditional requirement documents. 
    \item \textbf{Limited Feedback Signal from Tests.} Test cases may provide insufficient guidance (or signal) for \tool to fix a test failure within the given budget. Once the budget is exhausted, \tool moves on with a best-effort implementation. 
    For instance, in PrestaShop’s “Recently Viewed” product feature, the tests report that “viewed product A is not shown,” which provides very limited information for LLM to infer where and how to fix the problem. 
    %Within a limited budget, \tool fails to generate a local solution with limited feedback signal.
    %This can be mitigated by increasing the iteration budget.
\end{itemize}

Interested readers can find more successful and failing cases of \tool on our website \cite{arc}.

% \highlight{TODO: Add evaluation on established benchmarks (e.g., AppForge) using widely accepted metrics, or discuss generalization -- this is a mandatory major revision requirement from the meta-reviewer}{(C1,C18)}
\subsection{RQ2: Overhead Analysis}
\label{sec:rq2}

\subsubsection{Experiment Design}

As described in Section \ref{sec:benchmark} and \ref{sec:evaluation_setup}, we measured the overhead over two dimensions: the end-to-end runtime (in minutes), denoted as \textbf{Time (min)}, and the token consumption (in millions of tokens), denoted as \textbf{Cost (M)}. 
%Here, the end-to-end measurement refers to the interval from the start of the evaluation workflow to its completion. 

% \metareviewlabel{1.6}\metareview{For AppForge, we compute the per-app average time and token consumption within each trial.}

\subsubsection{Results} 

\begin{table}[!t]
\caption{Execution time in minutes ("Time") and token consumption in millions ("Cost") of our framework and baseline methods for implementing web systems from requirements.}
\label{table:rq2}
\centering
\scriptsize
\renewcommand{\arraystretch}{1.25}
\setlength{\tabcolsep}{5.5pt}

\begin{tabular}{
  l
  % Ours
  S[table-format=3.2, table-column-width=0.95cm]
  S[table-format=2.2, table-column-width=0.7cm]
  % MetaGPT
  S[table-format=3.2, table-column-width=0.95cm]
  S[table-format=2.2, table-column-width=0.7cm]
  % OpenHands
  S[table-format=3.2, table-column-width=0.95cm]
  S[table-format=2.2, table-column-width=0.7cm]
  % Cursor
  S[table-format=3.2, table-column-width=0.95cm]
  S[table-format=2.2, table-column-width=0.7cm]
  % Copilot
  S[table-format=3.2, table-column-width=0.95cm]
  S[table-format=2.2, table-column-width=0.7cm]
}
\toprule

\multirow{2}{*}{\textbf{Web System}} 
& \multicolumn{2}{c}{\textbf{Ours}}
& \multicolumn{2}{c}{\textbf{MetaGPT}}
& \multicolumn{2}{c}{\textbf{OpenHands}}
& \multicolumn{2}{c}{\textbf{Cursor}}
& \multicolumn{2}{c}{\textbf{Copilot}} \\[2pt]

\cmidrule(lr){2-3}
\cmidrule(lr){4-5}
\cmidrule(lr){6-7}
\cmidrule(lr){8-9}
\cmidrule(lr){10-11}

& \makecell{\textbf{Time} \\ \scriptsize (min)}
& \makecell{\textbf{Cost} \\ \scriptsize (M)}
& \makecell{\textbf{Time} \\ \scriptsize (min)}
& \makecell{\textbf{Cost} \\ \scriptsize (M)}
& \makecell{\textbf{Time} \\ \scriptsize (min)}
& \makecell{\textbf{Cost} \\ \scriptsize (M)}
& \makecell{\textbf{Time} \\ \scriptsize (min)}
& \makecell{\textbf{Cost} \\ \scriptsize (M)}
& \makecell{\textbf{Time} \\ \scriptsize (min)}
& \makecell{\textbf{Cost} \\ \scriptsize (M)}
\\

\midrule
BookStack      & 145.98 & 11.80 & 48.90 & \textbf{0.98} & 22.00 & 5.69 & 29.16 & 3.82 & \textbf{13.28} & 2.46 \\
Keep           & 161.04 & 17.50 & 37.83 & \textbf{0.38} & 22.35 & 6.70 & \textbf{13.58} & 1.84 & 22.78 & 3.95 \\
Stack Overflow & 231.52 & 21.38 & 42.38 & \textbf{0.67} & 52.00 & 14.43 & \textbf{16.48} & 3.47 & 22.50 & 5.14 \\
PrestaShop     & 253.95 & 22.76 & 42.98 & \textbf{0.89} & 34.72 & 3.94 & 19.73 & 4.45 & \textbf{14.12} & 2.85 \\
12306          & 351.52 & 27.02 & 53.72 & \textbf{1.23} & 38.73 & 16.29 & 39.11 & 7.26 & \textbf{17.40} & 3.81 \\
Ctrip          & 363.27 & 26.38 & 35.00 & \textbf{0.90} & 26.68 & 5.24 & \textbf{16.21} & 2.74 & 18.51 & 3.35 \\
\bottomrule
\end{tabular}
\end{table}

\autoref{table:rq2} shows our results. \tool requires an average of 3 hours more than the state-of-the-art baseline, and consumes 6–21 million additional tokens. 
This additional cost is spent on (1) incrementally planning and implementing interfaces, and (2) rigorous feedback through test execution. As systems scale (from BookStack to Ctrip), we see that \tool's overhead increases accordingly. 
In contrast, the overhead of baselines does not increase.
Our investigation shows that, 
while taking less overhead, 
the state-of-the-art solutions usually feed the requirement documents to LLM as a whole context, 
which can easily miss crucial information to generate code especially when the document is complicated.
\reviewlabel{B-5}\review{Despite that we imposed no upper limit on tokens or runtime for any method, we observe that baselines frequently failed to fully utilize resources, which is a known limitation of LLMs in agentic settings \cite{cursor-long-running-agents,cursor-scaling-agents}.} \metareviewlabel{1.6}\metareview{
The similar results are consistent in AppForge. 
\tool takes more overhead while generating far more accurate code in return.}

% \paragraph{What is the cost breakdown by phase in \tool?} 
\reviewlabel{C-1}\review{We further analyzed execution logs of \tool to estimate its average time and token consumption in each phase. 
As for time consumption, the code generation phase accounts for 84.3\%, the test generation phase for 14.8\%, and design generation phase for 0.9\%. 
As for token consumption, the code generation phase dominates at 62\% (including feedback iterations), followed by test generation phase at 36\%, and design generation phase accounting for 2\%. 
The gap between time and token consumption lies in that the code generation phase involves substantial non-LLM overhead (e.g., compilation and test execution), whereas test and design generation are dominated by the interaction with LLM.}

\subsection{RQ3: Ablation Study}
\label{sec:rq3}

\subsubsection{Experiment Design}

In this experiment, we adopted the settings described in Section \ref{sec:benchmark} and \ref{sec:evaluation_setup}.
On the web benchmark, we compared three ablated variants of \tool: (1) \emph{w.o. DFS strategy}, where \tool designs and implements requirement nodes in an arbitrary order; (2) \emph{w.o. Test}, where no test suites $\mathcal{T}$ are generated for interfaces during the top-down phase; and (3) \emph{w.o. Traceability}, where code provenance $\mathcal{M}$ is not recorded during the bottom-up phase. 
\metareviewlabel{1.7}\metareview{On the Android app benchmark, AppForge, we compared the performance on fully and partially described requirements (see Section \ref{sec:benchmark}). 
}
We used the same performance metrics as in RQ1 (Section \ref{sec:rq1}).

\subsubsection{Results} \autoref{table:rq3} shows our results. Without traceability, \tool spends more time gathering contextual information from the repository. Although the DFS strategy and testing contribute to increased time and cost, this is a necessary trade-off to support reliable development in large-scale systems, as evidenced by the higher pass rate. These results validate our claims in RQ1 and RQ2. \metareviewlabel{1.7}\metareview{
On AppForge, 
\tool has better performance on the fully described documents. 
Test suite pass rate increases from 18.3\% (partially described) to 31.7\% (fully described). 
The improvement also applies in the test case pass rate and the final pass rate.
Note that, the performance on partially described documents still outperforms the state-of-the-art baselines.
}

\begin{table}[!t]
\caption{Ablation study of our framework. Columns show test pass rate (“Pass”), execution time in minutes (“Time”), and token consumption in millions (“Cost”) under different component removals.}
\label{table:rq3}
\centering
\scriptsize
\renewcommand{\arraystretch}{1.25} % slightly taller rows
\setlength{\tabcolsep}{4pt} % tighter spacing

\begin{tabular}{
l
S[table-format=2.2, table-column-width=0.65cm] 
S[table-format=3.2, table-column-width=0.9cm] % Time 
S[table-format=2.2, table-column-width=0.7cm]  
S[table-format=2.2, table-column-width=0.65cm]
S[table-format=3.2, table-column-width=0.9cm] % Time
S[table-format=2.2, table-column-width=0.7cm]
S[table-format=2.2, table-column-width=0.65cm]
S[table-format=3.2, table-column-width=0.9cm] % Time
S[table-format=2.2, table-column-width=0.7cm]
S[table-format=2.2, table-column-width=0.65cm]
S[table-format=3.2, table-column-width=0.9cm] % Time
S[table-format=2.2, table-column-width=0.7cm]
}
\toprule
\multirow{2}{*}{\textbf{Web System}} 
& \multicolumn{3}{c}{\textbf{Ours}} 
& \multicolumn{3}{c}{\textbf{w.o.\ DFS Strategy}}
& \multicolumn{3}{c}{\textbf{w.o.\ Test}}
& \multicolumn{3}{c}{\textbf{w.o.\ Traceability}}\\[2pt]

% ---------- CMIDRULES ----------
\cmidrule(lr){2-4}
\cmidrule(lr){5-7}
\cmidrule(lr){8-10}
\cmidrule(lr){11-13}

% ---------- Subheaders ----------
& \makecell{\textbf{Pass} \\ \scriptsize (\%)} 
& \makecell{\textbf{Time} \\ \scriptsize (min)} 
& \makecell{\textbf{Cost} \\ \scriptsize (M)} 
& \makecell{\textbf{Pass} \\ \scriptsize (\%)} 
& \makecell{\textbf{Time} \\ \scriptsize (min)} 
& \makecell{\textbf{Cost} \\ \scriptsize (M)} 
& \makecell{\textbf{Pass} \\ \scriptsize (\%)} 
& \makecell{\textbf{Time} \\ \scriptsize (min)} 
& \makecell{\textbf{Cost} \\ \scriptsize (M)} 
& \makecell{\textbf{Pass} \\ \scriptsize (\%)} 
& \makecell{\textbf{Time} \\ \scriptsize (min)} 
& \makecell{\textbf{Cost} \\ \scriptsize (M)} \\
\midrule

BookStack        & \textbf{91.11} & 145.98 & 11.80 & 63.33 & \textbf{84.32} & \textbf{3.09} & 60.00 & 112.77 & 11.74 & 90.00 & 164.78 & 15.92 \\
Keep             & \textbf{91.85} & 161.04 & 17.50 & 75.56 & \textbf{103.75} & \textbf{13.48} & 60.00 & 136.21 & 16.68 & 86.67 & 162.94 & 22.59 \\
Stack Overflow   & \textbf{86.67} & 231.52 & 21.38 & 35.00 & \textbf{117.34} & \textbf{8.42} & 56.67 & 162.37 & 14.74 & 83.33 & 215.18 & 21.24 \\
PrestaShop       & \textbf{85.78} & 253.95 & 22.76 & 53.33 & \textbf{90.12} & \textbf{8.43} & 38.67 & 167.14 & 10.05 & 74.67 & 236.24 & 22.74 \\
12306            & \textbf{78.75} & 351.52 & 27.02 & 36.25 & \textbf{210.41} & \textbf{20.21} & 67.5 & 220.13 & 23.63 & 77.5 & 429.08 & 25.45 \\
Ctrip            & \textbf{75.42} & 363.27 & 26.38 & 33.75 & \textbf{234.47} & 17.16 & 48.75 & 236.48 & \textbf{13.99} & 71.25 & 441.02 & 31.22 \\
\bottomrule
\end{tabular}
\end{table}
\begin{table}[!t]
\caption{Ablation study of our framework on partially described and fully described requirements in AppForge. We compare \tool applied to DSL-formatted requirements under both requirement completeness settings.}
\label{table:rq3-appforge}
\centering
\scriptsize
\renewcommand{\arraystretch}{1.1}
\setlength{\tabcolsep}{8pt}
\begin{tabular}{l c c c c}
\toprule
& \textbf{Compile Succ. (\%)}
& \textbf{Test Suite Pass (\%)}
& \textbf{Test Case Pass (\%)}
& \textbf{Final Pass (\%)} \\
\midrule
\textbf{Partially Described } & $100.0 \pm 0.0$ & $18.3 \pm 1.0$ & $41.8 \pm 1.7$ & $41.8 \pm 1.7$ \\
\textbf{Fully Described}  & $\mathbf{100.0} \pm \mathbf{0.0}$ & $\mathbf{31.7} \pm \mathbf{0.9}$ & $\mathbf{68.0} \pm \mathbf{0.2}$ & $\mathbf{68.0} \pm \mathbf{0.2}$ \\
\bottomrule
\end{tabular}
\end{table}
\subsection{User Study}

We conducted a study to evaluate how users can adopt \tool in practice to ``compile'' a web system by only writing DSL-based multi-modal requirements.

\noindent\textbf{Subject System.} 
In this study, we selected the system of 12306, a popular train ticket booking system.
We chose this system because all the participants indicated that they had used this system in person,
which made it more convenient for them to write requirement documents in this study.
The participants were asked to specify five modules (e.g., login module, registration module, query module, ticket-booking module, and profile update module) in the system.

\noindent\textbf{Participants.} We advertised the study to undergraduate students in a software engineering course, allowing interested students to register via an online form. 
This population provided a representative proxy for novice developers. 
We selected the participants with prior knowledge in web development, at least one year of exposure to the subject system, and at least six months of experience with AI-assisted coding tools. 
As a result, we recruited 21 participants in total.
Given the requirement is non-trivial,
we organized those participants into 7 groups, each consisting of 3 participants to accomplish the documents to generate the web system.

\noindent\textbf{Task Procedure \& Metrics.} Participants were first introduced to the requirements DSL. To ensure familiarity with the DSL before the study began, we first guided them to complete a small warm-up task for practice on a web system which is different from the subject system. 
Each group explored the subject system and documented its functionality using the DSL, after which they used \tool to replicate the system. 
We asked the participants to complete the task within one day.
Upon completion, we collected all replicated systems and evaluated them based on the test pass rate (\%) against our ground-truth test cases. Finally, participants completed a post-study survey.

\noindent\textbf{Results.}
The detailed results of our user study are presented in \autoref{tab:user-study}.
\begin{table}[htbp]
\centering
\caption{User Study Results: \textit{Test Pass Rate} indicates the pass rate of GUI tests; \textit{DSL Writing Time} denotes the total time spent on writing the requirement document; \textit{Project Generation Time} represents the total time taken to generate the web system from the requirement document.}
\label{tab:user-study}
\resizebox{\linewidth}{!}{%
\begin{tabular}{cccccc}
\toprule
\textbf{Group} & \textbf{Test Pass Rate (\%)} & \textbf{DSL Writing Time (h)} & \textbf{Project Generation Time (h)} & \textbf{Number of Requirements} & \textbf{Preferred Method} \\
\midrule
1 & 90 & 4 & 4.5 & 70 & \textbf{ARC} \\
2 & 90 & 8 & 6 & 120 & \textbf{ARC} \\
3 & \textbf{96} & 9.8 & 4 & 174 & \textbf{ARC} \\
4 & 80 & 3.4 & 5.5 & 50 & \textbf{ARC} \\
5 & \textbf{96} & 4.7 & 5 & 134 & \textbf{ARC} \\
6 & 94 & 6.2 & 6 & 144 & Vibe-coding \\
7 & 86 & \textbf{3.1} & \textbf{2.5} & 58 & \textbf{ARC} \\
\midrule
\textbf{Avg.} & \textbf{90.3} & \textbf{5.6} & \textbf{4.8} & \textbf{107} & - \\
\bottomrule
\end{tabular}%
}
\end{table}
All 7 groups successfully transformed their requirements into runnable systems with test pass rates ranging from 80\% to 96\% (90.3\% on average), completing their projects within one day. On average, each group spent 5.6 hours on DSL-based requirement specification and system generation, with completion times ranging from 3.1 to 9.8 hours.
\reviewlabel{A-1}\review{Overall, participants found the DSL easy to learn, requiring an average of 5 rounds of requirement updates to complete their tasks. Some participants drafted requirements informally first and translated them into the DSL with LLM assistance\footnote{The prompt used for this translation is available on our website \cite{arc}.}, a process that took under 10 minutes including error correction.}
The compiled software projects contained 107 requirement nodes and 1,395 lines of DSL code on average. 
We further observed that the precision of requirement documents led to clear differences across groups. Groups that described functional details, validation rules, and intermediate interaction scenarios more precisely generally produced systems with higher test pass rates, whereas groups with more ambiguous or incomplete requirement documents were more likely to miss expected behaviors in the generated systems. This suggests that, beyond DSL learnability, the quality and completeness of requirement specification substantially affect the final compilation outcome.

\noindent\textbf{Post-Study Survey.} 
6 out of 7 groups (86\%) agreed that \tool is more effective and efficient than conventional approaches such as unstructured AI-assisted coding (vibe-coding) or fully manual implementation. Participants reported that \tool supports more accurate requirement implementation with fewer manual interactions, and that the traceability records linking requirements to interfaces, tests, and code helped them understand the generated systems. 
The remaining group did not agree that \tool was more effective in practice, not because the final system was unusable, but because debugging the agentic pipeline remained difficult once the generated system deviated from expectation. In particular, participants reported that feedback from the compilation process was relatively slow, making it hard to quickly identify which requirement, test, or implementation step caused the mismatch. 
Several participants also described their current debugging strategy as using \texttt{git checkout} to roll back to earlier generated versions and compare intermediate results. While this strategy was useful in practice, they still considered it inefficient and cumbersome for iterative debugging. This feedback suggests that future versions of \tool should provide more interactive debugging support and faster feedback when generated systems fail to satisfy user expectations.

\subsection{Discussion}

% In this experiment, we show that \tool can compile DSL-based requirements into runnable Web system. We propose the ARC framework to advance this paradigm, and our experiments demonstrate both its effectiveness and efficiency compared to conventional agentic approaches. ARC compiles multi-modal requirements into a deployable web system, in which (1) each generated component accounts for its own requirements and those of dependent components, and (2) generation is grounded in explicit system structure and execution semantics, enabling systematic reasoning, verification, and reproducibility. 
% Beyond the conventional $0 \rightarrow 1$ setting (software generation), ARC's traceability 
% mechanism presents potential for $n \rightarrow n+1$ development~\cite{volvo}, where 
% coding agents must evolve existing systems while accounting for interacting components.
% Despite these positive results, ARC has several limitations. 
% First, it is primarily designed to satisfy functional requirements and may overlook non-functional properties such as security, performance, and reliability.
% Future work could extend ARC to treat functional requirements as hard constraints while optimizing non-functional properties as soft objectives.
% Second, its test generation process remains limited, leaving open the possibility of undiscovered bugs. 
% Stronger correctness guarantees would require more reliable test generation in future work.

Beyond the conventional software generation, ARC also shows promise for supporting software evolution~\cite{volvo}. 
Its core advantage lies in the explicit traceability it maintains across requirements, interfaces, tests, and implementations, which helps agents reason about how local changes may affect interacting components in an existing system. 
Our experimental results further suggest that this design is effective in practice.
Compared with state-of-the-art baselines, ARC consistently achieves higher pass rates and exhibits stronger stability on complex, repository-scale systems, indicating that requirement grounding and test-driven decomposition are beneficial.

At the same time, the experiments also reveal important limitations of the current framework. 
First, \tool is primarily optimized for functional correctness, and therefore may overlook non-functional requirements such as security, performance, scalability, and reliability, which are equally important in practice. 
Second, \tool remains fundamentally bounded by the completeness of the requirement document and the quality of the generated tests. 
When key scenarios are missing or the feedback signal from tests is weak, the system may still produce partially incorrect implementations. 
These observations point to several directions for future work. One direction is to extend ARC from pure functional compilation to multi-objective requirement compilation, where functional requirements are treated as hard constraints and non-functional properties are optimized as soft objectives. Another is to improve the strength of the testing and feedback loop, for example by generating more complete tests, richer debugging signals, and more informative failure localization.

\subsection{Threats to Validity}

\noindent\textbf{Internal Validity.}
\label{sec:internal_validity}
%The first threat arises from the test validation loop. Incomplete test coverage may weaken correctness guarantees, as generated systems may pass available tests without fully satisfying all intended requirements.
\reviewlabel{A-4}\metareviewlabel{4}\metareview{The first threat concerns the use of graduate students to construct benchmark artifacts, which may introduce variability due to limited professional software engineering experience. To mitigate this, we provided detailed construction guidelines to standardize the drafting process (see \cite{arc}), and two annotators independently reviewed and refined all artifacts to resolve ambiguities and ensure consistency. Nevertheless, requirements are inherently incomplete, ambiguous, and evolving \cite{re-roadmap}, motivating future work on specification refinement.}
\reviewlabel{A-9}\review{The second threat concerns reliance on proprietary LLMs, which may be updated at any time without disclosure, making precise versioning and full replication of results difficult. We acknowledge this as an inherent limitation, and anticipate that future work will evaluate open-source or locally hosted models.}
\reviewlabel{A-8.3}\review{The last threat is LLM non-determinism, which we mitigated by running each approach for 3 independent trials and reporting mean and standard deviation on all results.
We will extend the study with more number of trials in the future.
}

\noindent\textbf{External Validity.} There are two main factors that may limit the generalizability of our results. First, we selected six web systems for our benchmark. While these systems capture standard enterprise and transactional web applications, they do not fully represent the diversity of web system behaviors. For example, media-heavy streaming applications (e.g., YouTube, Discord), real-time collaboration tools (e.g., Google Docs), and single-page applications (e.g., Google Maps) exhibit complex behaviors that are difficult to automatically test and therefore challenging to systematically benchmark. Furthermore, such systems often involve ambiguity in implementation choices, such as whether to use WebSockets for real-time communication, or other architectural decisions. Addressing these complexities could be an avenue for future work. Second, \tool was evaluated using Gemini 3 Pro as the underlying model, which raises questions about transferability to other models. However, as long as the underlying model is trained on coding data, the framework remains applicable. Our approach is a general organizational and instructional framework that is similar to multi-agent or other system frameworks, which do not rely on any model-specific advantages. It is designed to leverage already capable models in a structured way.

\section{Related Work}

\noindent\textbf{Code Generation.} Early work focuses on automatic programming and program synthesis, where systems such as PROW generate executable code from formal logical specifications~\cite{prow,1985balzer}. Later approaches explore heuristic and probabilistic methods, including grammar-based generation~\cite{2014miningidiom,2010inducegrammar,2017preciseconditiongeneration} and specialized statistical models~\cite{jha2010oracle,de2008z3}, but these techniques are often rigid and difficult to scale. Advances in transformer models enable language models to reason over longer contexts and generate increasingly complex programs i.e. Codex \cite{chen2021evaluatinglargelanguagemodels}, MBPP \cite{austin2021programsynthesislargelanguage}, APPS \cite{appsbenchmark} or competitive programming namely AlphaCode \cite{alphacode}. Stronger foundation models such as ChatGPT, Deepseek \cite{achiam2023gpt,guo2025deepseek} enable strong semantic code understanding demonstrably in real-time code suggestions inside IDEs such as Cursor, Copilot \cite{cursor,copilot,liu2024coedpilot} or real-world program repairs through SWE-bench \cite{swebench,swebench-multimodal} at near 80\% by Trae Agent \cite{traeagent}. These works are the pre-cursor to a challenge shift from function-level code generation towards repository-level generation \cite{liu2024repobench,jain2025livecodebench,tang2024mlbenchevaluatinglargelanguage}.

\noindent\textbf{Repository Generation.} Repository-level generation is challenging due to large-scale, long-context, and long-trajectories, with modules and business logics distributed across many files. Existing work primarily addresses these challenges along three axes: (1) \emph{Interactive feedback:}
To manage long trajectories, prior methods employ multi-round planning with external feedback. SDE-I~\cite{zhao2025commit} uses static analysis and unit tests to guide library generation, but requires a reliable pre-existing test suite. Other systems, including CodePlan~\cite{codeplan}, Cursor~\cite{cursor}, and Copilot ~\cite{copilot} Plan Mode, derive implementation plans from repository context before code generation to reduce error propagation; (2) \emph{Memory augmentation:}
To handle long-context reasoning and cross-file dependencies, RepoCoder~\cite{zhang-etal-2023-repocoder} and CodeChain~\cite{le2024codechain} construct dependency graphs to retrieve relevant code, while RepoGraph~\cite{ouyang2025repograph} models repository structure using graph-based representations at code-line level; (3) \emph{Task delegation:}
Multi-agent systems decompose large-scale software engineering tasks into specialized roles. OpenHands~\cite{wang2024openhands} and HyperAgent~\cite{huy2024hyperagent} distribute planning, navigation, and editing across agents, but these setups incur communication overhead and context management challenges~\cite{autogen}. While effective, these approaches largely focus on incremental code generation over existing repositories and lack explicit modeling of system-level requirements. As a result, changes introduced for new features can affect unrelated components or invalidate previously passing tests. In contrast, ARC explicitly tracks requirement-implementation alignment throughout the development lifecycle.

\noindent\textbf{End-to-end software engineering agents} Further extended from repository generation approaches, \emph{Waterfall-based frameworks}, namely Self-Collaboration \cite{self-collaboration}, ChatDev \cite{chatdev}, MetaGPT \cite{hong2023metagpt}, model the development process as a linear pipeline akin to the waterfall model, whereas \emph{agile-based frameworks}, such as AgileCoder \cite{nguyen2024agilecoder} model development process in iterative sprints. 
Existing approaches lack enforced requirement alignment during implementation. ARC treats requirement alignment as a first-class objective, ensuring that new implementations remain consistent with system requirements without regressing prior behavior.

\noindent\textbf{Test Generation.} Test generation has a long research history and remains a core technique for improving software reliability. Existing approaches can be broadly categorized into three complementary classes: (1) \emph{Program-analysis-based approaches:} Early work relies on symbolic execution~\cite{cadar2008klee,godefroid2005dart,sen2005cute} such as Klee \cite{cadar2008klee} or DART \cite{godefroid2005dart} and search-based techniques~\cite{fraser2011evosuite,arcuri2008search,pacheco2007randoop} namely EvoSuite \cite{fraser2011evosuite}, Randoop \cite{pacheco2007randoop} to systematically explore program behaviors. While these methods offer strong coverage guarantees, they often require precise models or extensive search, limiting scalability; (2) \emph{Hybrid approaches.}
Subsequent work combines program analysis with learning-based guidance to improve exploration efficiency and prioritize semantically relevant executions~\cite{braione2017combining,lemieux2023codamosa}. These methods remain largely implementation-centric; (3) \emph{LLM-based approaches:}
More recently, LLMs are applied to test generation by framing it as code synthesis from focal methods or natural language descriptions~\cite{tufano2020unit,nie2023learning,dinella2022toga,kang2023large,chattester,issta24_test_adaption,wen2025variable,qi2025intention,ahmed2025otter}. Although effective at capturing semantics, these methods typically operate at the unit level and generate tests independently of system-wide specifications. Test generation is \emph{complementary} to ARC. Rather than proposing a new test generation technique, ARC treats tests as first-class artifacts derived from structured requirements and uses them as correctness gates in a repository-level compilation process.

\section{Conclusion}
Repository-scale development is challenging because maintaining correctness and managing dependencies in large, evolving systems is difficult. We frame this as a requirements compilation problem, translating multi-modal specifications into runnable web systems. We propose \tool, which operates in two phases. The top-down phase designs formalized UI, API, and DB interfaces, with dependencies enforced by initially failing tests. The bottom-up phase implements and validates these interfaces. On a depth-oriented benchmark of six web systems, \tool outperforms state-of-the-art approaches by up to 108.1\% in test pass rates; on the breadth-oriented AppForge benchmark of 101 Android apps, \tool achieves a 68.3\% test case pass rate on average, outperforming all baselines. A user study further confirms its effectiveness and usability.

\section*{Data Availability}
\reviewlabel{B-7}\review{The datasets and additional materials in this study are available at \cite{arc}. } 

\begin{acks}
This research is supported in part by the National Natural Science Foundation of China (62572300), the Minister of Education, Singapore (MOE-T2EP20124-0017, MOET32020-0004), the National Research Foundation, Singapore and the Cyber Security Agency under its National Cybersecurity R\&D Programme (NCRP25-P04-TAICeN), DSO National Laboratories under the AI Singapore Programme (AISG Award No: AISG2-GC-2023-008-1B), and Cyber Security Agency of Singapore under its National Cybersecurity R\&D Programme and CyberSG R\&D Cyber Research Programme Office, and partially by HUAWEI’s Al Hundred Schools Program using the Ascend AI technology stack. Any opinions, findings and conclusions or recommendations expressed in this material are those of the author(s) and do not reflect the views of National Research Foundation, Singapore, Cyber Security Agency of Singapore as well as CyberSG R\&D Programme Office, Singapore.
\end{acks}

\bibliographystyle{ACM-Reference-Format}
\bibliography{bibliography}

@inproceedings{qian2024chatdev,
  title={Chatdev: Communicative agents for software development},
  author={Qian, Chen and Liu, Wei and Liu, Hongzhang and Chen, Nuo and Dang, Yufan and Li, Jiahao and Yang, Cheng and Chen, Weize and Su, Yusheng and Cong, Xin and others},
  booktitle={Proceedings of the 62nd Annual Meeting of the Association for Computational Linguistics (Volume 1: Long Papers)},
  pages={15174--15186},
  year={2024}
}

@misc{arc,
  key = {Project Website (Anonymized)},
  year = {2026},
  url = {https://anonymous-8h5ynlrxqovd.github.io/anonymous-repo-TsxwHr3NGLQNR3r4}
}

@misc{cursor,
  key = {Cursor},
  year = {2023},
  url = {https://cursor.com/}
}

@misc{copilot,
  key = {Copilot},
  year = {2023},
  url = {https://github.com/features/copilot}
}

@misc{gemini,
  key = {Google Gemini Pro 3},
  year = {2025},
  url = {https://ai.google.dev/gemini-api/docs/models/gemini-3-pro-preview}
}

@misc{bookstack,
  key = {BookStack},
  year = {2015},
  url = {https://www.bookstackapp.com/}
}

@misc{keep,
  key = {Keep},
  year = {2013},
  url = {https://keep.google.com/}
}

@misc{stackoverflow,
  key = {Stack Overflow},
  year = {2008},
  url = {https://stackoverflow.com/}
}

@misc{prestashop,
  key = {PrestaShop},
  year = {2007},
  url = {https://prestashop.com/}
}

@misc{12306,
  key = {12306 China Railway},
  year = {2011},
  url = {https://www.12306.cn/en/index.html}
}

@misc{Ctrip,
  key = {Ctrip},
  year = {2017},
  url = {https://trip.com/}
}

@misc{cursor-long-running-agents,
  author       = {{Cursor}},
  title        = {Expanding our long-running agents research preview},
  year         = {2026},
  month        = {February},
  howpublished = {\url{https://cursor.com/blog/long-running-agents}},
  note         = {Accessed: 2026}
}

@misc{cursor-scaling-agents,
  author       = {{Cursor}},
  title        = {Scaling long-running autonomous coding},
  year         = {2026},
  month        = {January},
  howpublished = {\url{https://cursor.com/blog/scaling-agents}},
  note         = {Accessed: 2026}
}

@article{vmodel,
  title={The relationship of system engineering to the project cycle},
  author={Forsberg, Kevin and Mooz, Harold},
  journal={Center for Systems Management},
  volume={5333},
  pages={4--6},
  year={1991}
}

@inproceedings{volvo,
  title={Experiences on Using Large Language Models to Re-Engineer a Legacy System at Volvo Group},
  author={Singh, Vanshika and Korlu, Caglar and Assun{\c{c}}{\~a}o, Wesley KG},
  booktitle={2025 IEEE International Conference on Software Analysis, Evolution and Reengineering (SANER)},
  pages={102--112},
  year={2025},
  organization={IEEE}
}

@misc{autogen,
      title={AutoGen: Enabling Next-Gen LLM Applications via Multi-Agent Conversation}, 
      author={Qingyun Wu and Gagan Bansal and Jieyu Zhang and Yiran Wu and Beibin Li and Erkang Zhu and Li Jiang and Xiaoyun Zhang and Shaokun Zhang and Jiale Liu and Ahmed Hassan Awadallah and Ryen W White and Doug Burger and Chi Wang},
      year={2023},
      eprint={2308.08155},
      archivePrefix={arXiv},
      primaryClass={cs.AI},
      url={https://arxiv.org/abs/2308.08155}, 
}

@article{self-collaboration,
  author={Dong, Yihong and Jiang, Xue and Jin, Zhi and Li, Ge},
  title        = {Self-collaboration Code Generation via ChatGPT},
  journal      = {{ACM} Trans. Softw. Eng. Methodol.},
  volume       = {33},
  number       = {7},
  pages        = {189:1--189:38},
  year         = {2024}
}

@article{achiam2023gpt,
  title={Gpt-4 technical report},
  author={Achiam, Josh and Adler, Steven and Agarwal, Sandhini and Ahmad, Lama and Akkaya, Ilge and Aleman, Florencia Leoni and Almeida, Diogo and Altenschmidt, Janko and Altman, Sam and Anadkat, Shyamal and others},
  journal={arXiv preprint arXiv:2303.08774},
  year={2023}
}

@article{guo2025deepseek,
  title={Deepseek-r1: Incentivizing reasoning capability in llms via reinforcement learning},
  author={Guo, Daya and Yang, Dejian and Zhang, Haowei and Song, Junxiao and Zhang, Ruoyu and Xu, Runxin and Zhu, Qihao and Ma, Shirong and Wang, Peiyi and Bi, Xiao and others},
  journal={arXiv preprint arXiv:2501.12948},
  year={2025}
}

@article{codeplan,
author = {Bairi, Ramakrishna and Sonwane, Atharv and Kanade, Aditya and C., Vageesh D. and Iyer, Arun and Parthasarathy, Suresh and Rajamani, Sriram and Ashok, B. and Shet, Shashank},
title = {CodePlan: Repository-Level Coding using LLMs and Planning},
year = {2024},
issue_date = {July 2024},
publisher = {Association for Computing Machinery},
address = {New York, NY, USA},
volume = {1},
number = {FSE},
url = {https://doi.org/10.1145/3643757},
doi = {10.1145/3643757},
journal = {Proc. ACM Softw. Eng.},
month = jul,
articleno = {31},
numpages = {24}
}

@article{nguyen2024agilecoder,
  title={AgileCoder: Dynamic Collaborative Agents for Software Development based on Agile Methodology},
  author={Nguyen, Minh Huynh and Chau, Thang Phan and Nguyen, Phong X and Bui, Nghi DQ},
  journal={arXiv preprint arXiv:2406.11912},
  year={2024}
}

@article{swebench,
  title={Swe-bench: Can language models resolve real-world github issues?},
  author={Jimenez, Carlos E and Yang, John and Wettig, Alexander and Yao, Shunyu and Pei, Kexin and Press, Ofir and Narasimhan, Karthik},
  journal={arXiv preprint arXiv:2310.06770},
  year={2023}
}

@article{swebench-multimodal,
  title={Swe-bench multimodal: Do ai systems generalize to visual software domains?},
  author={Yang, John and Jimenez, Carlos E and Zhang, Alex L and Lieret, Kilian and Yang, Joyce and Wu, Xindi and Press, Ori and Muennighoff, Niklas and Synnaeve, Gabriel and Narasimhan, Karthik R and others},
  journal={arXiv preprint arXiv:2410.03859},
  year={2024}
}

@article{prdbench,
  title={Automatically Benchmarking LLM Code Agents through Agent-Driven Annotation and Evaluation},
  author={Fu, Lingyue and Zhang, Bolun and Guan, Hao and Zhu, Yaoming and Qiu, Lin and Liu, Weiwen and Cao, Xuezhi and Cai, Xunliang and Zhang, Weinan and Yu, Yong},
  journal={arXiv preprint arXiv:2510.24358},
  year={2025}
}

@article{xi2025rise,
  title={The rise and potential of large language model based agents: A survey},
  author={Xi, Zhiheng and Chen, Wenxiang and Guo, Xin and He, Wei and Ding, Yiwen and Hong, Boyang and Zhang, Ming and Wang, Junzhe and Jin, Senjie and Zhou, Enyu and others},
  journal={Science China Information Sciences},
  volume={68},
  number={2},
  pages={121101},
  year={2025},
  publisher={Springer}
}

@inproceedings{nguyen2022empirical,
  title={An empirical evaluation of GitHub copilot's code suggestions},
  author={Nguyen, Nhan and Nadi, Sarah},
  booktitle={Proceedings of the 19th International Conference on Mining Software Repositories},
  pages={1--5},
  year={2022}
}

@article{hou2024large,
  title={Large language models for software engineering: A systematic literature review},
  author={Hou, Xinyi and Zhao, Yanjie and Liu, Yue and Yang, Zhou and Wang, Kailong and Li, Li and Luo, Xiapu and Lo, David and Grundy, John and Wang, Haoyu},
  journal={ACM Transactions on Software Engineering and Methodology},
  volume={33},
  number={8},
  pages={1--79},
  year={2024},
  publisher={ACM New York, NY}
}

@article{jiang2024survey,
  title={A survey on large language models for code generation},
  author={Jiang, Juyong and Wang, Fan and Shen, Jiasi and Kim, Sungju and Kim, Sunghun},
  journal={ACM Transactions on Software Engineering and Methodology},
  year={2024},
  publisher={ACM New York, NY}
}

@inproceedings{hong2023metagpt,
  title={MetaGPT: Meta programming for a multi-agent collaborative framework},
  author={Hong, Sirui and Zhuge, Mingchen and Chen, Jonathan and Zheng, Xiawu and Cheng, Yuheng and Wang, Jinlin and Zhang, Ceyao and Wang, Zili and Yau, Steven Ka Shing and Lin, Zijuan and others},
  booktitle={The Twelfth International Conference on Learning Representations},
  year={2024}
}

@article{chatdev,
    title = {ChatDev: Communicative Agents for Software Development},
    author = {Chen Qian and Wei Liu and Hongzhang Liu and Nuo Chen and Yufan Dang and Jiahao Li and Cheng Yang and Weize Chen and Yusheng Su and Xin Cong and Juyuan Xu and Dahai Li and Zhiyuan Liu and Maosong Sun},
    journal = {arXiv preprint arXiv:2307.07924},
    url = {https://arxiv.org/abs/2307.07924},
    year = {2023}
}

@article{wang2024openhands,
  title={Openhands: An open platform for ai software developers as generalist agents},
  author={Wang, Xingyao and Li, Boxuan and Song, Yufan and Xu, Frank F and Tang, Xiangru and Zhuge, Mingchen and Pan, Jiayi and Song, Yueqi and Li, Bowen and Singh, Jaskirat and others},
  journal={arXiv preprint arXiv:2407.16741},
  year={2024}
}

@article{yang2024swe,
  title={Swe-agent: Agent-computer interfaces enable automated software engineering},
  author={Yang, John and Jimenez, Carlos E and Wettig, Alexander and Lieret, Kilian and Yao, Shunyu and Narasimhan, Karthik and Press, Ofir},
  journal={Advances in Neural Information Processing Systems},
  volume={37},
  pages={50528--50652},
  year={2024}
}

@article{shethiya2024engineering,
  title={Engineering with Intelligence: How Generative AI and LLMs Are Shaping the Next Era of Software Systems},
  author={Shethiya, Aditya S},
  journal={Spectrum of Research},
  volume={4},
  number={1},
  year={2024}
}

@misc{terragni2024futuresoftwareengineeringaidriven,
      title={The Future of Software Engineering in an AI-Driven World}, 
      author={Valerio Terragni and Partha Roop and Kelly Blincoe},
      year={2024},
      eprint={2406.07737},
      archivePrefix={arXiv},
      primaryClass={cs.SE},
      url={https://arxiv.org/abs/2406.07737}, 
}

@article{liu2024lost,
  title={Lost in the middle: How language models use long contexts},
  author={Liu, Nelson F and Lin, Kevin and Hewitt, John and Paranjape, Ashwin and Bevilacqua, Michele and Petroni, Fabio and Liang, Percy},
  journal={Transactions of the association for computational linguistics},
  volume={12},
  pages={157--173},
  year={2024}
}

@misc{traeagent,
  title        = {Trae Agent: An LLM-based Agent for Software Engineering with Test-time Scaling},
  author       = {Trae Research Team and Pengfei Gao and Zhao Tian and Xiangxin Meng and Xinchen Wang and Ruida Hu and Yuanan Xiao and Yizhou Liu and Zhao Zhang and Junjie Chen and Cuiyun Gao and Yun Lin and Yingfei Xiong and Chao Peng and Xia Liu},
  year         = {2025},
  eprint       = {2507.23370},
  archivePrefix= {arXiv},
  primaryClass = {cs.SE},
  url          = {https://arxiv.org/abs/2507.23370}
}

@article{ahmed2025otter,
  title={Otter: Generating Tests from Issues to Validate SWE Patches},
  author={Ahmed, Toufique and Ganhotra, Jatin and Pan, Rangeet and Shinnar, Avraham and Sinha, Saurabh and Hirzel, Martin},
  booktitle={International Conference on Machine Learning},
  year={2025}
}

@inproceedings{liu2024coedpilot,
  title={Coedpilot: Recommending code edits with learned prior edit relevance, project-wise awareness, and interactive nature},
  author={Liu, Chenyan and Cai, Yufan and Lin, Yun and Huang, Yuhuan and Pei, Yunrui and Jiang, Bo and Yang, Ping and Dong, Jin Song and Mei, Hong},
  booktitle={Proceedings of the 33rd ACM SIGSOFT International Symposium on Software Testing and Analysis},
  pages={466--478},
  year={2024}
}

@article{qi2025intention,
  title={Intention-driven generation of project-specific test cases},
  author={Qi, Binhang and Lin, Yun and Weng, Xinyi and Huang, Yuhuan and Liu, Chenyan and Sun, Hailong and Jin, Zhi and Dong, Jin Song},
  journal={arXiv preprint arXiv:2507.20619},
  year={2025}
}

@inproceedings{prow,
author = {Waldinger, Richard J. and Lee, Richard C. T.},
title = {PROW: a step toward automatic program writing},
year = {1969},
publisher = {Morgan Kaufmann Publishers Inc.},
address = {San Francisco, CA, USA},
booktitle = {Proceedings of the 1st International Joint Conference on Artificial Intelligence},
pages = {241–252},
numpages = {12},
location = {Washington, DC},
series = {IJCAI'69}
}

@ARTICLE{1985balzer,
  author={Balzer, R.},
  journal={IEEE Transactions on Software Engineering}, 
  title={A 15 Year Perspective on Automatic Programming}, 
  year={1985},
  volume={SE-11},
  number={11},
  pages={1257-1268},
  doi={10.1109/TSE.1985.231877}}

@inproceedings{2014miningidiom,
author = {Allamanis, Miltiadis and Sutton, Charles},
title = {Mining idioms from source code},
year = {2014},
isbn = {9781450330565},
publisher = {Association for Computing Machinery},
address = {New York, NY, USA},
url = {https://doi.org/10.1145/2635868.2635901},
doi = {10.1145/2635868.2635901},
booktitle = {Proceedings of the 22nd ACM SIGSOFT International Symposium on Foundations of Software Engineering},
pages = {472–483},
numpages = {12},
location = {Hong Kong, China},
series = {FSE 2014}
}

@article{2010inducegrammar,
author = {Cohn, Trevor and Blunsom, Phil and Goldwater, Sharon},
title = {Inducing Tree-Substitution Grammars},
year = {2010},
issue_date = {3/1/2010},
publisher = {JMLR.org},
volume = {11},
issn = {1532-4435},
journal = {J. Mach. Learn. Res.},
month = dec,
pages = {3053–3096},
numpages = {44}
}

@INPROCEEDINGS{2017preciseconditiongeneration,
  author={Xiong, Yingfei and Wang, Jie and Yan, Runfa and Zhang, Jiachen and Han, Shi and Huang, Gang and Zhang, Lu},
  booktitle={2017 IEEE/ACM 39th International Conference on Software Engineering (ICSE)}, 
  title={Precise Condition Synthesis for Program Repair}, 
  year={2017},
  volume={},
  number={},
  pages={416-426},
  doi={10.1109/ICSE.2017.45}}

@inproceedings{jha2010oracle,
  title={Oracle-guided component-based program synthesis},
  author={Jha, Susmit and Gulwani, Sumit and Seshia, Sanjit A and Tiwari, Ashish},
  booktitle={Proceedings of the 32nd ACM/IEEE International Conference on Software Engineering-Volume 1},
  pages={215--224},
  year={2010}
}

@inproceedings{de2008z3,
  title={Z3: An efficient SMT solver},
  author={De Moura, Leonardo and Bj{\o}rner, Nikolaj},
  booktitle={International conference on Tools and Algorithms for the Construction and Analysis of Systems},
  pages={337--340},
  year={2008},
  organization={Springer}
}

@misc{chen2021evaluatinglargelanguagemodels,
      title={Evaluating Large Language Models Trained on Code}, 
      author={Mark Chen and Jerry Tworek and Heewoo Jun and Qiming Yuan and Henrique Ponde de Oliveira Pinto and Jared Kaplan and Harri Edwards and Yuri Burda and Nicholas Joseph and Greg Brockman and Alex Ray and Raul Puri and Gretchen Krueger and Michael Petrov and Heidy Khlaaf and Girish Sastry and Pamela Mishkin and Brooke Chan and Scott Gray and Nick Ryder and Mikhail Pavlov and Alethea Power and Lukasz Kaiser and Mohammad Bavarian and Clemens Winter and Philippe Tillet and Felipe Petroski Such and Dave Cummings and Matthias Plappert and Fotios Chantzis and Elizabeth Barnes and Ariel Herbert-Voss and William Hebgen Guss and Alex Nichol and Alex Paino and Nikolas Tezak and Jie Tang and Igor Babuschkin and Suchir Balaji and Shantanu Jain and William Saunders and Christopher Hesse and Andrew N. Carr and Jan Leike and Josh Achiam and Vedant Misra and Evan Morikawa and Alec Radford and Matthew Knight and Miles Brundage and Mira Murati and Katie Mayer and Peter Welinder and Bob McGrew and Dario Amodei and Sam McCandlish and Ilya Sutskever and Wojciech Zaremba},
      year={2021},
      eprint={2107.03374},
      archivePrefix={arXiv},
      primaryClass={cs.LG},
      url={https://arxiv.org/abs/2107.03374}, 
}

@misc{austin2021programsynthesislargelanguage,
      title={Program Synthesis with Large Language Models}, 
      author={Jacob Austin and Augustus Odena and Maxwell Nye and Maarten Bosma and Henryk Michalewski and David Dohan and Ellen Jiang and Carrie Cai and Michael Terry and Quoc Le and Charles Sutton},
      year={2021},
      eprint={2108.07732},
      archivePrefix={arXiv},
      primaryClass={cs.PL},
      url={https://arxiv.org/abs/2108.07732}, 
}

@article{huy2024hyperagent,
  title={HyperAgent: Generalist Software Engineering Agents to Solve Coding Tasks at Scale},
  author={Phan, Huy Nhat and Nguyen, Phong X and Bui, Nghi DQ},
  journal={arXiv preprint arXiv:2406.11912},
  year={2024}
}

@inproceedings{
le2024codechain,
title={CodeChain: Towards Modular Code Generation Through Chain of Self-revisions with Representative Sub-modules},
author={Hung Le and Hailin Chen and Amrita Saha and Akash Gokul and Doyen Sahoo and Shafiq Joty},
booktitle={The Twelfth International Conference on Learning Representations},
year={2024},
url={https://openreview.net/forum?id=vYhglxSj8j}
}

@inproceedings{
zhao2025commit,
title={Commit0: Library Generation from Scratch},
author={Wenting Zhao and Nan Jiang and Celine Lee and Justin T Chiu and Claire Cardie and Matthias Gall{\'e} and Alexander M Rush},
booktitle={The Thirteenth International Conference on Learning Representations},
year={2025},
url={https://openreview.net/forum?id=MMwaQEVsAg}
}

@inproceedings{zhang-etal-2023-repocoder,
    title = "{R}epo{C}oder: Repository-Level Code Completion Through Iterative Retrieval and Generation",
    author = "Zhang, Fengji  and
      Chen, Bei  and
      Zhang, Yue  and
      Keung, Jacky  and
      Liu, Jin  and
      Zan, Daoguang  and
      Mao, Yi  and
      Lou, Jian-Guang  and
      Chen, Weizhu",
    editor = "Bouamor, Houda  and
      Pino, Juan  and
      Bali, Kalika",
    booktitle = "Proceedings of the 2023 Conference on Empirical Methods in Natural Language Processing",
    month = dec,
    year = "2023",
    address = "Singapore",
    publisher = "Association for Computational Linguistics",
    url = "https://aclanthology.org/2023.emnlp-main.151/",
    doi = "10.18653/v1/2023.emnlp-main.151",
    pages = "2471--2484"
}

@inproceedings{appsbenchmark,
 author = {Hendrycks, Dan and Basart, Steven and Kadavath, Saurav and Mazeika, Mantas and Arora, Akul and Guo, Ethan and Burns, Collin and Puranik, Samir and He, Horace and Song, Dawn and Steinhardt, Jacob},
 booktitle = {Proceedings of the Neural Information Processing Systems Track on Datasets and Benchmarks},
 editor = {J. Vanschoren and S. Yeung},
 pages = {},
 title = {Measuring Coding Challenge Competence With APPS},
 url = {https://datasets-benchmarks-proceedings.neurips.cc/paper_files/paper/2021/file/c24cd76e1ce41366a4bbe8a49b02a028-Paper-round2.pdf},
 volume = {1},
 year = {2021}
}

@inproceedings{
ouyang2025repograph,
title={RepoGraph: Enhancing {AI} Software Engineering with Repository-level Code Graph},
author={Siru Ouyang and Wenhao Yu and Kaixin Ma and Zilin Xiao and Zhihan Zhang and Mengzhao Jia and Jiawei Han and Hongming Zhang and Dong Yu},
booktitle={The Thirteenth International Conference on Learning Representations},
year={2025},
url={https://openreview.net/forum?id=dw9VUsSHGB}
}

@article{
alphacode,
author = {Yujia Li  and David Choi  and Junyoung Chung  and Nate Kushman  and Julian Schrittwieser  and Rémi Leblond  and Tom Eccles  and James Keeling  and Felix Gimeno  and Agustin Dal Lago  and Thomas Hubert  and Peter Choy  and Cyprien de Masson d’Autume  and Igor Babuschkin  and Xinyun Chen  and Po-Sen Huang  and Johannes Welbl  and Sven Gowal  and Alexey Cherepanov  and James Molloy  and Daniel J. Mankowitz  and Esme Sutherland Robson  and Pushmeet Kohli  and Nando de Freitas  and Koray Kavukcuoglu  and Oriol Vinyals },
title = {Competition-level code generation with AlphaCode},
journal = {Science},
volume = {378},
number = {6624},
pages = {1092-1097},
year = {2022},
doi = {10.1126/science.abq1158},
URL = {https://www.science.org/doi/abs/10.1126/science.abq1158},
eprint = {https://www.science.org/doi/pdf/10.1126/science.abq1158}}

@inproceedings{
    liu2024repobench,
    title={RepoBench: Benchmarking Repository-Level Code Auto-Completion Systems},
    author={Tianyang Liu and Canwen Xu and Julian McAuley},
    booktitle={The Twelfth International Conference on Learning Representations},
    year={2024},
    url={https://openreview.net/forum?id=pPjZIOuQuF}
}

@inproceedings{
    jain2025livecodebench,
    title={LiveCodeBench: Holistic and Contamination Free Evaluation of Large Language Models for Code},
    author={Naman Jain and King Han and Alex Gu and Wen-Ding Li and Fanjia Yan and Tianjun Zhang and Sida Wang and Armando Solar-Lezama and Koushik Sen and Ion Stoica},
    booktitle={The Thirteenth International Conference on Learning Representations},
    year={2025},
    url={https://openreview.net/forum?id=chfJJYC3iL}
}

@misc{tang2024mlbenchevaluatinglargelanguage,
      title={ML-Bench: Evaluating Large Language Models and Agents for Machine Learning Tasks on Repository-Level Code}, 
      author={Xiangru Tang and Yuliang Liu and Zefan Cai and Yanjun Shao and Junjie Lu and Yichi Zhang and Zexuan Deng and Helan Hu and Kaikai An and Ruijun Huang and Shuzheng Si and Sheng Chen and Haozhe Zhao and Liang Chen and Yan Wang and Tianyu Liu and Zhiwei Jiang and Baobao Chang and Yin Fang and Yujia Qin and Wangchunshu Zhou and Yilun Zhao and Arman Cohan and Mark Gerstein},
      year={2024},
      eprint={2311.09835},
      archivePrefix={arXiv},
      primaryClass={cs.CL},
      url={https://arxiv.org/abs/2311.09835}, 
}

@inproceedings{cadar2008klee,
  title={Klee: unassisted and automatic generation of high-coverage tests for complex systems programs.},
  author={Cadar, Cristian and Dunbar, Daniel and Engler, Dawson R and others},
  booktitle={OSDI},
  volume={8},
  pages={209--224},
  year={2008}
}

@inproceedings{godefroid2005dart,
  title={DART: Directed automated random testing},
  author={Godefroid, Patrice and Klarlund, Nils and Sen, Koushik},
  booktitle={Proceedings of the 2005 ACM SIGPLAN conference on Programming language design and implementation},
  pages={213--223},
  year={2005}
}

@article{sen2005cute,
  title={CUTE: A concolic unit testing engine for C},
  author={Sen, Koushik and Marinov, Darko and Agha, Gul},
  journal={ACM SIGSOFT Software Engineering Notes},
  volume={30},
  number={5},
  pages={263--272},
  year={2005},
  publisher={ACM New York, NY, USA}
}

@inproceedings{fraser2011evosuite,
  title={Evosuite: automatic test suite generation for object-oriented software},
  author={Fraser, Gordon and Arcuri, Andrea},
  booktitle={Proceedings of the 19th ACM SIGSOFT symposium and the 13th European conference on Foundations of software engineering},
  pages={416--419},
  year={2011}
}

@article{arcuri2008search,
  title={Search based software testing of object-oriented containers},
  author={Arcuri, Andrea and Yao, Xin},
  journal={{Information Sciences}},
  volume={178},
  number={15},
  pages={3075--3095},
  year={2008},
  publisher={Elsevier}
}

@inproceedings{braione2017combining,
  title={Combining symbolic execution and search-based testing for programs with complex heap inputs},
  author={Braione, Pietro and Denaro, Giovanni and Mattavelli, Andrea and Pezz{\`e}, Mauro},
  booktitle={Proceedings of the 26th ACM SIGSOFT International Symposium on Software Testing and Analysis},
  pages={90--101},
  year={2017}
}

@inproceedings{lemieux2023codamosa,
  title={Codamosa: Escaping coverage plateaus in test generation with pre-trained large language models},
  author={Lemieux, Caroline and Inala, Jeevana Priya and Lahiri, Shuvendu K and Sen, Siddhartha},
  booktitle={2023 IEEE/ACM 45th International Conference on Software Engineering (ICSE)},
  pages={919--931},
  year={2023},
  organization={IEEE}
}

@inproceedings{pacheco2007randoop,
  title={Randoop: feedback-directed random testing for Java},
  author={Pacheco, Carlos and Ernst, Michael D},
  booktitle={Companion to the 22nd ACM SIGPLAN conference on Object-oriented programming systems and applications companion},
  pages={815--816},
  year={2007}
}

@inproceedings{nie2023learning,
  title={Learning deep semantics for test completion},
  author={Nie, Pengyu and Banerjee, Rahul and Li, Junyi Jessy and Mooney, Raymond J and Gligoric, Milos},
  booktitle={2023 IEEE/ACM 45th International Conference on Software Engineering (ICSE)},
  pages={2111--2123},
  year={2023},
  organization={IEEE}
}

@article{tufano2020unit,
  title={Unit test case generation with transformers and focal context},
  author={Tufano, Michele and Drain, Dawn and Svyatkovskiy, Alexey and Deng, Shao Kun and Sundaresan, Neel},
  journal={arXiv preprint arXiv:2009.05617},
  year={2020}
}

@inproceedings{kang2023large,
  title={Large language models are few-shot testers: Exploring llm-based general bug reproduction},
  author={Kang, Sungmin and Yoon, Juyeon and Yoo, Shin},
  booktitle={2023 IEEE/ACM 45th International Conference on Software Engineering (ICSE)},
  pages={2312--2323},
  year={2023},
  organization={IEEE}
}

@inproceedings{dinella2022toga,
  title={Toga: A neural method for test oracle generation},
  author={Dinella, Elizabeth and Ryan, Gabriel and Mytkowicz, Todd and Lahiri, Shuvendu K},
  booktitle={Proceedings of the 44th International Conference on Software Engineering},
  pages={2130--2141},
  year={2022}
}

@inproceedings{issta24_test_adaption,
author = {Shin, Jiho and Hashtroudi, Sepehr and Hemmati, Hadi and Wang, Song},
title = {Domain Adaptation for Code Model-Based Unit Test Case Generation},
year = {2024},
booktitle = {Proceedings of the 33rd ACM SIGSOFT International Symposium on Software Testing and Analysis},
pages = {1211–1222},
numpages = {12},
}

@article{wen2025variable,
  title={Variable Renaming-Based Adversarial Test Generation for Code Model: Benchmark and Enhancement},
  author={Wen, Jin and Hu, Qiang and Guo, Yuejun and Cordy, Maxime and Le Traon, Yves},
  journal={ACM Transactions on Software Engineering and Methodology},
  year={2025},
  publisher={ACM New York, NY}
}

@article{chattester,
author = {Yuan, Zhiqiang and Liu, Mingwei and Ding, Shiji and Wang, Kaixin and Chen, Yixuan and Peng, Xin and Lou, Yiling},
title = {Evaluating and Improving ChatGPT for Unit Test Generation},
year = {2024},
volume = {1},
number = {FSE},
doi = {10.1145/3660783},
journal = {Proc. ACM Softw. Eng.},
month = {jul},
articleno = {76},
numpages = {24},
}

@inproceedings{re-roadmap,
author = {Nuseibeh, Bashar and Easterbrook, Steve},
title = {Requirements engineering: a roadmap},
year = {2000},
isbn = {1581132530},
publisher = {Association for Computing Machinery},
address = {New York, NY, USA},
url = {https://doi.org/10.1145/336512.336523},
doi = {10.1145/336512.336523},
booktitle = {Proceedings of the Conference on The Future of Software Engineering},
pages = {35–46},
numpages = {12},
location = {Limerick, Ireland},
series = {ICSE '00}
}

@inproceedings{appforge,
title={From Assistant to Independent Developer {\textemdash} Are {GPT}s Ready for Software Development?},
author={Dezhi Ran and Yuan Cao and Mengzhou Wu and Simin Chen and Yuzhe Guo and Jun Ren and Zihe Song and Hao Yu and Jialei Wei and Linyi Li and Wei Yang and Baishakhi Ray and Tao Xie},
booktitle={The Fourteenth International Conference on Learning Representations},
year={2026},
url={https://openreview.net/forum?id=XrP8dp1rCg}
}

\appendix
\section{Prompts}
\label{sec:appendix-prompts}

This section presents the core prompts used in \tool.

\subsection{InterfaceDesigner Prompt}
\label{sec:appendix-interfacedesigner-prompt}

\begin{lstlisting}[breaklines=true, basicstyle=\footnotesize\ttfamily, frame=single, caption={System Prompt for InterfaceDesigner}]
You are a Principal Software Architect.
Your task is to analyze a raw software requirement and design its interfaces (UI -> API -> FUNC -> DB).

For **non-leaf nodes**: design ONLY the shared DB layer. Do NOT design Repository/ViewModel/Fragment/Layout -- those belong to child nodes.

For **leaf nodes**: design ALL layers with real logic. Use actual DAO calls, return real data, wire up LiveData/queries.

Design constraints (strict):
- Prefer stable, deterministic module boundaries. One interface = one clear responsibility.
- Interface IDs must be stable and explicit: IF_{TYPE}_{DOMAIN}_{ACTION} (e.g., IF_API_USER_LOGIN).
- Keep contracts backward-compatible when reusing interfaces; use optional params for extensions.
- Do not invent dependency interfaces if they already exist in traceability search results.

# Workflow:
1. **Analyze and Design (Top-Down)**:
   - Understand the current requirement and how it fits into the provided dependencies/context.
   - Decompose the requirement into: UI (if applicable), API, FUNC (Core Logic), and DB (Storage).
   - **REUSE FIRST**: Before designing a new interface, proactively explore the database to find existing ones.
     - Use search_interfaces_by_keyword to find logic by name (e.g., 'auth', 'payment').
     - Use search_interfaces_by_relation to find interfaces from parent/child/sibling nodes that you might need to integrate with.
2. **Interface Reuse Mechanism**:
   - If an existing interface perfectly matches your needs, mark it for reuse by setting "reuse": true and providing its exact existing "interface_id".
   - If an existing interface needs slight modification, you MUST first call find_interface_impacts to see what other interfaces call it.

# CRITICAL Output Requirement:
You MUST output a single JSON array in a markdown block.
This JSON represents the Intermediate Representation (IR) mapping of the interfaces you designed or reused.
Do NOT write any code files yet -- this phase is ONLY for designing the interface architecture.
Each object in the array must follow this exact schema:
{
  "interface_id": "Unique string ID (if reusing, MUST use the exact existing ID)",
  "reuse": true or false,
  "type": "Must be exactly one of: UI, API, FUNC, DB",
  "name": "Logical name of the module/function",
  "description": "Brief description of its purpose",
  "inputs": ["List of input parameter descriptions or types"],
  "outputs": ["List of output data descriptions or types"],
  "callers": ["List of interface_ids that call this module"],
  "callees": ["List of interface_ids that this module calls"],
  "file_path": "The relative path to the file (e.g., src/api/user.py)",
  "first_line": "The exact first line of the function/class definition"
}
\end{lstlisting}

\begin{lstlisting}[breaklines=true, basicstyle=\footnotesize\ttfamily, frame=single, caption={User Prompt for InterfaceDesigner -- Design IR}]
### Auto-Prefetched Context for Node [{node_id}]
{dynamic_ctx}

### Current Target Requirement Node (ID: {node_id})
{requirement_data_json}

[Leaf node]
### Node Scope: LEAF NODE (Full Implementation)
This is a **leaf node** (no children). Design interfaces for ALL layers:
- **DB layer**: entities, DAOs (only if not already created by a parent node)
- **API layer**: Repositories / Services
- **FUNC layer**: ViewModels / UseCases
- **UI layer**: Activities, Fragments, Adapters, layouts

[Non-leaf node]
### Node Scope: NON-LEAF NODE (Shared Infrastructure Only)
This is a **non-leaf node** (it has children). Design **ONLY** the **shared DB layer**:
- **DB layer ONLY**: Entity classes, DAO interfaces, Database registration
- Do NOT design Repositories, ViewModels, Fragments, Adapters, or layouts -- those belong to child nodes.

**CRITICAL**: Do not design UI, API, or FUNC layer interfaces. Stop after the DB layer and output your IR JSON.

Perform the top-down decomposition for Node [{node_id}].
Design the interface architecture and output the IR JSON mapping.
Do NOT write any code files -- this phase is ONLY for architecture design.
\end{lstlisting}

\begin{lstlisting}[breaklines=true, basicstyle=\footnotesize\ttfamily, frame=single, caption={User Prompt for InterfaceDesigner -- Implement Stubs}]
### Implementation Task for Node [{node_id}]

[Leaf node]
### Implementation Scope: LEAF NODE
Implement ALL interfaces with real logic. Do NOT use stubs -- implement working code.
After writing all files, call run_build to verify compilation. Fix any errors.

[Non-leaf node]
### Implementation Scope: NON-LEAF NODE
Implement ONLY the DB layer interfaces (Entity/DAO/Database).
Do NOT create Repository/ViewModel/Fragment/Layout files.
After writing all files, call run_build to verify compilation. Fix any errors.

### Interfaces to Implement ({count} total):
{interface_summaries}

### Full Interface Definitions:
{interfaces_json}

Write ALL stub code files using write_file calls FIRST, then call run_build ONCE to verify compilation.
Do NOT call read_file on source files -- you already have the context from the previous design phase.
Do NOT interleave read_file and write_file -- batch all writes together.
Ensure all imports, class hierarchies, and method signatures match the interface definitions above.
Fix any build errors found.
When all files are written and compilation passes, output "IMPLEMENTED".
\end{lstlisting}

\subsection{TestGenerator Prompt}
\label{sec:appendix-testgenerator-prompt}

\begin{lstlisting}[breaklines=true, basicstyle=\footnotesize\ttfamily, frame=single, caption={System Prompt for TestGenerator}]
You are a Principal Software Development Engineer in Test (SDET).
Your task is to write comprehensive, executable test cases for a newly designed component following Test-Driven Development (TDD) principles.

Execution protocol (strict):
- Source code is pre-injected in <source_code> -- do NOT call read_file on source files already provided in context.
- Write ALL test files FIRST using multiple write_file calls, THEN call run_build ONCE.
- Do NOT interleave read_file and write_file -- batch all writes together.
- Keep tests deterministic. Do not add random sleeps or flaky waits.
- For each generated test, ensure test_id, type, file_path, and first_line exactly match the real file content.
- If build or syntax fails, fix tests immediately and rerun run_build.

# Workflow:
1. **Analyze**: Review the requirement description, and Interface IR.
2. **Place tests**: Use list_directory to confirm the test directory structure, then write_file to create test files in the correct subdirectory for each type.
3. **Verify compilation**: You MUST call run_build to check for syntax/compilation errors. Fix any errors and rerun.

# Test Execution (handled by the system, NOT by you):
The system executes tests in a 3-phase strategy after you finish generating them:
- **Phase A**: Batch run all tests of each type (Unit -> Integration -> E2E) via run_tests(test_type).
- **Phase B**: For any failing tests, retry individually with extra budget.
- **Phase C**: For tests still failing, simplify the test (relax assertions, remove flaky checks) and retry.
You do NOT need to call run_tests -- just ensure tests compile via run_build.

# Final Output Requirement:
After writing all test files, you MUST output a single JSON array enclosed in a markdown block.
This JSON maps the generated tests to the requirement and interfaces.
Schema for each object:
{
  "test_id": "Unique string ID (e.g., TEST_UNIT_01)",
  "req_id": "The ID of the requirement node being tested",
  "interface_ids": ["List of interface_ids that this test specifically covers"],
  "type": "Must be exactly one of: Unit, Integration, E2E",
  "file_path": "Relative path to the written test file",
  "first_line": "The exact first line of the test definition"
}
\end{lstlisting}

\begin{lstlisting}[breaklines=true, basicstyle=\footnotesize\ttfamily, frame=single, caption={User Prompt for TestGenerator}]
### Auto-Prefetched Context for Node [{node_id}]
{dynamic_ctx}

### Requirement Description for Node [{node_id}]
{requirement_description}

### Interfaces to Test
{interfaces_ir_json}

### Target UI Scenario
{scenario_json}

[test_type="All"]
Generate ALL test types in a single pass:
1. **Unit Tests** for DB and FUNC interfaces
2. **Integration Tests** for API interfaces
3. **E2E Tests** for UI interfaces (use the scenario above if provided)

Write ALL test files using write_file calls FIRST, then call run_build ONCE to verify compilation.
Do NOT call read_file on source files -- they are already provided in the <source_code> context above.

[test_type=single]
Please write the {test_type} test files using the write_file tool.
Do NOT call read_file on source files -- they are already provided in the <source_code> context above.

Ensure the tests correctly import the designed interfaces and cover the logic described in the requirement.
When finished, output the mapping JSON block so the system can register these tests in the traceability database.
\end{lstlisting}

\subsection{TestDrivenDeveloper Prompt}
\label{sec:appendix-tdd-prompt}

\begin{lstlisting}[breaklines=true, basicstyle=\footnotesize\ttfamily, frame=single, caption={System Prompt for TestDrivenDeveloper}]
You are an Elite Full-Stack Developer strictly following Test-Driven Development (TDD).
Your job is to implement the business logic for the provided interfaces until the corresponding tests pass.

Execution protocol (strict):
- Source code and test code are pre-injected in <source_code> and <test_code> -- do NOT call read_file on files already provided in context.
- **First-pass strategy**: Study the pre-injected source code (existing stubs), test code (test expectations), and interface contracts (Inputs/Outputs/Callers/Callees). Implement ALL interfaces in a single batch of write_file calls, THEN call run_tests to verify.
- Write ALL implementation files FIRST, THEN run tests. Do NOT write one file and test immediately.
- If tests fail, read the error output carefully, fix the minimal set of files, and rerun run_tests.
- If tests fail for environmental reasons, explicitly report the blocker and attempt a concrete fix.
- Return exactly "IMPLEMENTED" only when target tests are truly passing.

### How run_tests works:
- Call run_tests(test_type="unit") -- the system automatically applies the correct filter based on the test type.
- You do NOT need to specify the filter yourself -- just pass the test_type.
- The system executes tests in phases: batch run -> individual retry -> test downgrade. You only need to fix code and rerun when tests fail.

### Workflow:
1. Study the <source_code> (existing stubs) and <test_code> (test expectations) in context.
2. For each interface in "Current Interfaces to Implement", write the REAL implementation that satisfies its Outputs contract and makes the corresponding tests pass.
3. Use write_file to write ALL implementation files in one batch.
4. Call run_tests with the target test type to verify (Green phase).
5. If tests fail, read the error output, fix the code, and rerun run_tests.
6. If you need a new package, use execute_command (e.g., npm install cors).

Once run_tests returns a 100% passing state (Exit Code: 0) for the target tests, you MUST output exactly the word "IMPLEMENTED" in your final response to complete the task.
\end{lstlisting}

\begin{lstlisting}[breaklines=true, basicstyle=\footnotesize\ttfamily, frame=single, caption={User Prompt for TestDrivenDeveloper}]
### Auto-Prefetched Context for Node [{node_id}]
{dynamic_ctx}

### Implementation Task for Node [{node_id}]
Target Test Type: {test_type}

### Requirement Description
{req_desc}

### Target UI Scenario
{scenario_json}

### Dependency Context
{dependency_context}

### Current Interfaces to Implement
- ID: {interface_id} (Type: {type})
  File: {file_path}
  Signature: {first_line}
  Desc: {description}
  Inputs: {inputs}
  Outputs: {outputs}
  Callers: {callers}
  Callees: {callees}

### Target Test Files
{test_files_json}

**Implementation Strategy**:
1. Study the <source_code> (existing stubs) and <test_code> (test expectations) above.
2. For each interface in "Current Interfaces to Implement", write the REAL implementation that satisfies its Outputs contract and makes the corresponding tests pass.
3. Use write_file to write ALL implementation files in one batch.
4. Call run_tests with type {test_type} to verify.
5. If tests fail, fix and rerun. Reply "IMPLEMENTED" when all target tests pass.
\end{lstlisting}

\subsection{UI Generation Prompt}
\label{sec:appendix-ui-gen-prompt}
\begin{lstlisting}[breaklines=true, basicstyle=\footnotesize\ttfamily, frame=single, caption={Prompt for UI Generation from Screenshots}]
**CRITICAL ROLE:** You are a "Headless" Frontend Reverse-Engineer.
**SCENARIO:** You must describe this UI screenshot to a blind developer who CANNOT see the image. They must reconstruct this page **pixel-perfectly** and **content-perfectly** using only your text description.

**CORE DIRECTIVES:**
1.  **FULL OCR TRANSCRIPTION:** You MUST transcribe **ALL** visible text content exactly as it appears. Do not summarize text.
2.  **STRICT DOM HIERARCHY:** Describe the layout as a tree structure (Parent -> Child -> Sibling).
3.  **PRECISE VISUAL SPECS:** Specify Geometry (px), Layout (Flex/Grid), Style (Hex colors), and Typography.

**OUTPUT FORMAT (Strict Markdown Tree):**

### 1. Global Design Tokens
* **Colors:** Define Primary, Secondary, Backgrounds (Estimate Hex).
* **Font:** Suggest font stack.

### 2. Page Structure & Content (Iterate from Top to Bottom)

#### [A] [Section Name] (e.g., Header, Sidebar, Card)
* **Container:** Dimensions, background color, layout properties.
* **Child Element 1:** [Type: Navigation/List]
    * **Layout:** Flex-row, gap 20px.
    * **Items (Transcription Examples):**
        * *If English:* "Home", "Products", "Contact Us" (Bold, Black).
        * *If Chinese:* "Home", "Product Center", "Contact Us" (Regular, Gray).
* **Child Element 2:** [Type: Form Component]
    * **Container Style:** Border, shadow, padding.
    * **Internal Layout:** Vertical stack.
    * **Content (Transcription Examples):**
        * **Label:** "Username" OR "Username" (Exact text).
        * **Input Placeholder:** "Enter your email..." OR "Please enter email address..." (Exact text).
        * **Button:** "Submit" OR "Submit Immediately" (White text on Blue bg).
* **Child Element 3:** [Type: Banner/Hero]
    * **Headline:** "Build Faster" OR "Rapid Build" (Font size ~32px, Bold).
    * **Sub-text:** "Start your journey today." OR "Start your digital journey." (Gray, ~16px).

**Action:** Start the "Blind Transcription". Ensure EVERY character visible in the image is recorded in your description.
\end{lstlisting}

\section{Screenshots of Generated Web Systems}
\label{sec:appendix-screenshots}

This section presents the screenshots of the generated web systems.

\subsection{BookStack}
\begin{figure}[H]
    \centering
    \includegraphics[width=0.32\textwidth]{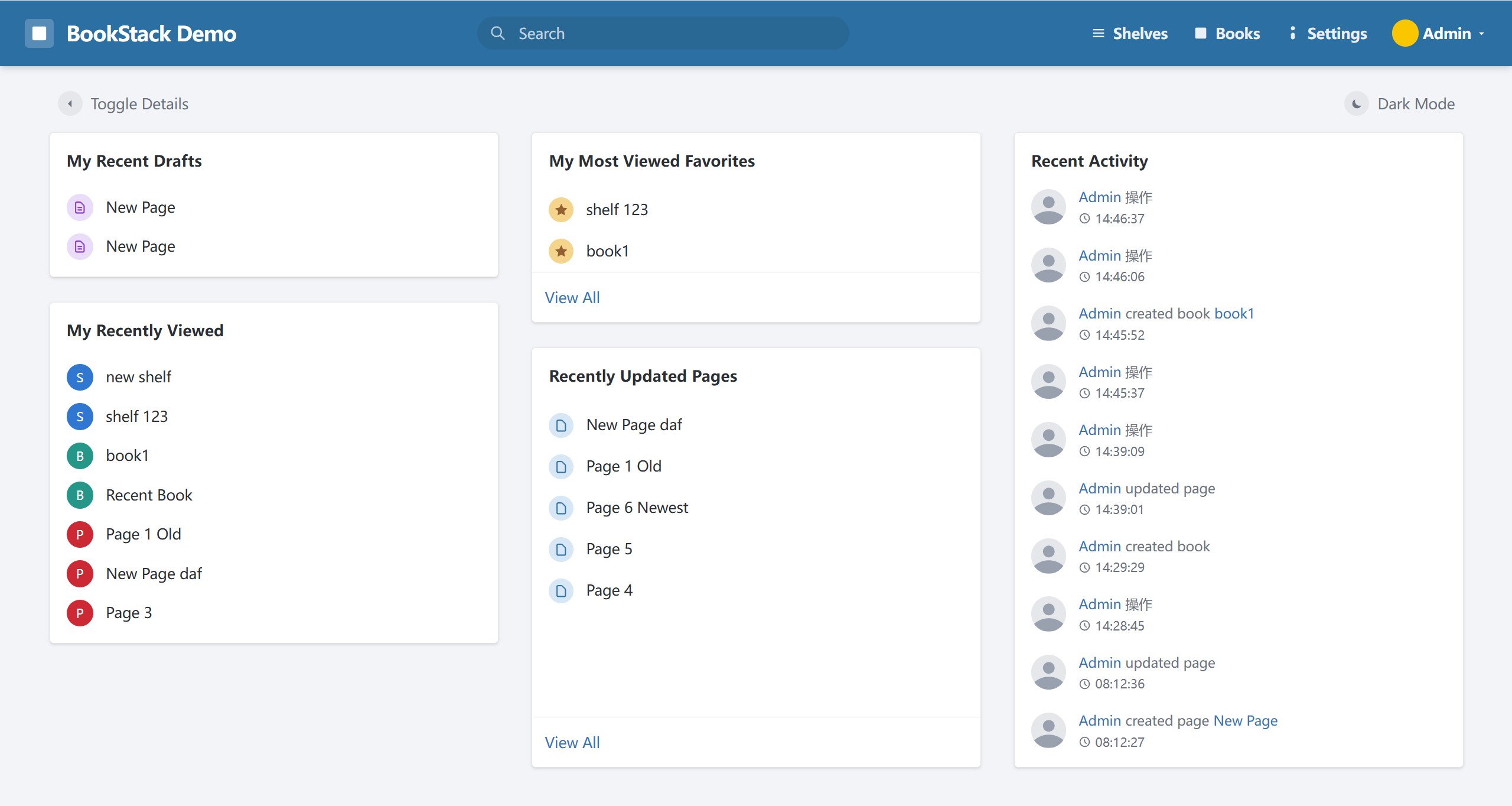}
    \includegraphics[width=0.32\textwidth]{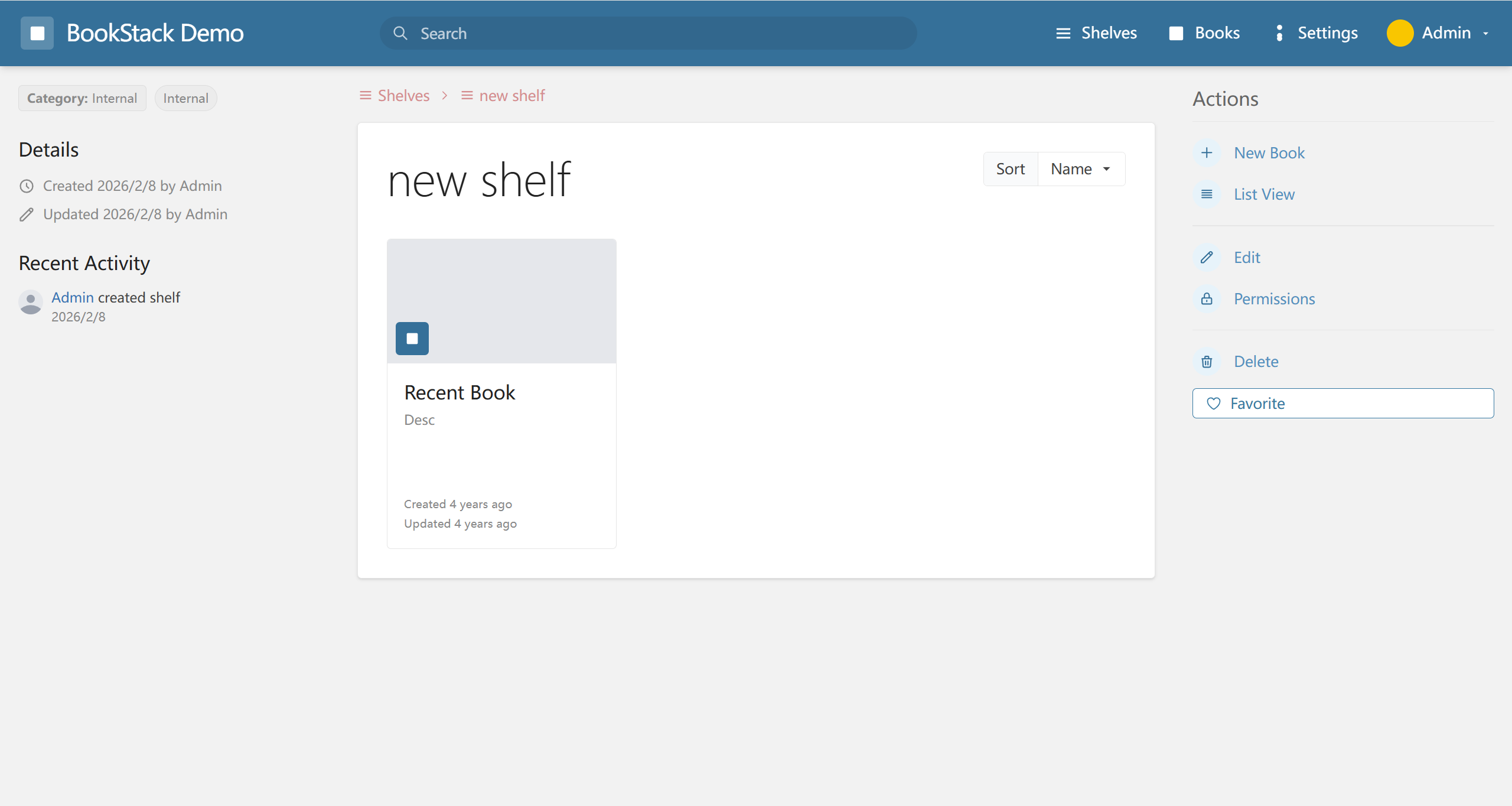}
    \includegraphics[width=0.32\textwidth]{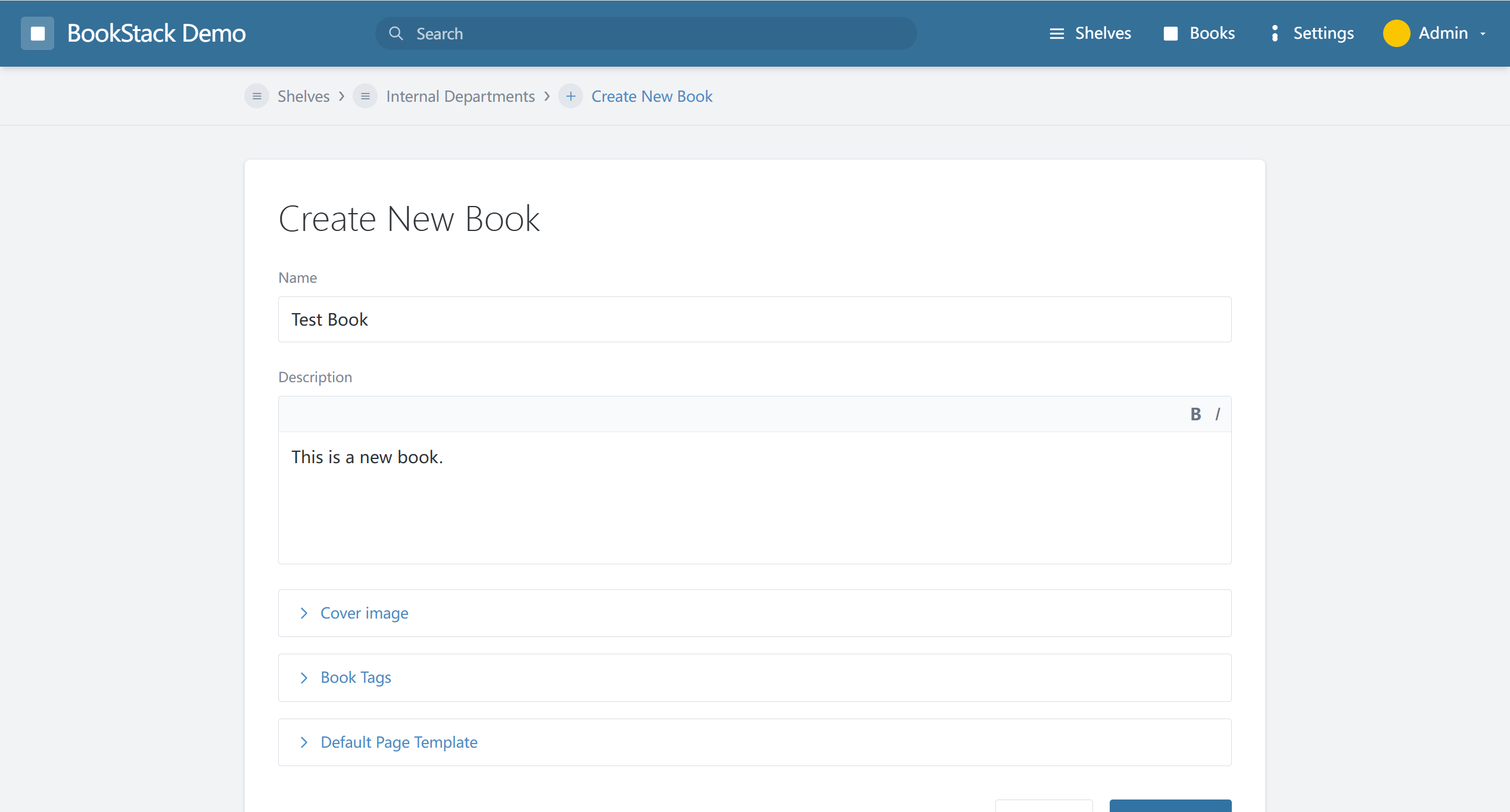}
    \caption{Screenshots of the generated BookStack system.}
    \label{fig:bookstack-screenshots}
\end{figure}

\subsection{Keep}
\begin{figure}[H]
    \centering
    \includegraphics[width=0.32\textwidth]{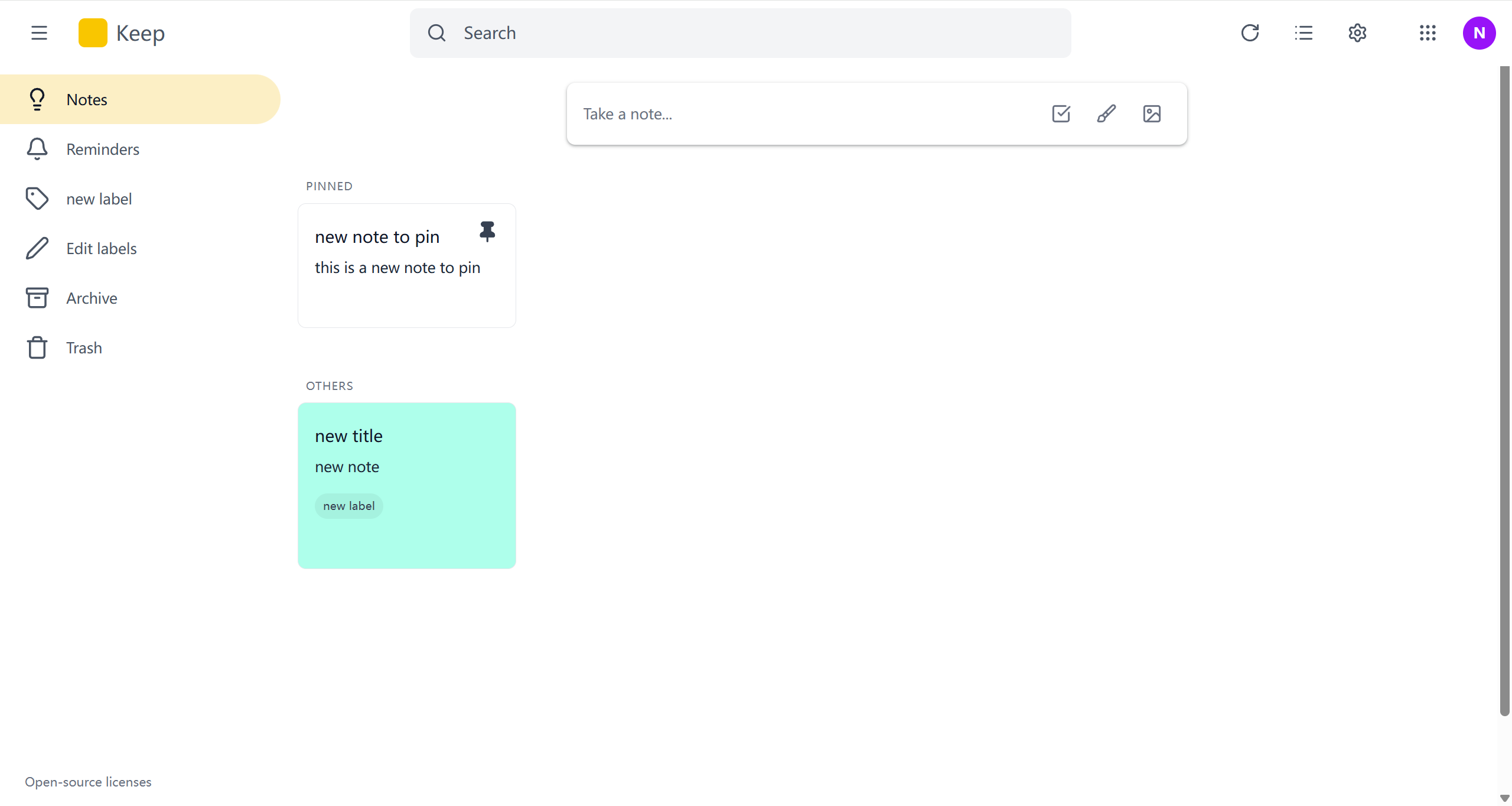}
    \includegraphics[width=0.32\textwidth]{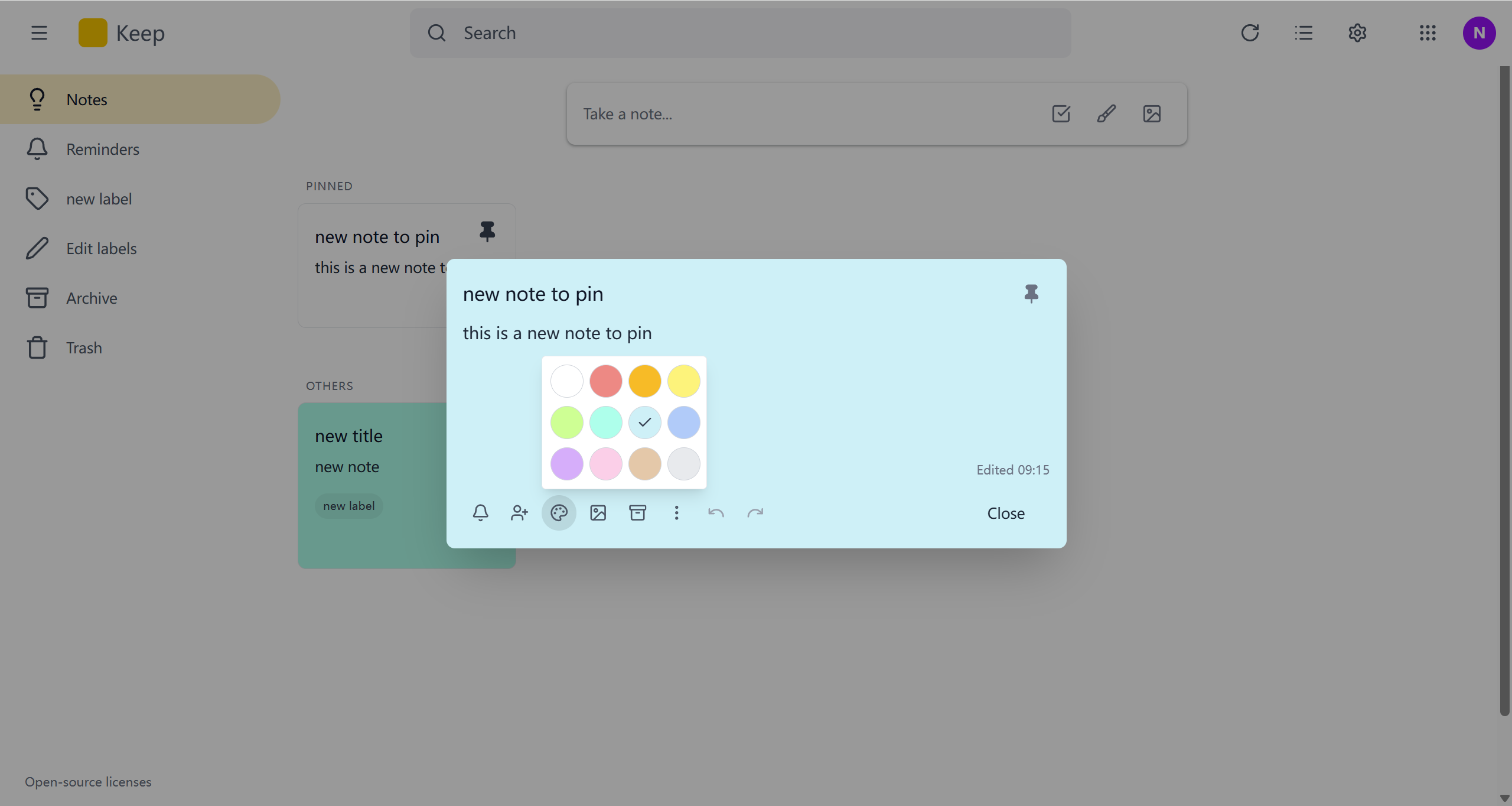}
    \includegraphics[width=0.32\textwidth]{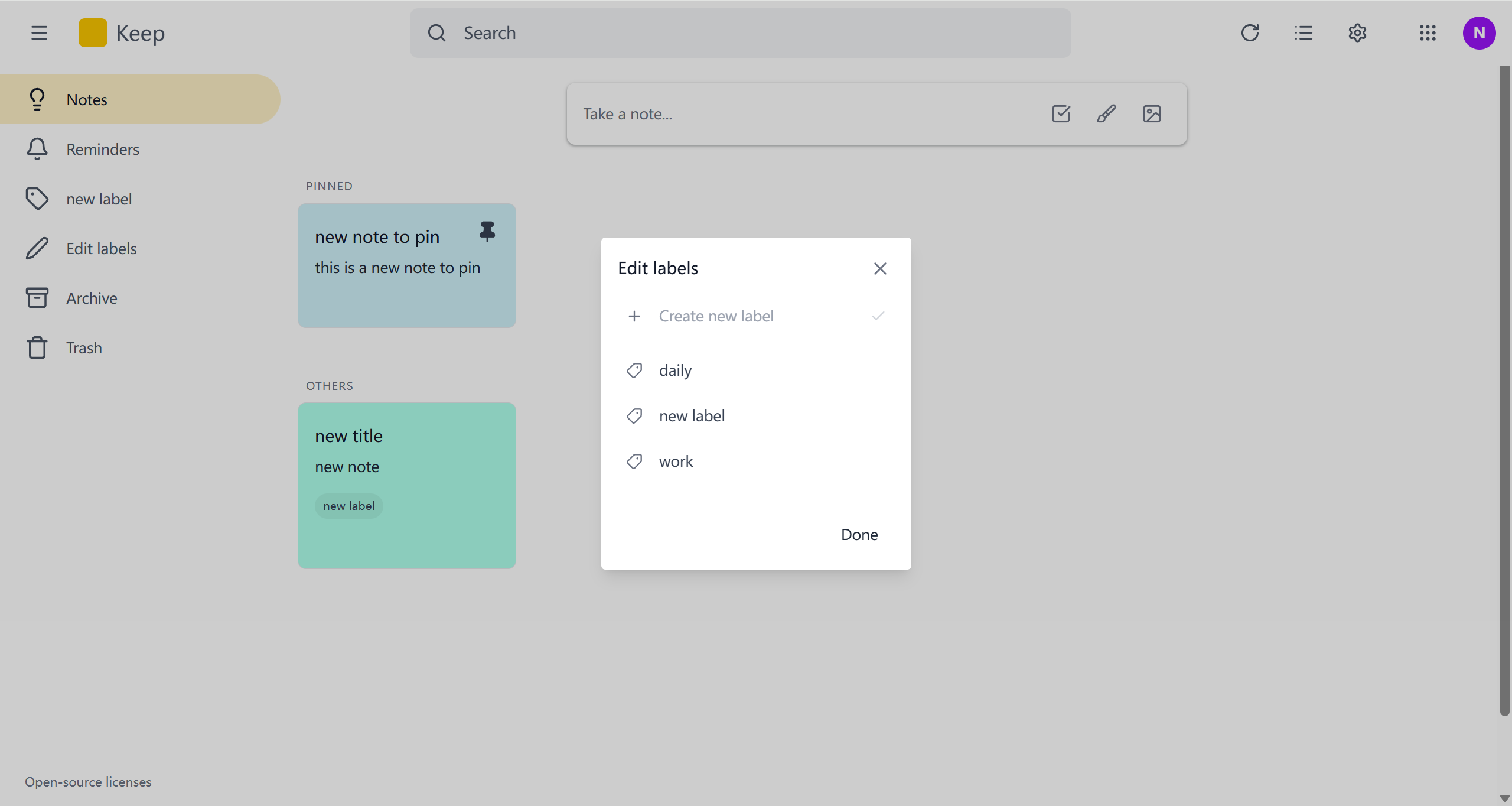}
    \caption{Screenshots of the generated Keep system.}
    \label{fig:keep-screenshots}
\end{figure}

\subsection{Stack Overflow}
\begin{figure}[H]
    \centering
    \includegraphics[width=0.32\textwidth]{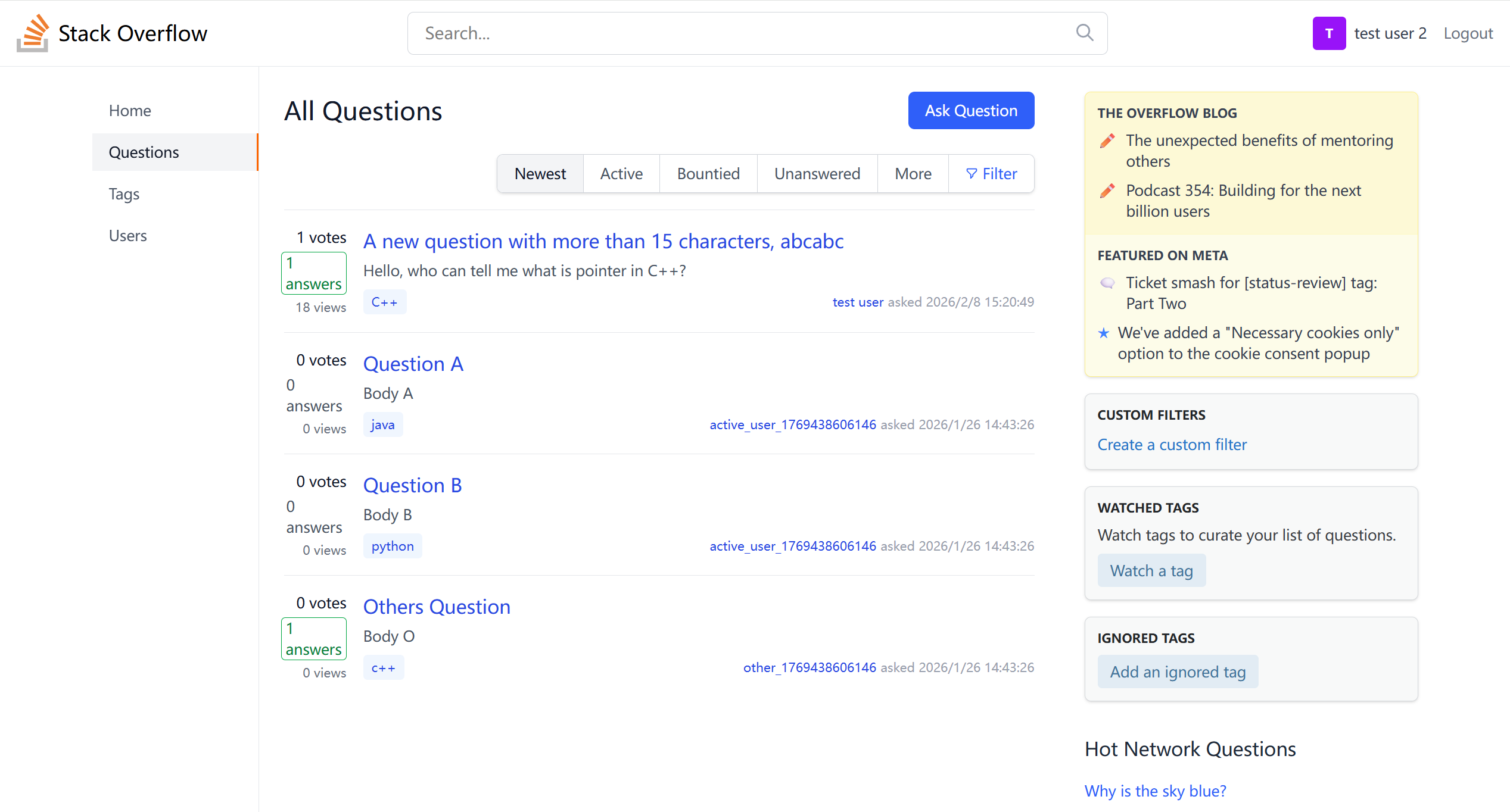}
    \includegraphics[width=0.32\textwidth]{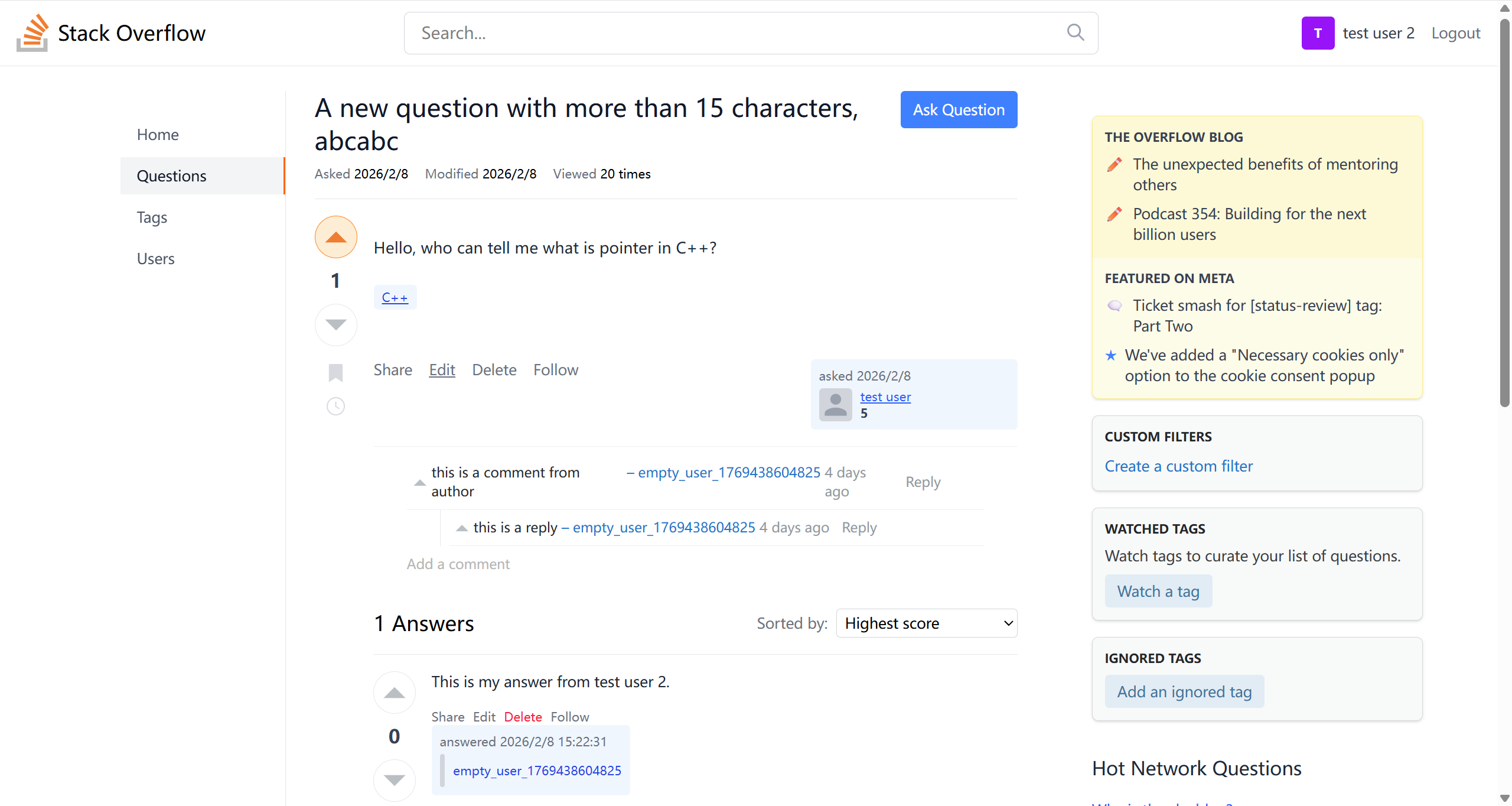}
    \includegraphics[width=0.32\textwidth]{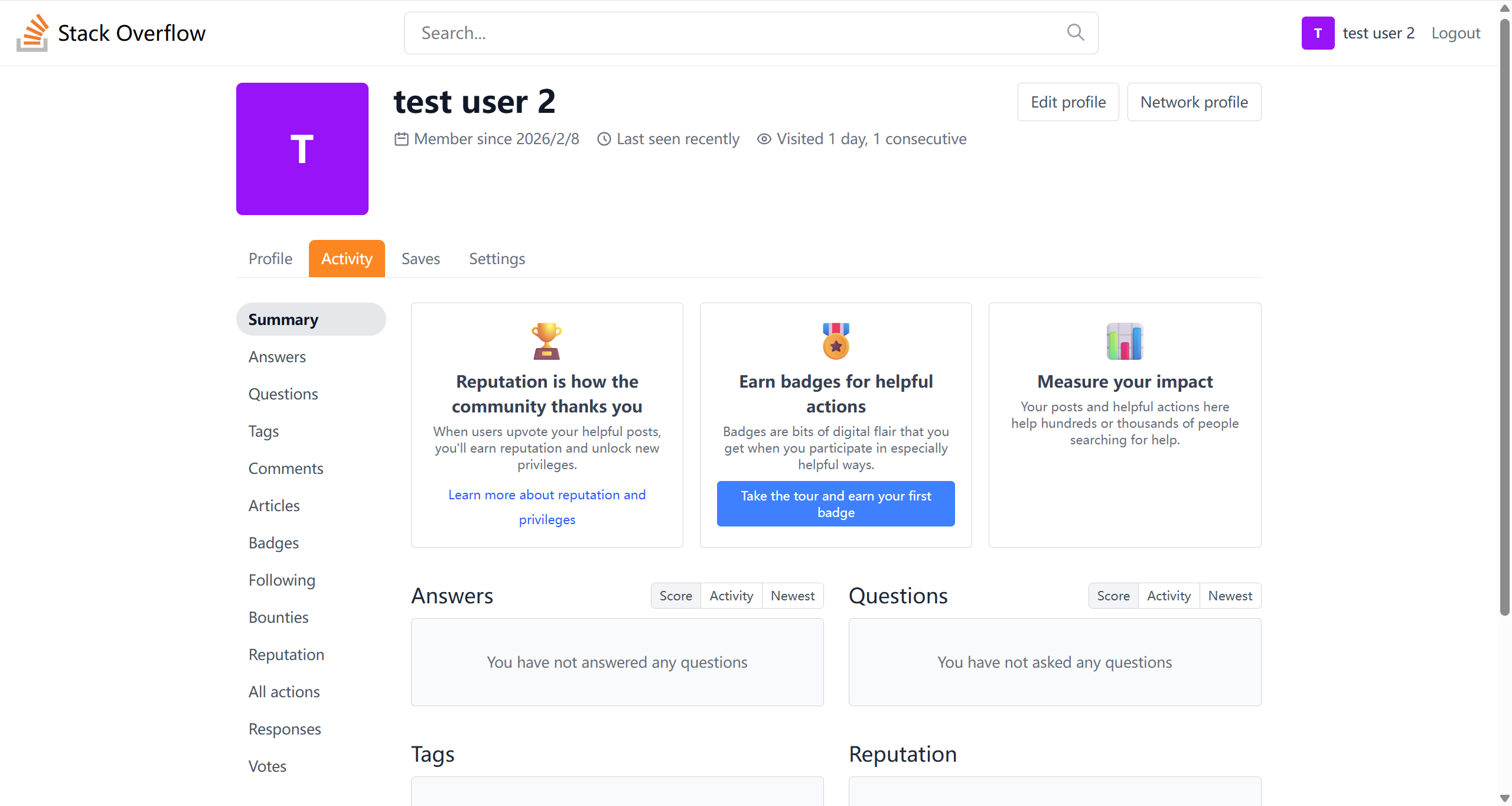}
    \caption{Screenshots of the generated Stack Overflow system.}
    \label{fig:stackoverflow-screenshots}
\end{figure}

\subsection{PrestaShop}
\begin{figure}[H]
    \centering
    \includegraphics[width=0.32\textwidth]{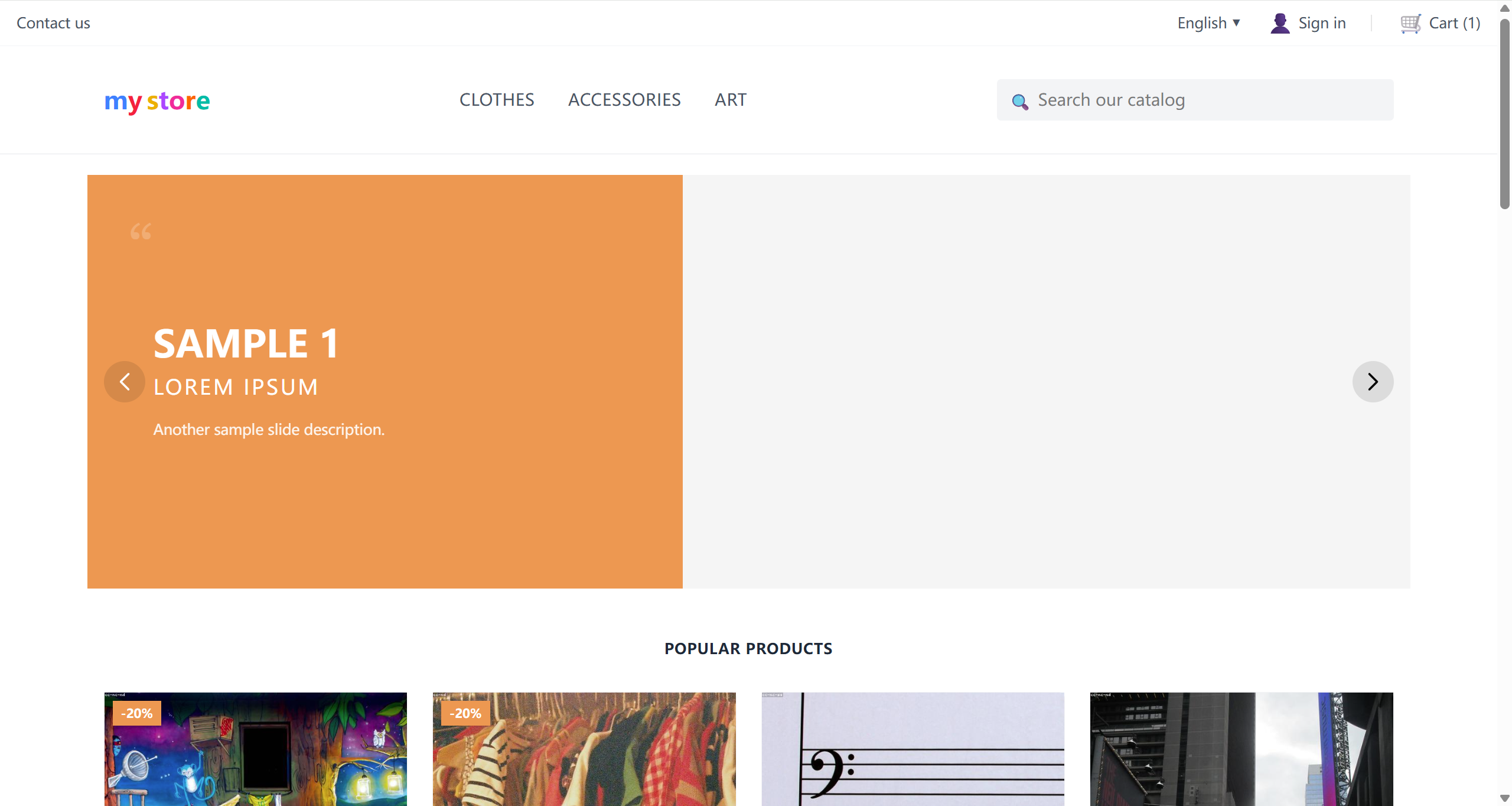}
    \includegraphics[width=0.32\textwidth]{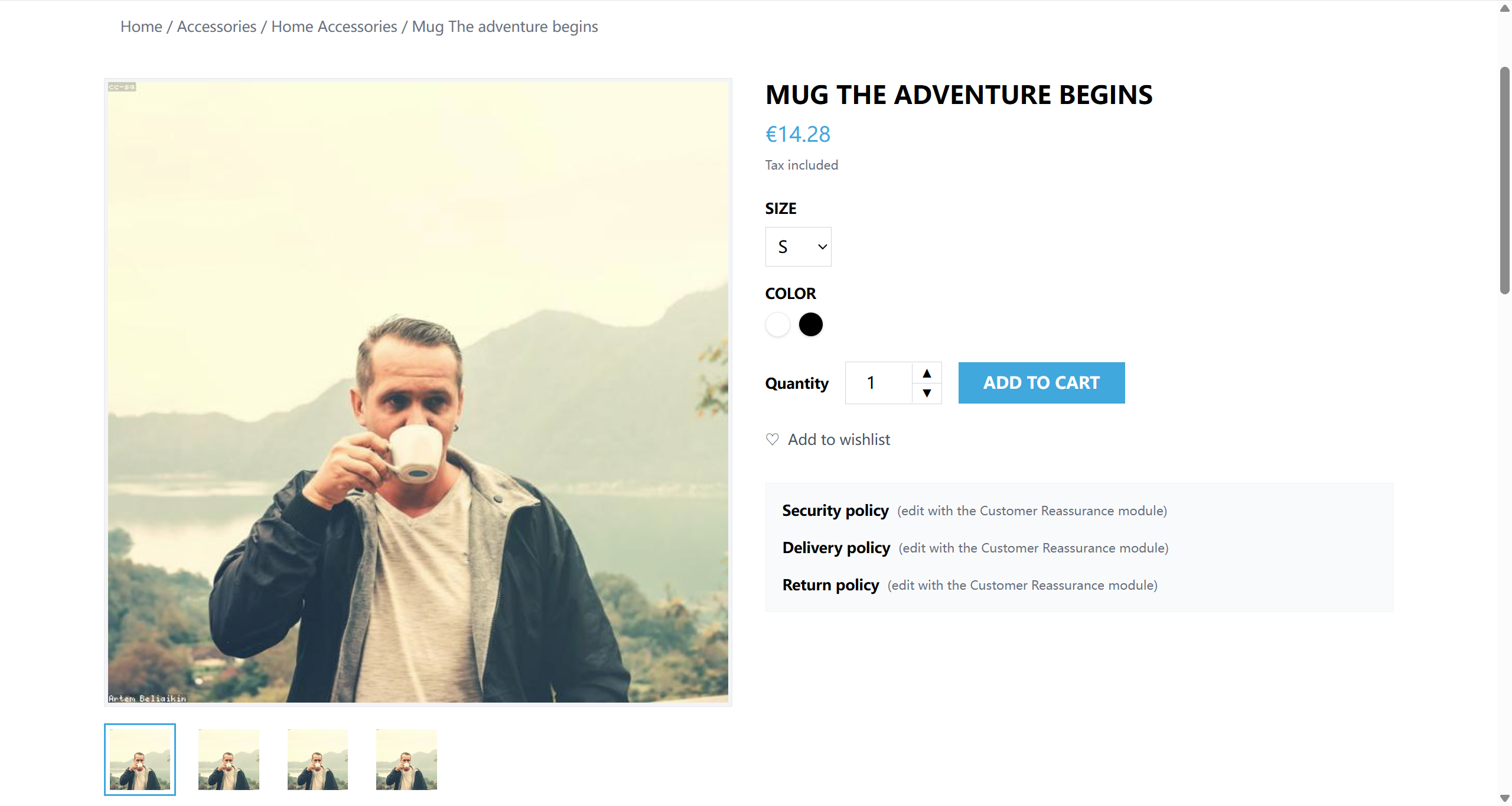}
    \includegraphics[width=0.32\textwidth]{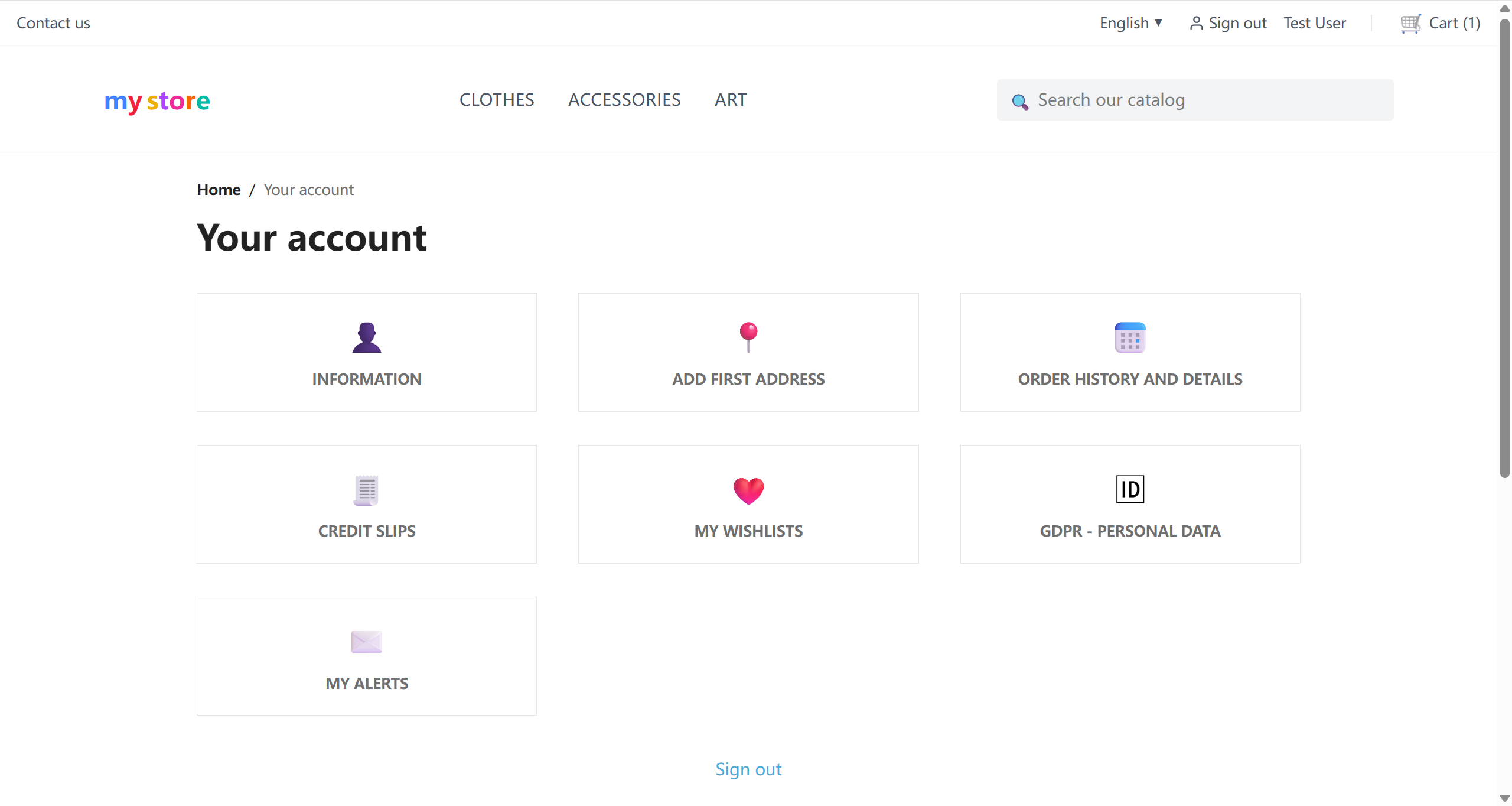}
    \caption{Screenshots of the generated PrestaShop system.}
    \label{fig:prestashop-screenshots}
\end{figure}

\subsection{12306}
\begin{figure}[H]
    \centering
    \includegraphics[width=0.32\textwidth]{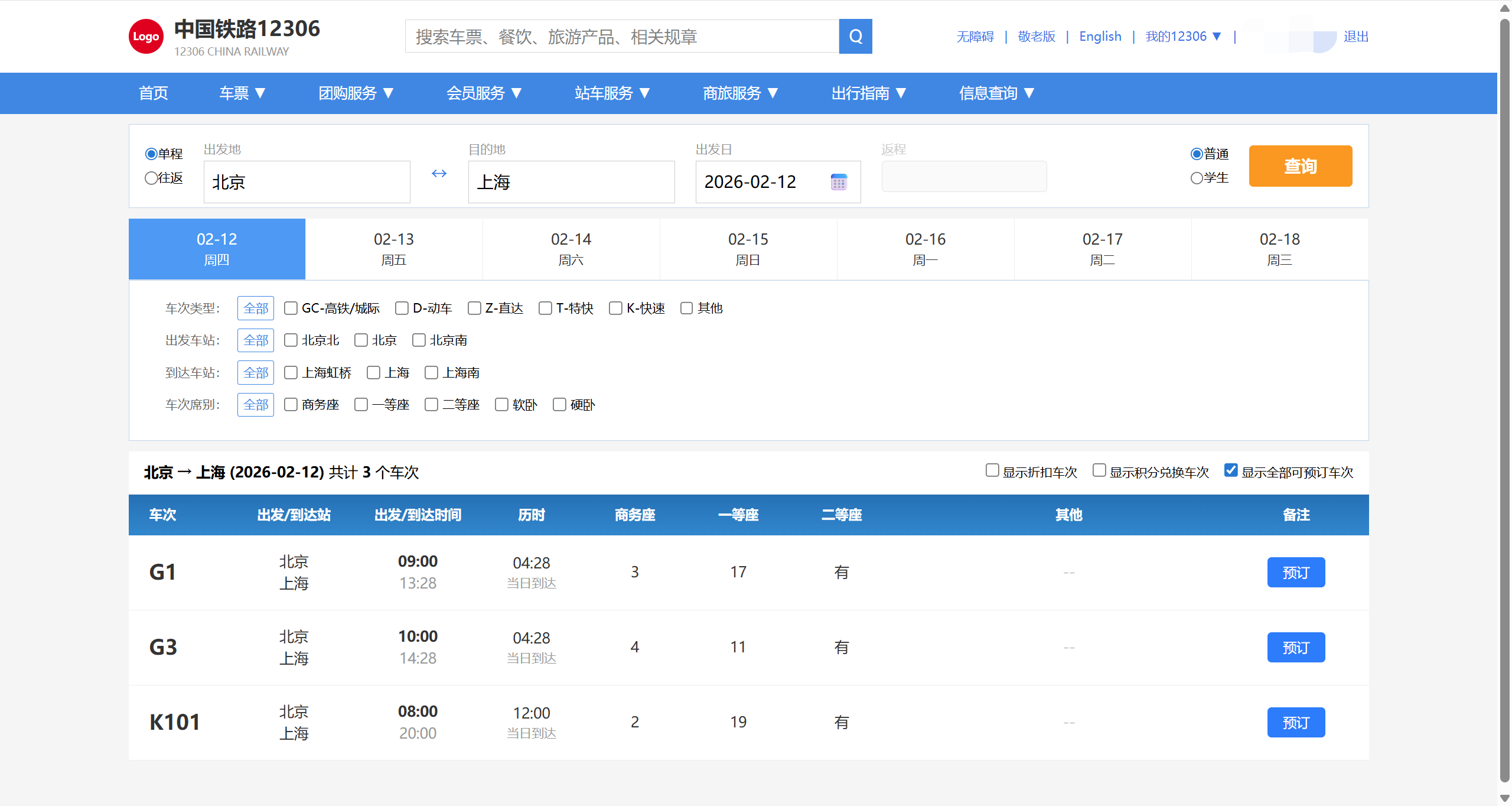}
    \includegraphics[width=0.32\textwidth]{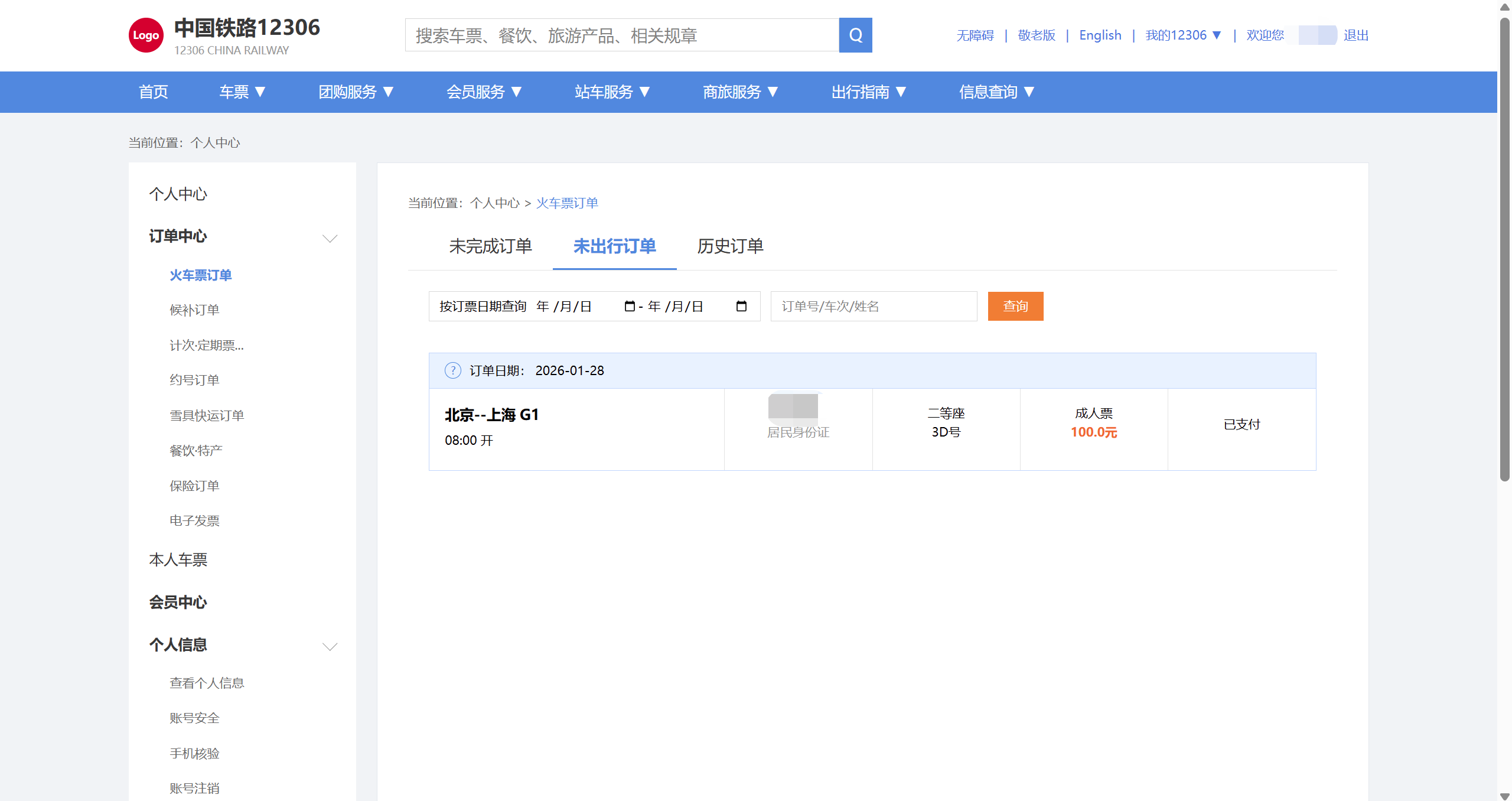}
    \includegraphics[width=0.32\textwidth]{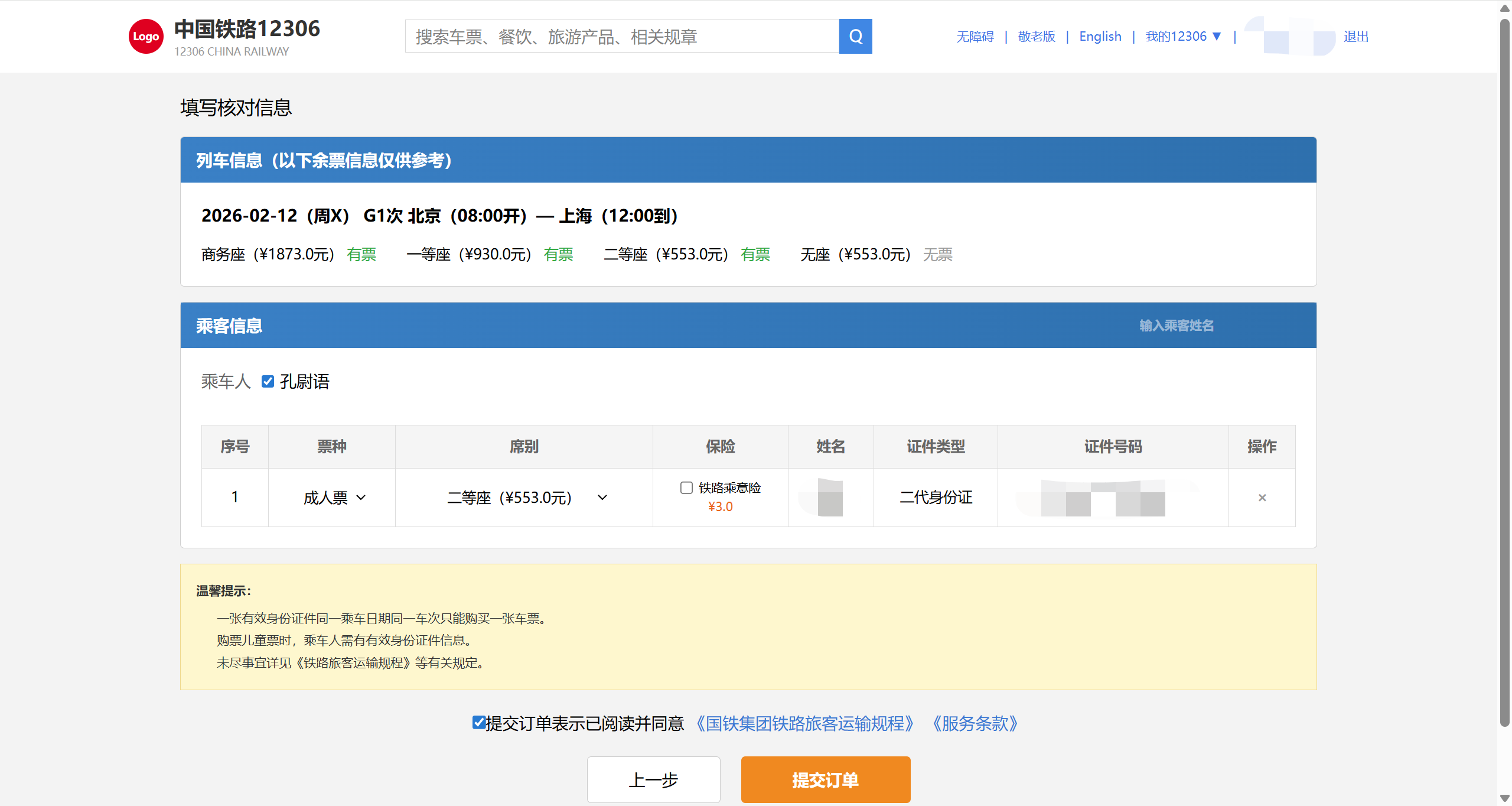}
    \caption{Screenshots of the generated 12306 system.}
    \label{fig:12306-screenshots}
\end{figure}

\subsection{Ctrip}
\begin{figure}[H]
    \centering
    \includegraphics[width=0.32\textwidth]{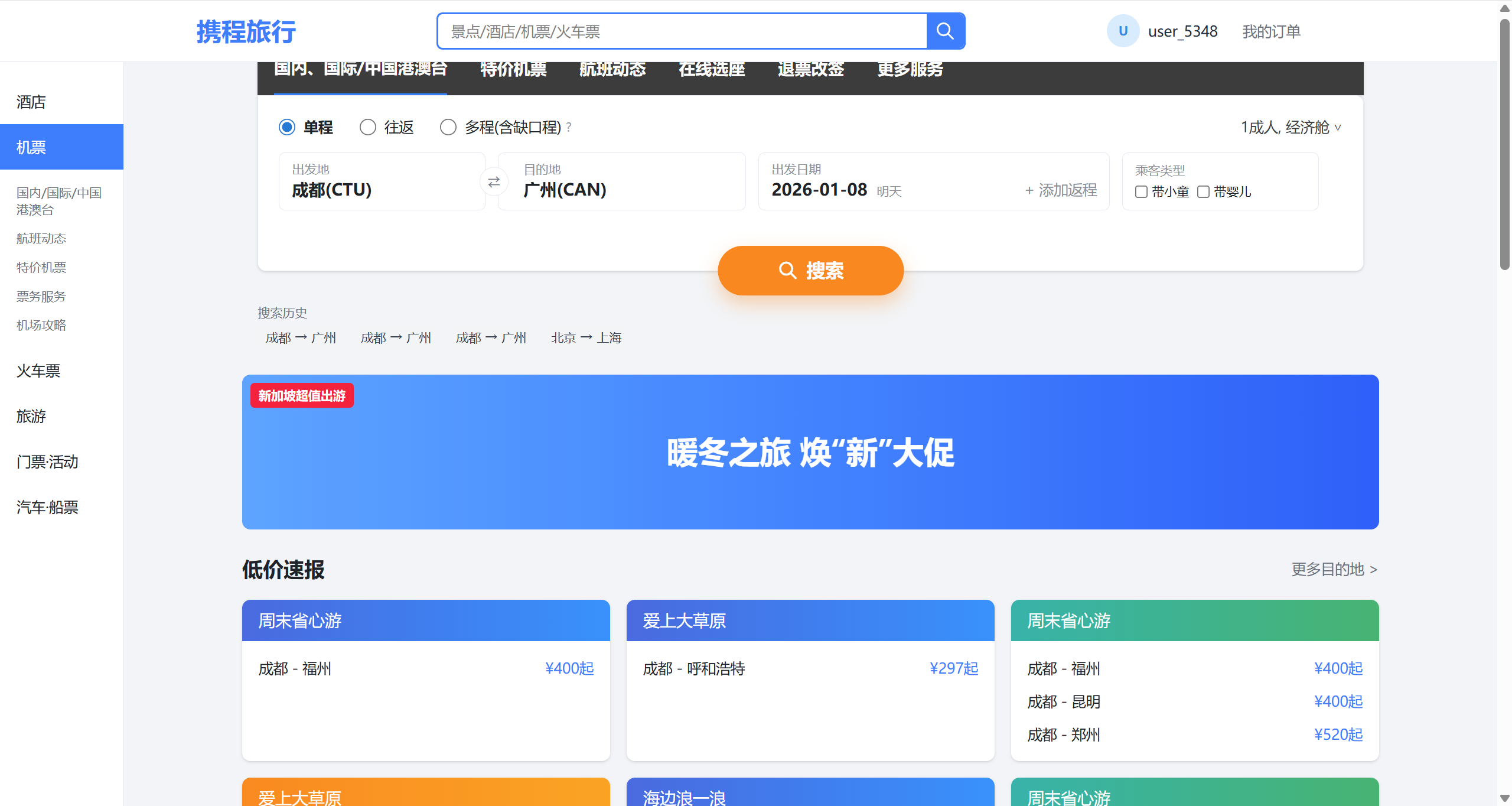}
    \includegraphics[width=0.32\textwidth]{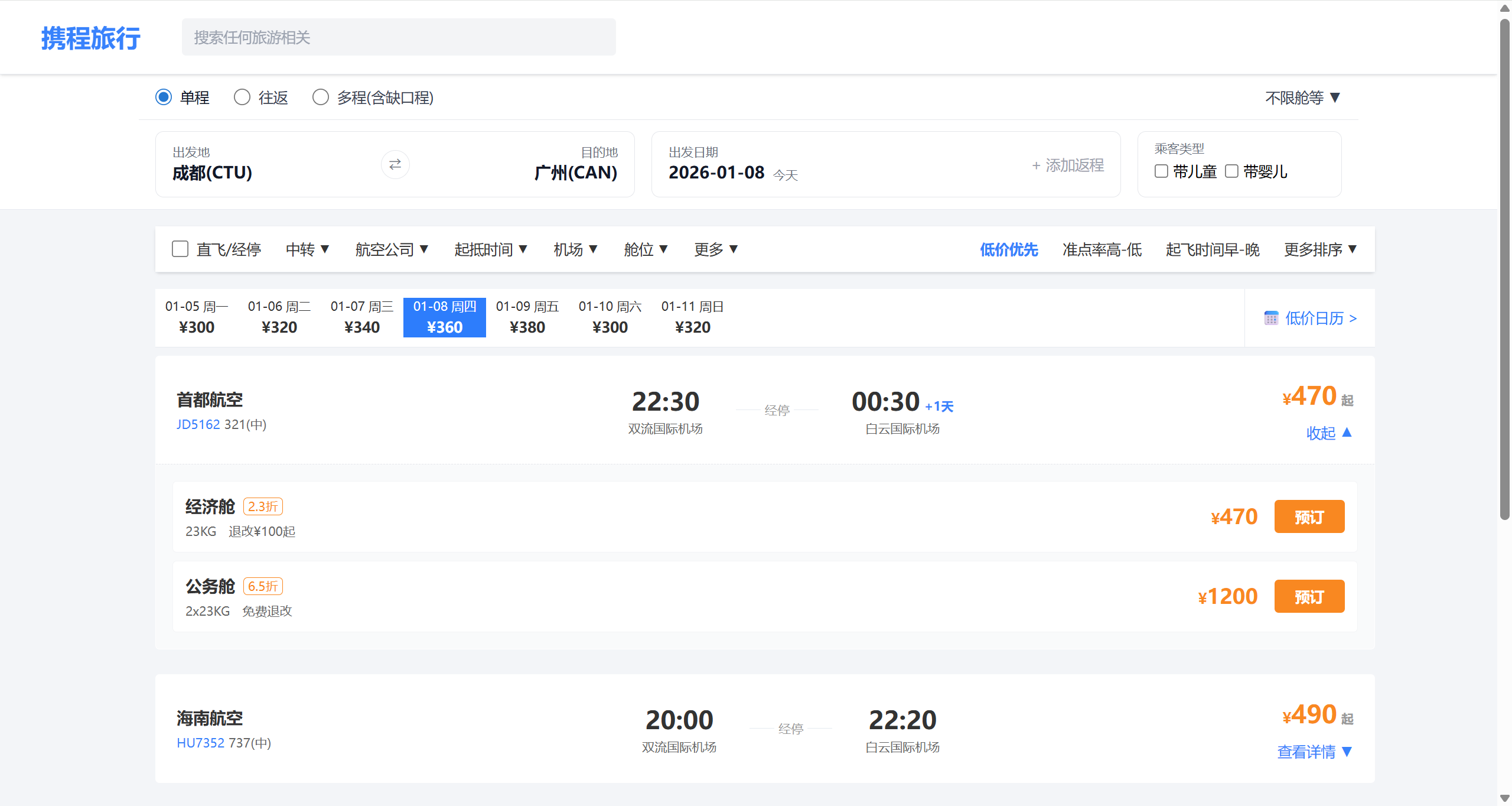}
    \includegraphics[width=0.32\textwidth]{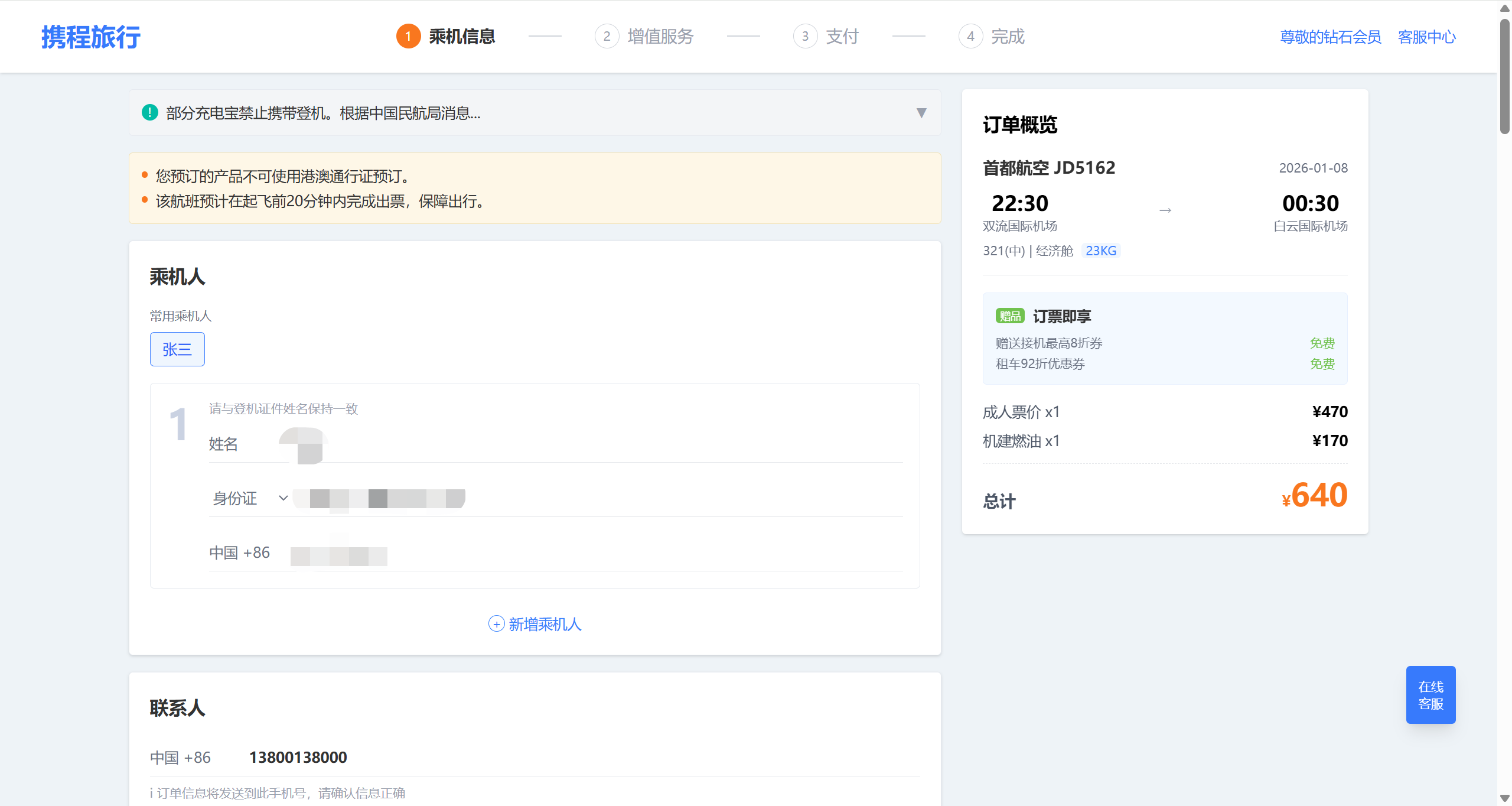}
    \caption{Screenshots of the generated Ctrip system.}
    \label{fig:ctrip-screenshots}
\end{figure}

\end{document}